\documentclass{pinchcr}

\usepackage{amsmath,amssymb,graphicx}
\usepackage{subcaption}
\usepackage[utf8]{inputenc}
\usepackage[english,russian,english]{babel}

\usepackage{multirow} 
\usepackage{fancybox}

\newcommand{\ens}[0]{\ensuremath}
\newcommand{\ii}[0]{\ens{\mathrm{i}}}
\newcommand{\e}[0]{\ens{\mathrm{e}}}
\newcommand{\dd}[0]{\ens{\mathrm{d}}}
\newcommand{\gen}[1]{\ens{\langle #1 \rangle}}
\newcommand{\ket}[1]{\ensuremath{\vert#1\rangle}}
\newcommand{\bra}[1]{\ensuremath{\langle #1|}}
\newcommand{\N}[0]{\ens{\mathbb{N}}}

\usepackage{changes}

\title{Temporal and spectral properties of quantum light}

\author{Birgit Stiller, Ulrich Seyfarth and Gerd Leuchs}
\affiliation{Max-Planck-Institut für die Physik des Lichts, 91058 Erlangen, Germany}

\begin{document}

\maketitle

\acknowledgements

The authors would like to thank Gunnar Björk, Luis Sanchez-Soto and Lev Plimak for helpful discussions and proof reading of parts of the chapter.

\chapter{Introduction}
The modes of the electromagnetic field are solutions of Maxwell's equations taking into account
the material boundary conditions. The field modes of classical optics -- properly normalized -- are also
the mode functions of quantum optics. Quantum physics adds that the excitation within each mode is quantized in close analogy to the harmonic oscillator. A complete set of mode functions forms a basis with which any
new modes can be reconstructed. In full generality each electromagnetic mode function in the four dimensional space-time is
mathematically equivalent to a harmonic oscillator.

The quantization of the electromagnetic field defines the excitation per mode and the correlation between
modes. In classical optics there can be oscillations and stochastic fluctuations of amplitude, phase,
polarization et cetera. In quantum optics there are in addition uncertain quantum  field components, quantum
correlations and quantized energies. This range of topics is discussed in the book by Mandel and Wolf~\cite{Mandel}. Here, we present selected topics from classical to quantum optics.

The spectrum of the light field is determined by its temporal evolution. In quantum optics this corresponds
to the time dependent part of the mode function. The different properties of the light field which can be measured are the frequency, the intensity, as well as phase differences. In particular one can measure the spectral densities
which are associated with these parameters. These different spectral densities are related in a non-trivial way.
Therefore, we start in the second chapter with the classical optics description of a light field and its spectral densities, their
measurement and their interpretation. In the third chapter the quantum properties of a single light mode are reviewed as well as ways to measure these
quantum properties. Gaussian states of a light mode are emphasized, i.\,e. states for which the Wigner function
has a two dimensional Gaussian shape. The fourth chapter will be concerned with more than one mode presenting a
unifying approach to quadratic Hamiltonians including phase conjugation which is related to time reversal.

\chapter{Temporal and spectral properties of classical light}

\section{Classical light fields}\label{sec:classlight}

\begin{figure}[h!]
 \center
 \includegraphics[height=100pt]{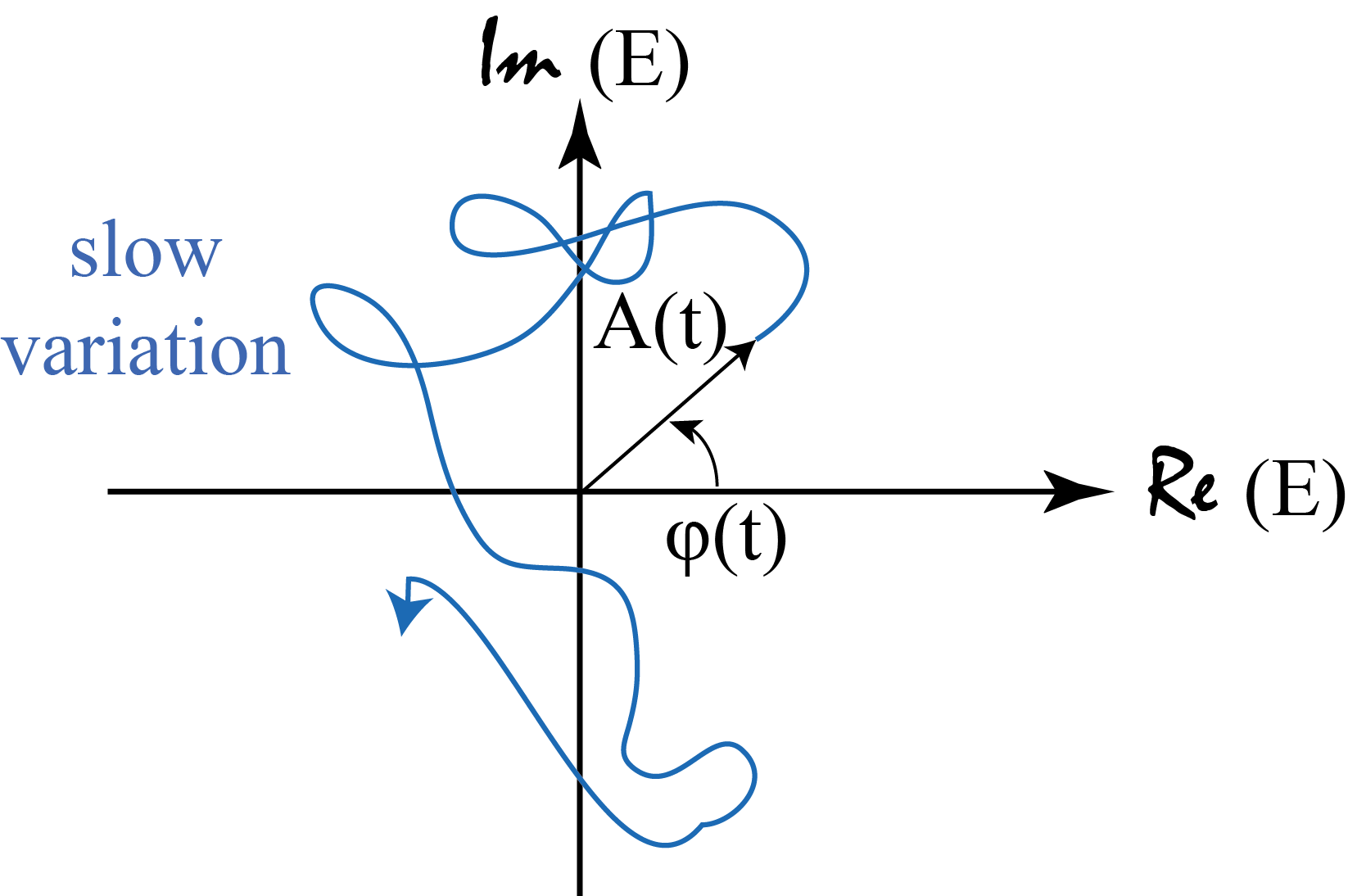}
 \caption{Electric field diagram ($\equiv$ phase space).}\label{img:ElectricField1}
\end{figure}

A spatial mode of a given frequency $\omega$ as defined by Maxwell's equations can be written as
\begin{align}
  \vec{E} ( \vec{r}, t) = \vec{u} (\vec r) \varepsilon(t) 
 \e^{\ii \omega   t} + c.c.,
\end{align}
where $\vec u (\vec r)$ represents the normalized mode function and
$\varepsilon(t) = A(t) \e^{\ii \phi(t)}$ is equivalent to the slowly varying optical field (Fig.~\ref{img:ElectricField1})

\begin{figure}[h!]
 \center
 \includegraphics[height=200pt]{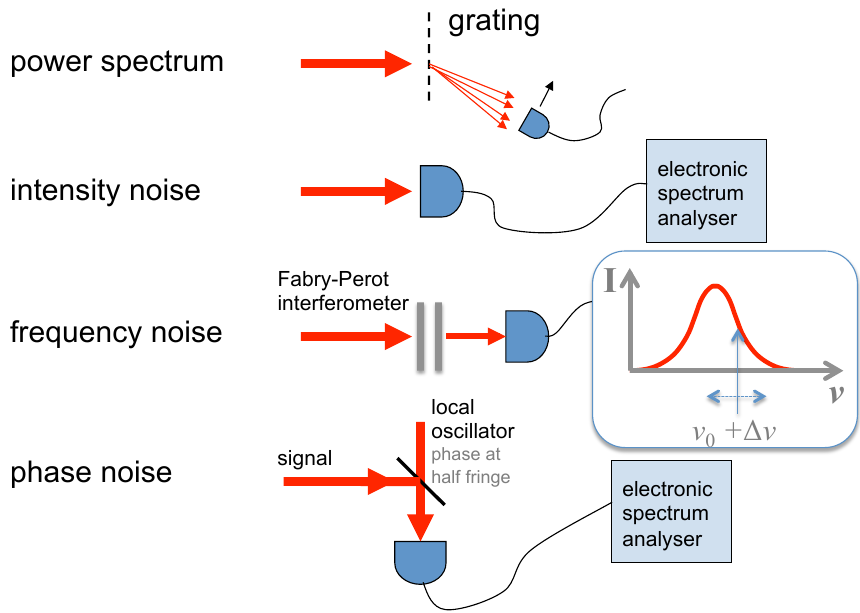}
 \caption{Sketches of experimental setups for measuring different parameters of a light field. Note that the measurement shown for phase noise only works if the phase noise is small enough.}
 \label{img:measurement1}
\end{figure}

Assuming that the light field is stationary, the
outcome of the measurement does not depend on \emph{when} the
measurement is made. It is difficult to measure all aspects of the optical field directly. Different types of measurements
respectively measurement devices are helpful:
 \begin{itemize}
  \item Spectrometer: spectrum $\tilde I(\omega)$,
  \item Fabry-Perot-Interferometer: $(\omega - \omega_0)(t) = \Delta \omega (t)$,
  \item Homodyne measurement: $\Delta \varphi (t)$,
  \item Direct detection $I(t) = (c \varepsilon \varepsilon_0) E^* (t) E(t)$, where $A(t) = \sqrt{ I(t) }$.
 \end{itemize}
The possible measurements on a monochromatic field are summarized in Fig.~\ref{img:measurement1}.
The quantity $\tilde I(\omega)$ measured in a spectrometer can be interpreted as a result of interference.
For one particular frequency $\omega$ at the output one finds
\begin{align}
 \tilde I(\omega) \sim \tilde E^*(\omega) \tilde E(\omega).
\end{align}
\noindent It is worth noting that $\tilde E^*(\omega) \tilde E(\omega)$ is not the Fourier transform of $I(t) = \vert E(t)\vert^2$.

$\tilde I(\omega)$ is closely related to the spectral density of the electric field:

\begin{align}
 S_E (\omega) \sim
  \int\limits_{-\infty}^\infty E^* (t) \e^{-\ii \omega t} \dd t
  \int\limits_{-\infty}^\infty
  E(t') \e^{\ii \omega t'} \dd t'
\end{align}
However, in an experiment the Fourier transform will always be based on a time series of finite length. Thus we rewrite

\begin{align}
  S_E (\omega) = & \lim_{T\rightarrow \infty} \frac{1}{2T}
  \int\limits_{-T}^T E^* (t) \e^{-\ii \omega t} \dd t
  \int\limits_{-T}^T
  E(t') \e^{\ii \omega t'} \dd t' \nonumber\\
  = & \lim_{T\rightarrow \infty} \frac{1}{2T} \iint\limits_{2T} E^*(t) E(t') \e^{\ii \omega (t'-t)} \dd t \dd t'\nonumber\\
  =& \int\limits_{-\infty}^{\infty} \left[ \lim_{T\rightarrow \infty}
  \frac{1}{2T} \int\limits_{-T}^T \dd t \frac{1}{2} \{ E^* (t) E(t+\tau) +E^* (t+\tau)
  E(t) \} \right] \e^{\ii \omega \tau} \dd \tau.\label{eqn:1.4}
\end{align}
Note that $S_E(\omega)$ is real as is $\tilde I(\omega)$. The last step is achieved by enforcing the symmetry between $t$ and $t'$, i.\,e. by averaging over the substitutions
$t^\prime \rightarrow t + \tau$ and $t \rightarrow t^\prime + \tau$. By using 
$\gen{E^*(t) E(t+\tau)} \equiv \lim_{T\rightarrow \infty} \frac{1}{2T} \int_{-T}^T \dd t E^*(t) E(t+\tau)$
we can write
\begin{align}
  S_E (\omega) = \mathcal{R} \left\{ \int\limits_{-\infty}^{\infty} \langle
  E^* (t) E(t+\tau) \rangle \e^{\ii \omega \tau} \dd \tau \right\},
\end{align}
which is the \emph{Wiener-Khinchin theorem} (\foreignlanguage{russian}{Хинчин}). The quantity $\langle E^* (t) E(t+\tau) \rangle$ can be
measured by a Michelson interferometer. The spectrum $S_E(\omega)$ is obtained by Fourier transformation of the field correlation
function $\langle E^* (t) E(t+\tau) \rangle$, which itself is obtained by moving one of the interferometer mirrors. Such a spectrometer is thus
called a \emph{Fourier transform spectrometer}.

\section{The spectrum of intensity noise}
\begin{figure}[h!]
 \center
 \includegraphics[height=70pt]{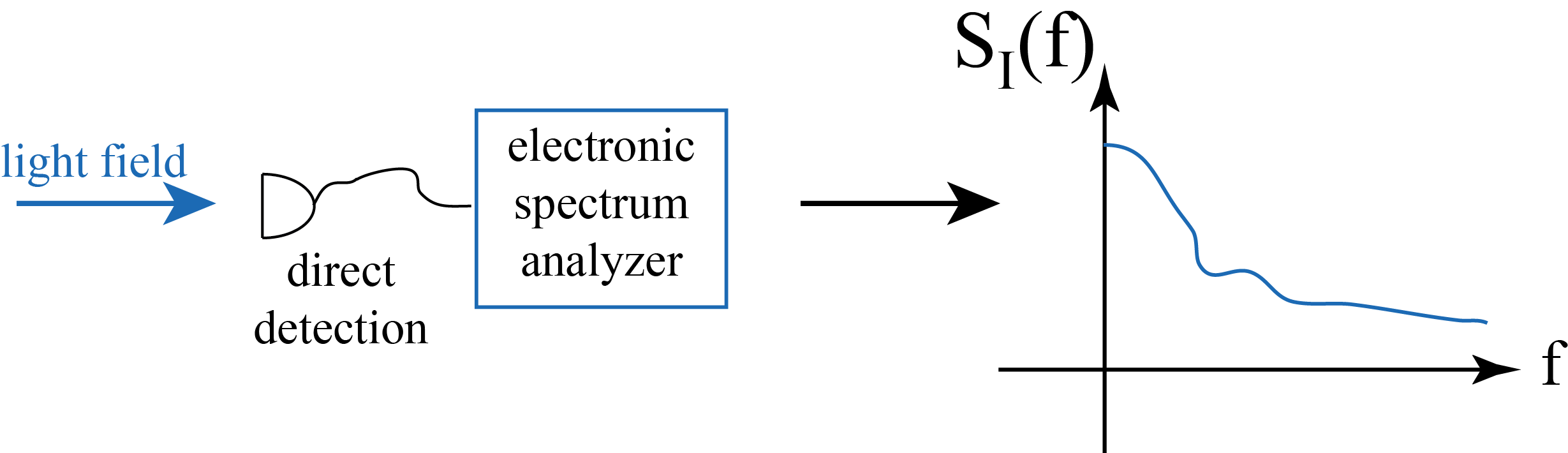}
 \caption{Direct detection scheme.}\label{img:measurement2}
\end{figure}

Taking the Fourier transform of $I(t)$ directly gives a different quantity. It corresponds to the experiment
shown in Fig.~\ref{img:measurement2}.
with
\begin{align}
 S_I(f) = \int\limits_{-\infty}^{\infty} \underbrace{\langle I(t) I(t+\tau) \rangle}_{\equiv G^{(2)} (\tau)} \cos ( 2\pi f \tau) \dd \tau,
\end{align}
where the parameter $f$ indicates a low frequency in the radio frequency range, see also~\cite{Fabre}. The correlation function of a real valued function is an even function of time.
Hence, $\exp(\ii 2 \pi f \tau)$ can be replaced by $\cos(2 \pi f \tau)$.
The spectral density $S_I(f)$ is given by $(\tilde I (f) )^2$, much like $S_E(\omega)$ 
is given by $\vert \tilde E (\omega) \vert^2$. $S_I(f)$ is closely related to the amplitude spectral density defined correspondingly
\begin{align}
 S_A(f) = \int\limits_{-\infty}^{\infty} \langle A(t) A(t+\tau) \rangle \cos ( 2\pi f \tau) \dd \tau.
\end{align}
Yet another quantity is the phase spectral density
\begin{align}
 S_{\varphi}(f) = \int\limits_{-\infty}^{\infty} \langle \varphi(t) \varphi(t+\tau) \rangle \cos ( 2\pi f \tau) \dd \tau.
\end{align}
The spectral density of the frequency noise is closely related by $\delta \nu = \frac{1}{2\pi} \dd \varphi / \dd t$ as
\begin{align}
 S_{\nu}(f) = f^2 S_{\varphi} (f)
\end{align}
and by using $\Omega = 2 \pi f$ we find

\begin{align}
 S_{\nu} (\Omega) = \left( \frac{\Omega}{2\pi}\right)^2 S_{\varphi}(\Omega).
\end{align}

A relation between $S_A(f)$ and $S_I(f)$ is obtained in a straightforward manner, as
%% offtopic 2
\begin{align}
 \gen{I(t) I(t+\tau)} =& \gen{A(t) A(t) A(t+\tau) A(t+\tau)}\nonumber\\
 =&\gen{A(t) A(t)} \gen{A(t+\tau) A(t+\tau)} + 2 \gen{ A(t) A(t+\tau) }^2\nonumber\\
 =& \gen{I }^2 + 2 \gen{ A(t) A(t+\tau) }^2,\nonumber
\end{align}
where the second step assumes Gaussian statistics relating the fourth moment to products of second moments
($\gen{ \sigma_1 \sigma_2 \sigma_3 \sigma_4} = \gen{\sigma_1 \sigma_2}\gen{\sigma_3 \sigma_4} + \gen{\sigma_1 \sigma_3} \gen{\sigma_2\sigma_4} + \gen{\sigma_1 \sigma_4} \gen{\sigma_2 \sigma_3}$).
Hence
\begin{align}\label{eqn:1:15}
 \langle A(t) A(t+\tau) \rangle =& \sqrt{ \frac{\langle I(t) I(t+\tau) \rangle - \langle I \rangle^2}{2}}.
\end{align}

\section{Absorption by an atom}\label{sec:absat}
\begin{figure}[b!]
 \center
 \includegraphics[height=80pt]{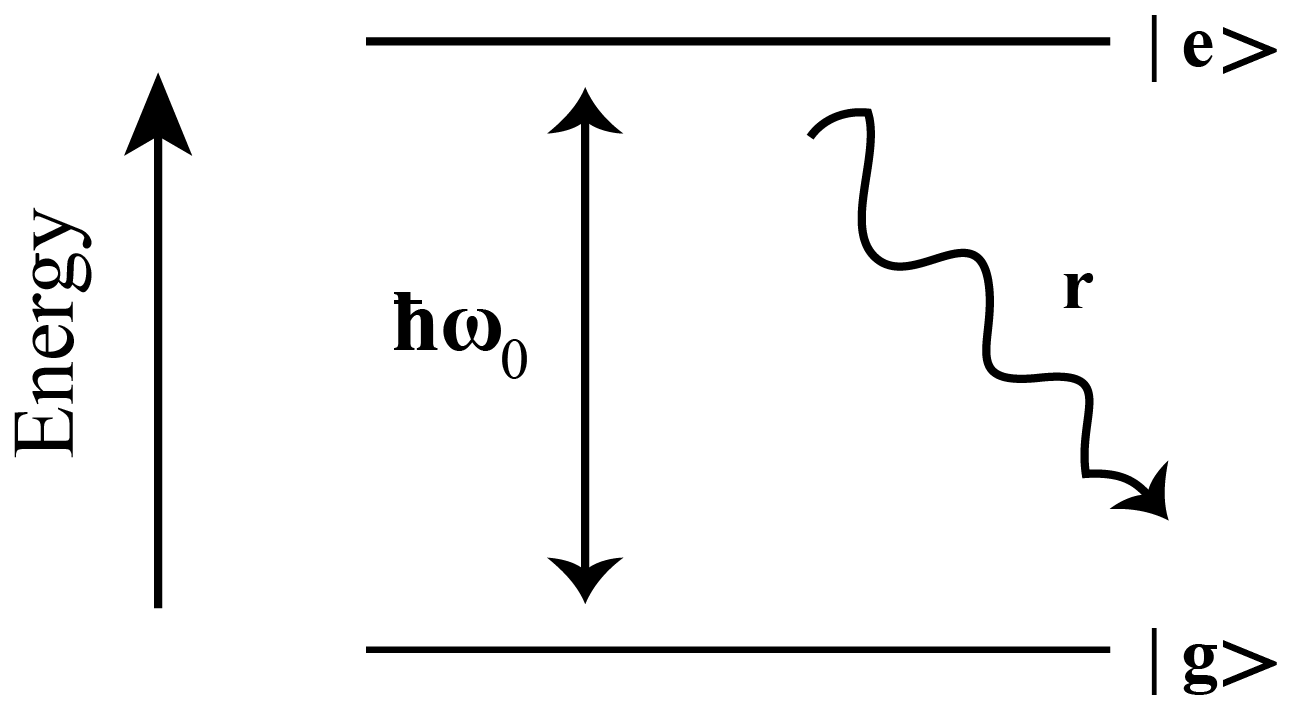}
 \caption{The diagram shows the energy level diagram of a two-level atom, denoted as \emph{ground state} $\ket{g}$ and \emph{excited state} $\ket{e}$, respectively.
 The difference between both levels is given by $\hbar \omega_0$ with the Rabi frequency $\omega_0$.}\label{img:diagram2}
\end{figure}

Here we will show how the rate of absorption by an atom behaves in dependence of the frequency of the field. We
can easily make the right guess, but going through the details will help later. The energy level diagram of the
two level atom is shown in Fig.~\ref{img:diagram2}, defining the notation, using $\ket{e(t)} = \e^{\ii \omega_0 t} \ket{e}$. 
The interaction Hamiltonian of the system is given by
\begin{align}
 V(t) = \hat d E(t),
\end{align}
where $\hat d$ is the operator for the atomic dipole moment $\hat d = e \hat r$. The transition probability is calculated as
\begin{align}
 P_{ge} (t) =& \frac{1}{\hbar^2} \left\vert \int\limits_0^t \e^{-\ii \omega_0 t'} \langle e \vert V(t') \vert g \rangle \dd t' \right\vert^2\nonumber\\
 =& \frac{1}{\hbar^2} \int\limits_0^t \dd t' \e^{-\ii \omega_0 t'} \langle e \vert V(t') \vert g \rangle \int\limits_0^t \dd t'' e^{\ii \omega_0 t''} \langle g \vert V^{\dagger} (t'') \vert e \rangle\nonumber\\
 =& \frac{1}{2\hbar^2} \int\limits_0^t \dd t'\int\limits_{-t'}^{t-t'} \dd \tau \e^{\ii \omega_0 \tau} \langle g \vert V^{\dagger} (t'+\tau) \vert e \rangle \langle e \vert V (t') \vert g \rangle + \left(\text{$t''$ \emph{and} $t'$ \emph{interchanged}}\right)
\end{align}
where $t'' = t' + \tau$ and $\dd t'' = \dd \tau$ (the interchange of $t''$ an $t'$ refers to Eq.~\eqref{eqn:1.4}). In order to account for the decay of $\ket{e}$ one replaces $\e^{\ii \omega_0 \tau}$ by $\e^{\ii \omega_0 \tau - \Gamma \tau}$

For $V(t) = \hat d E(t)$ the transition probability is given by
\begin{align}
 P_{ge} (t) = \frac{\vert \langle e \vert \hat d \vert g \rangle \vert^2}{\hbar^2} \mathcal{R} \left\{ \int\limits_{-t'}^{t-t'} \dd \tau \e^{\ii \omega_0 \tau-\Gamma \tau} \int\limits_0^t \dd t' E^*(t'+\tau) E(t')\right\}.
\end{align}
This translates to a transition rate of $R_{ge} (t) = \frac{\dd}{\dd t} P_{ge}(t)$
\begin{align}%TODO: check formula, mistake in slides
 R_{ge} (t) = \frac{\vert \langle e \vert \hat d \vert g \rangle \vert^2}{\hbar^2} \mathcal{R} \left\{ \int\limits_{0}^t \dd \tau \e^{\ii \omega_0 \tau - \Gamma \tau} \langle E^*(t+\tau) E(t) \rangle \right\}
\end{align}
where $\Gamma$ denotes the decay rate of the excited state. With $b$ we describe the bandwidth of the field
correlation function. We can consider two cases:
\begin{itemize}
 \item case 1: $\Gamma \ll b$, bandwidth of the optical field $\Rightarrow \Gamma \cong 0$\\
  $\Rightarrow R_{ge}$ is the Fourier transformation of the first order correlation function, in other words $R_{ge}$ is proportional to the spectrum (see Section \ref{sec:classlight}).
 \item case 2: $\Gamma \gg b$, the term $\e^{-\Gamma \tau}$ can be replaced by $\delta (\tau)$\\
  $\Rightarrow R_{ge} = \frac{\vert \langle e \vert \hat d \vert g \rangle \vert^2}{\hbar^2} \cdot \langle \vert E(t)\vert^2 \rangle \sim \langle I(t) \rangle$.
\end{itemize}

\section{Relationship between the different spectral densities}

The general relationship is obtained starting from the spectral density of the electric field

\begin{align}
 S_E (\omega) = \mathcal{R} \left\{ \int\limits_{-\infty}^{\infty} \langle E^*(t) E(t+\tau) \rangle \e^{\ii \omega \tau} \dd \tau \right\},
\end{align}
by inserting $E(t) = A(t) \e^{\ii \varphi (t)} \e^{-\ii \omega_0 t}$ with the amplitude $A(t)$, the phase $\varphi(t)$, the frequency $\nu_0 = \omega_0 / 2\pi$ and the
\emph{frequency excursion} $\Delta \nu(t) = \dot \varphi (t)$ leads to 
\begin{align}
 S_E(\omega) =& \mathcal{R} \left\{\int\limits_{-\infty}^{\infty} \dd \tau \langle A(t+\tau) \e^{\ii \varphi (t+\tau) - \ii \omega_0 (t+\tau)} A(t) \e^{-\ii \varphi (t) + \ii \omega_0 t} \rangle \e^{\ii \omega \tau}\right\}\nonumber\\
 =& \mathcal{R} \left\{\int\limits_{-\infty}^{\infty} \dd \tau \langle A(t+\tau) A(t) \e^{\ii (\varphi (t+\tau) - \varphi(t))} \rangle \e^{\ii (\omega-\omega_0) \tau}\right\}.\nonumber
\end{align}
Assuming no correlation between amplitude and phase leads to
\begin{align}
 S_E (\omega) =& \mathcal{R} \left\{ \int\limits_{-\infty}^{\infty} \dd \tau \left\langle A(t+\tau) A(t)\right\rangle \left\langle \e^{\ii (\varphi (t+\tau) -\varphi(t))} \right\rangle \e^{\ii (\omega-\omega_0) \tau}\label{eqn:SE}\right\},
\end{align}
where the averaging has been factorized. The second averaging term can be expanded, taking into account that
\begin{align}
 \e^x = 1 + x+ \frac{1}{2!} x^2 + \frac{1}{3!} x^3 + \frac{1}{4!} x^4 + \ldots
\end{align}
and with the averages of the odd order terms being zero
\begin{align}
 \langle \e^{\ii\Delta \varphi (t)} \rangle = 1 - \frac{1}{2} \langle (\Delta \varphi)^2 \rangle + \frac{1}{4!} \langle (\Delta \varphi)^4 \rangle - \frac{1}{6!} \langle (\Delta \varphi)^6 \rangle + \ldots.
\end{align}
For Gaussian variables (multivariate normal distributions see above) with
\begin{align}
 \gen{\varphi_1 \varphi_2 \varphi_3 \varphi_4} = \gen{\varphi_1 \varphi_2} \gen{\varphi_3 \varphi_4} + \gen{\varphi_1 \varphi_3} \gen{\varphi_2 \varphi_4} + \gen{\varphi_1 \varphi_4} \gen{\varphi_2 \varphi_3}
\end{align}
one obtains
\begin{align}
 \gen{\varphi^4} = 3 \gen{\varphi^2}^2\label{eqn:phi4}
\end{align}
or in general
\begin{align}
 \gen{x^n} =& N \int\limits_{-\infty}^{\infty} x^2 \e^{-ax^2} \dd x\\
  =& (n-1) \gen{x^2} \gen{x^{n-2}} \Rightarrow \gen{x^{2n}} = \frac{(2n-1)!}{2^{n-1} (n-1)!} \gen{x^2}^n,\label{eqn:xn}
\end{align}
thus

\begin{align}
 \left\langle \e^{\ii\Delta \varphi (\tau)} \right\rangle =& 1 - \frac{1}{2} \langle (\Delta \varphi)^2 \rangle + \frac{3}{4!} \langle (\Delta \varphi)^2 \rangle^2 - \frac{3\cdot 5}{6!} \langle (\Delta \varphi)^2 \rangle^3 + \ldots\\
 =& 1 + \left( -\frac{1}{2} \gen{(\Delta \varphi)^2}\right) + \frac{1}{2} \left( -\frac{1}{2} \gen{(\Delta \varphi)^2}\right)^2 + \frac{1}{3!} \left( -\frac{1}{2} \gen{(\Delta \varphi)^2}\right)^3+\ldots\\
 =& \e^{-\frac{1}{2} \gen{(\Delta \varphi)^2}} = \e^{-\frac{1}{2} \gen{\left( \varphi(t+\tau)-\varphi (t) \right)^2}}.\label{eqn:av}
\end{align}
Therefore, if higher order phase moments are Gaussian, the second averaged factor in Eq.~\eqref{eqn:SE} can be replaced by Eq.~\eqref{eqn:av},
$\frac{1}{2} \gen{\left( \varphi(t+\tau)-\varphi (t) \right)^2} = \gen{(\varphi(t))^2} - \gen{\varphi(t+\tau) \varphi(t)}$ with 
\begin{align}
 \gen{\varphi(t+\tau) \varphi(t)} = \int \dd \Omega S_{\varphi}(\Omega) \cos \Omega \tau.
\end{align}
This yields
\begin{align}
 S_E(\omega) = \mathcal{R} \left\{ \int\limits_{-\infty}^{\infty} \dd \tau \gen{A(t+\tau) A(t)} \e^{-\int \dd \Omega S_{\varphi}(\Omega) (1-\cos \Omega \tau)} \e^{\ii(\omega-\omega_0) \tau} \right\}.
\end{align}

Next we relate $\gen{A(t+\tau) A(t)}$ to $S_I(\Omega)$ using
\begin{align}
 \gen{I(t+\tau) I(t)} = \int \limits_0^{\infty} \dd \Omega S_I(\Omega) \cos (\Omega \tau),
\end{align}
and if higher amplitude moments are again Gaussian we can use Eq.~\eqref{eqn:1:15} to obtain 
\begin{align}
 \gen{A(t+\tau) A(t)} = \frac{1}{\sqrt{2}} \sqrt{ \int\limits_0^{\infty} \dd \Omega S_I (\Omega) \left(\cos (\Omega \tau) - \gen{I}^2\right)}.
\end{align}
Thus, we find
\begin{equation}
 \boxed{S_E(\omega) = 
 \mathcal{R} \left\{ 
 \int\limits_{-\infty}^{\infty} \dd \tau \frac{1}{\sqrt{2}} 
 \left(\sqrt{\int\limits_0^{\infty} \dd \Omega S_I (\Omega) \left(\cos (\Omega \tau) -\gen{I}^2\right)}\;\right) 
 \e^{-\int \dd \Omega S_{\varphi} (\Omega)(1-\cos \Omega \tau)} \e^{\ii(\omega-\omega_0)\tau} 
 \right\}
 }
\end{equation}

\subsection{Example: $\delta$ correlated amplitude}
Lets consider an example with constant phase and $\delta$-\emph{correlated} amplitudes, namely with
$\gen{A(t+\tau) A(t)} = S_A(0) \delta(\tau)$ and $\varphi(t+\tau) - \varphi(t) = 0$, 
it follows
\begin{align}
 S_E(\omega) =& \mathcal{R} \left\{ \int\limits_{-\infty}^{\infty} \dd \tau S_A (0) \delta(\tau) \e^{\ii (\omega -\omega_0) \tau} \right\} = S_A(0),
\end{align}
which describes a \emph{white spectrum}.

\subsection{Example: white frequency noise}
If the phase diffuses, the frequency $\nu(t)$ is $\delta$-correlated as
\begin{align}
 \gen{\nu(t) \nu(t+\tau)} = \delta (\tau) S_{\nu}(0).
\end{align}
Thus, $S_\nu(f)= \int\limits_{-\infty}^{\infty} \dd \tau \gen{ \nu(t) \nu(t+\tau)} \cos ( 2\pi f \tau) = S_\nu(0)$ is a constant. The phase
spectral density is thus
\begin{align}
 S_{\varphi}(\Omega) = 2\pi \frac{S_{\nu} (f)}{\Omega^2} \rightarrow 2\pi \frac{S_{\nu}(0)}{\Omega^2}.
\end{align}
This leads to
\begin{align}
 S_E(\omega) = \mathcal{R} \left\{ \int\limits_{-\infty}^{\infty} \dd \tau A^2 \e^{-\int \dd \Omega \frac{2\pi S_{\nu}(0)}{\Omega^2}(1-\cos \Omega \tau)} \e^{\ii(\omega-\omega_0)\tau}\right\}\label{eqn:SEfn}
\end{align}
We first evaluate the integral in the exponent
\begin{align}
 \int\limits_0^{\infty} \dd \Omega \frac{S_{\nu}(0)}{\Omega^2} (1-\cos \Omega \tau) =& S_{\nu}(0) \int\limits_0^{\infty} \dd \Omega \frac{1}{\Omega^2} (1-\cos \Omega \tau)\\
  =& S_{\nu}(0) \left(\int\limits_0^{\infty} \dd \Omega \frac{1}{\Omega^2} - \int\limits_0^{\infty} \dd \Omega \frac{1}{\Omega^2} \cos \Omega \tau\right).
\end{align}
Substituting $u' = 1/\Omega^2$ and $v=\cos \Omega \tau$, with $u=-1/\Omega$ and $v'=-\tau \sin (\Omega \tau)$. By partial integration one finds:
\begin{align}
 \int\limits_0^{\infty} \dd \Omega \frac{S_{\nu}(f)}{\Omega^2} (1-\cos \Omega \tau)
    =& S_{\nu}(0) \left( \left[ -\frac{1}{\Omega} + \frac{1}{\Omega} \cos (\Omega \tau) \right]_{0}^{\infty} + \int\limits_0^{\infty} \dd \Omega \frac{\sin (\Omega \tau)}{\Omega} \tau \right)\\
    =& \frac{\pi}{2} \tau S_{\nu}(0),
\end{align}
where the result is obtained by recognizing $\int_{0}^{\infty} \dd x \frac{\sin x}{x} = \pi/2$. Thus, the spectral density is given by
\begin{align}
 S_E(\omega) =& \mathcal{R} \left\{\int\limits_{-\infty}^{\infty} \dd \tau A^2 \e^{-\pi^2 S_{\nu}(0) \tau} \e^{\ii (\omega - \omega_0) \tau}\right\}\\
 =& \mathcal{R} \left\{ A^2 \frac{1}{-\ii (\omega - \omega_0) + \pi^2 S_{\nu}(0)}\right\}\\
 =& \frac{\frac{1}{4} S_{\nu}(0)}{({\nu}-{\nu}_0)^2 + \left( \frac{\pi}{2} S_{\nu}(0) \right)^2}.
\end{align}
The half width at half maximum is given by
\begin{align}
 \Delta {\nu}_{1/2} = \frac{\pi}{2} S_{\nu} (0),
\end{align}
full width at half maximum by
\begin{align}
 \Delta {\nu} = \pi S_{\nu} (0)\label{eqn:1:69}
\end{align}
(see also Schawlow-Townes line width of the ideal laser).

\subsection{Relation between frequency noise spectrum and spectral linewidth}

\begin{figure}[h!]
 \center
 \includegraphics[height=60pt]{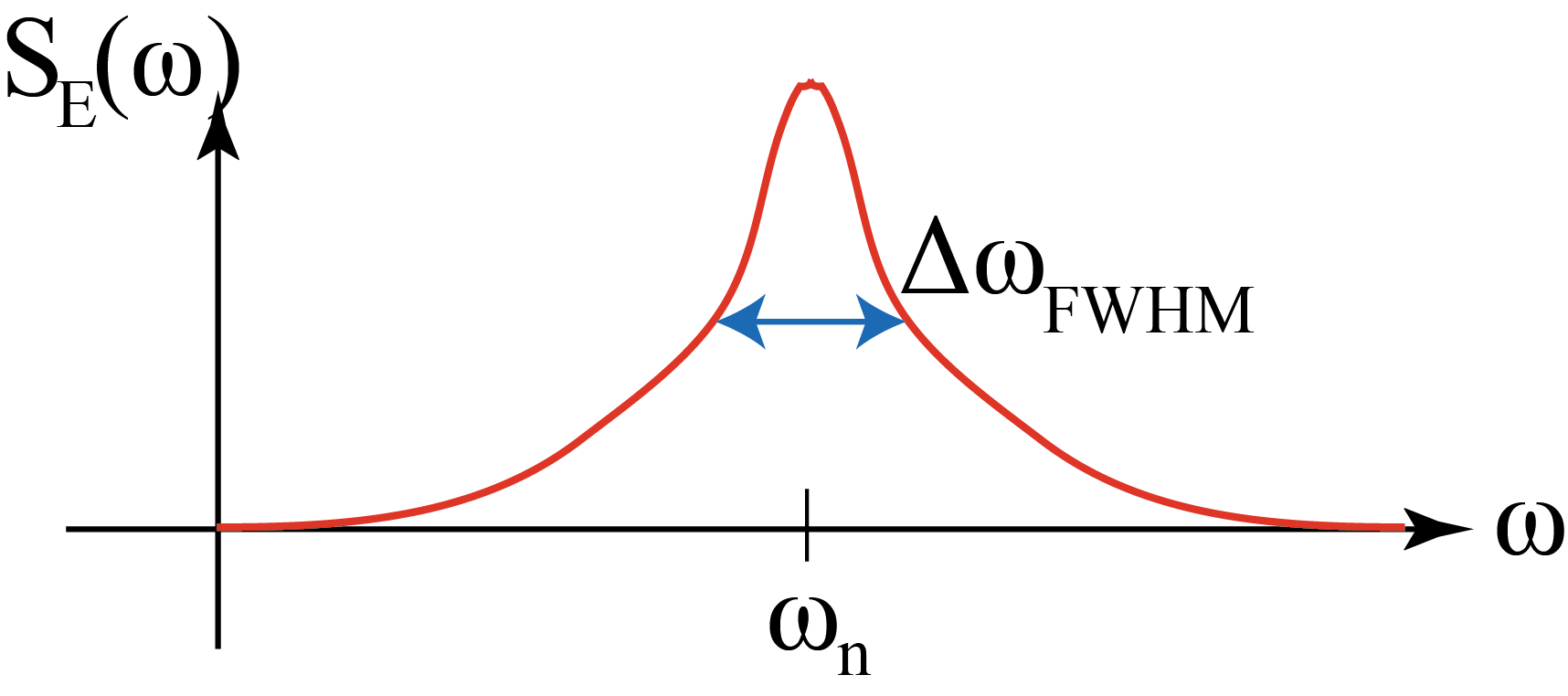}
 \caption{$\gen{\Delta \varphi^2}_{\Omega > \Omega_c}$ changes only the high frequency wings.}\label{img:diagram4}
\end{figure}

We now try to give a physical picture of this relation~\cite{Telle93,Telle96}. 
\begin{figure}[h!]
 \center
 \includegraphics[height=80pt]{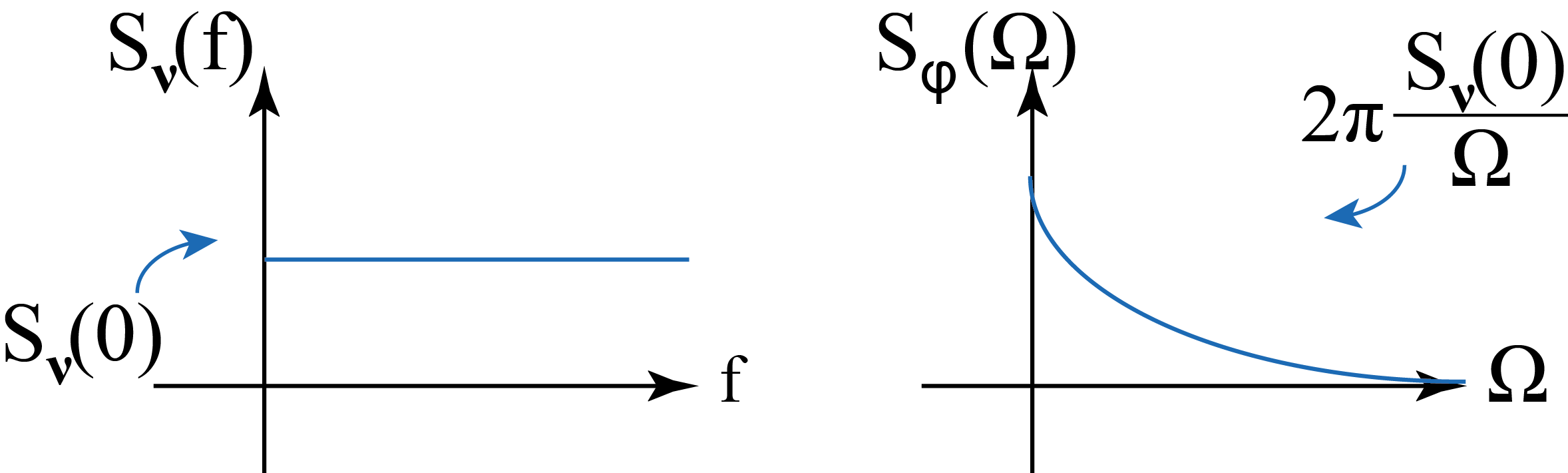}
 \caption{Noise spectrum.}
 \label{fig:diagram3}
\end{figure}
The value of $S_\nu (f)$ in a narrow frequency interval of width $b$ at noise frequency $f$ gives the variance of the frequency
fluctuations in this interval (Fig.~\ref{fig:diagram3}). In order to obtain the total variance one has to integrate $S_\nu (f)$ from zero to infinity.
For $S_\nu(f)$, the value diverges, of course. Therefore, we conclude that from a certain frequency $f_c$ onward (where c stands for cutoff) $S_\nu(f)$ does not contribute anymore to the total variance of the frequency fluctuation. The cutoff can be determined by setting the integral equal to the linewidth derived above (Eq.~\eqref{eqn:1:69}):
\begin{align}
 \int\limits_0^{f_c} S_{\nu} (f) \dd f \overset{!}{=} (\pi S_{\nu} (0))^2
 \end{align}
 It follows
 \begin{align}
 \rightarrow S_{\nu}(0) \cdot f_c = (\pi^2 S_{\nu}(0))^2
  \end{align}
 with
 \begin{align}
  f_c=\pi^2 S_{\nu}(0).
\end{align}
In order to understand the meaning of this cutoff we integrate $S_{\varphi}(\Omega)$ from $\Omega_c = 2\pi f_c$ to infinity.
\begin{align}
 \int\limits_{2\pi^3 S_{\nu} (0)}^{\infty} \dd\Omega\, 2\pi \frac{S_{\nu}(0)}{\Omega^2} = \frac{1}{\pi^2} = \gen{\Delta \varphi^2}_{\Omega > \Omega_c}
\end{align}
This means that the high frequency tail of the spectral density of the phase fluctuations does not contribute to the linewidth if the integrated
variance in this frequency range is significantly less than one radian (Fig. \ref{img:diagram4}). This gives us a recipe for the general case:
choose $f_c$ such that $\sqrt{\gen{\Delta \varphi^2}_{\Omega > \Omega_c}} \cong 1/\pi$ and then integrate $S_{\nu} (f)$ upto $f_c=\Omega_c/2\pi$.
This determines the contribution of the phase fluctuations (see right hand side of Fig.~\ref{fig:diagram3})to the spectral line width.

%%%%%%%%%%%%%%%%%%%%%%%%%%%%%%%%%%%%%%%%%%%%%%%%%%%%%%%%%%%%%%%%%%%%%%%%%%%%%%%%%%%%%%%%%%%%%%%%%%%%%%%%%%%%%%%%%%%%%
%%%%  2nd part 
%%%%%%%%%%%%%%%%%%%%%%%%%%%%%%%%%%%%%%%%%%%%%%%%%%%%%%%%%%%%%%%%%%%%%%%%%%%%%%%%%%%%%%%%%%%%%%%%%%%%%%%%%%%%%%%%%%%%%

\newpage
\section{Correlation functions}
Next we consider general electric field correlation functions
\begin{align}
 \left\langle\prod\limits_{j=1}^n \varepsilon^{(*)}(t_j)\right\rangle,
\end{align}
with $\gen{\ldots}$ denoting a time average, namely $\gen{\ldots} = \lim\limits_{T\rightarrow \infty} \frac{1}{2T}\int\limits_{-T}^T \ldots \dd t$. 
For $n$ odd the correlation functions of harmonic fields vanish, namely $\gen{\ldots} =0$. For $n$ even the non-zero correlation functions have the following general form with
an equal number of conjugated and unconjugated fields:
 \begin{align}
  G^{(n)} (t_1,t_2,\ldots, t_{2n}) \equiv \gen{ E^*(t_1) E^* (t_2) \ldots E^* (t_n) E(t_{n+1}) \ldots E(t_{2n})}
 \end{align}
 The order of the correlation function is given by $n$.\footnote{Some authors use $2n$. E.\,g. $\gen{E(t_2) E^* (t_2)}$ is sometimes called a second order correlation
 function since it involves two field amplitudes. Here we use $n$ in the superscript such that photon antibunching is described by a $G^{(2)}$ function.}
 For $n=1$ we have $\gen{E^*(t_1) E(t_2)} \equiv G^{(1)} (t_2 -t_1)$, where $G^{(1)}$ is second order
 in the electric field and first order in intensity. For $t_1 = t_2$ follows $G^{(1)} (0) = \gen{I(t)}$. The correlation function $G^{(1)} (t_2-t_1)$ determines the signal
 of the output of an interferometer. This was already discussed in Sections \ref{sec:classlight} and \ref{sec:absat}. Next we derive a similar
 relation for a correlation function with $n=2$ which is given by $\gen{E^*(t_1) E^*(t_2) E(t_3) E(t_4)}$
 \begin{paragraph}{1. special case} Choosing $t_1 = t_3$ and $t_2=t_4$ yields the intensity correlation function:
   \begin{align}
    \gen{\vert E(t_1)\vert^2 \vert E (t_2) \vert^2} = \gen{I(t_1) I(t_2)} \equiv G_I^{(2)}.
   \end{align}
The subscript ``I'' refers to the arrangement of a field describing the intensity correlation functions.
 \end{paragraph}
 \begin{paragraph}{2. special case} For two photon absorption we find
   \begin{align}
    V(t) =& \hat d \cdot E(t) = \sum\limits_i \frac{\hat d E(t) \ket{i} \bra{i} \hat d E(t)}{W_g + \hbar \omega_L - W_i}\\
     \sim& E^2(t),
   \end{align}
  where $W_g$ and $W_i$ denote energy of the atomic ground and intermediate state. In analogy to Section \ref{sec:absat}, i.\,e. by replacing $V(t) \sim E(t)$ by $V(t) \sim E^2(t)$, we have~\cite{Leuchs86}
  \begin{align}
   R_{ge} \sim& \int\limits_{0}^t \dd \tau \e^{\ii \omega_0 \tau} \gen{{E^*}^2 (t' + \tau) E^2 (t')}\\
   =& \int\limits_{0}^t \dd \tau \e^{\ii (\omega_0-2\omega_L) \tau} \underbrace{\gen{{E^*}^2 (t' + \tau) E^2 (t')}}_{G^{(2)}_{\mathrm{TP}}(\tau)}.
  \end{align}
\end{paragraph}
The subscript ``TP'' refers to the particular arrangement of the electric fields typical for two photon absorption.
  Thus, the second order correlation function
  \begin{align}
   \gen{{E^*}^2 (t_1) E^2 (t_3)} \equiv G^{(2)}_{\mathrm{TP}} (t_3 - t_1)
  \end{align}
determines a two photon absorption process

In general the $G^{(n)} (t_1, \ldots, t_{2n})$ are complex valued functions. The full set of all $G^{(n)} (t_1, \ldots, t_{2n})$ fully characterizes the field.
Compare: All higher moments fully characterize a distribution.\par

\begin{paragraph}{Example}
The three fields $E(t)$ shown in Fig. \ref{img:diagram5} have the same spectrum, but they are not identical.
\begin{figure}[b!]
 \center
 \includegraphics[height=120pt]{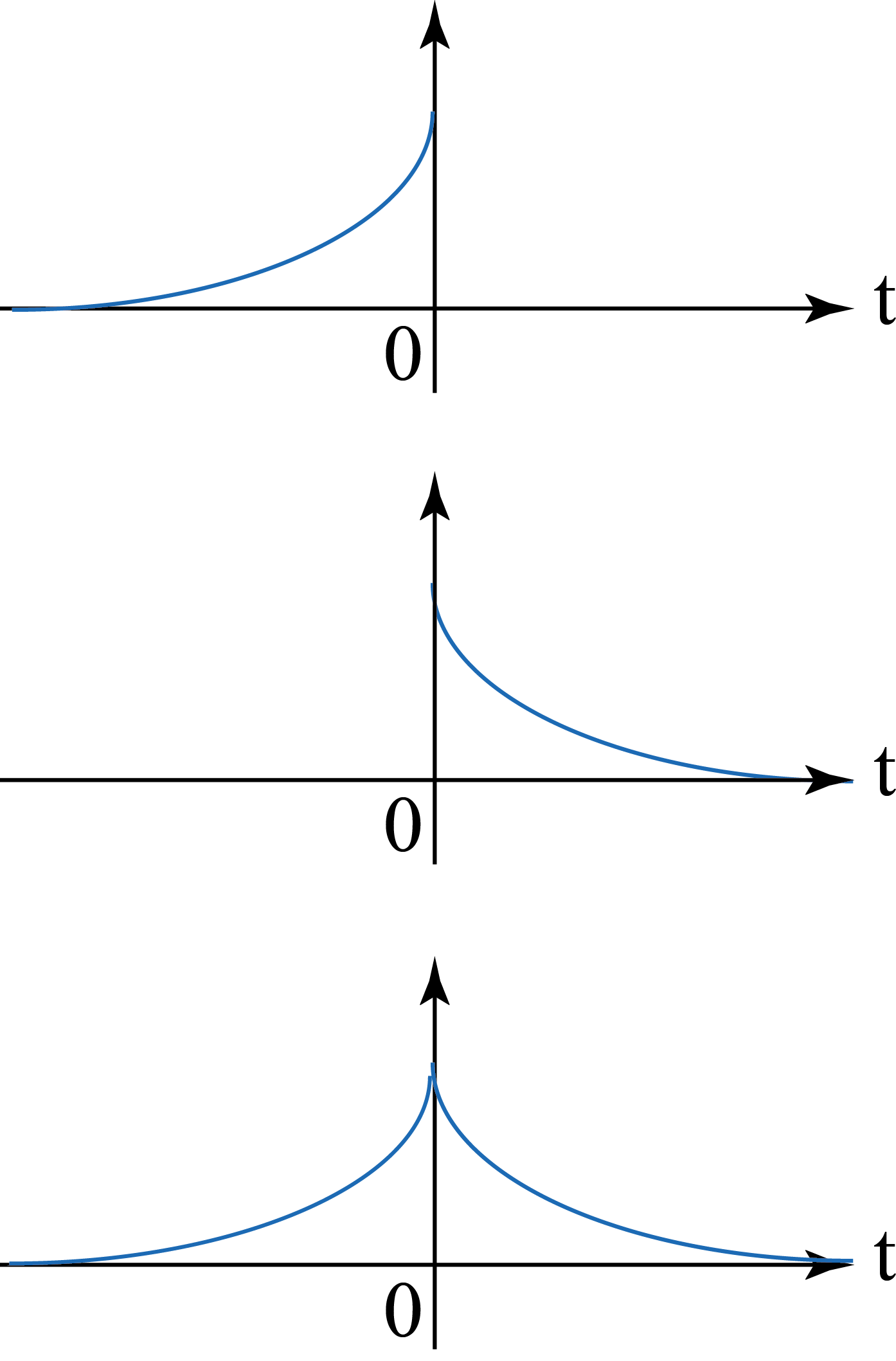}
 \caption{The diagrams show the cases, $E_1(t) = \e^{\Gamma t} \theta (-t)$, $E_2(t) = \e^{-\Gamma t} \theta (t)$, and $E_3(t) = E_1(t)+E_2(t)$,
 respectively. ($\theta(t)=1$ for $t \leq 0$ and $\theta(t) =0$ for $t<0$)}\label{img:diagram5}
\end{figure}
The Fourier transforms of $E_1(t)$ and $E_2(t)$ are complex and the Fourier transform of $E_3(t)$ is real:
\begin{align}
 \tilde E_1 (\omega) =& \mathrm{F.T.}\{ \varepsilon_1(t) \e^{-\ii \omega_L t} \} = \int\limits_{-\infty}^{\infty} \dd t \e^{\Gamma t} \theta (-t) \e^{-\ii \omega_L t} \e^{\ii \omega t}\\
 =& \int\limits_{-\infty}^{0} \dd t \e^{[\Gamma + \ii (\omega - \omega_L)] t} = \frac{1}{\Gamma + \ii ( \omega - \omega_L)} = \frac{\Gamma - \ii (\omega - \omega_K)}{\Gamma^2 + (\omega - \omega_L)^2}\\
 \tilde E_2 (\omega) =& \mathrm{F.T.}\{ \varepsilon_2(t) \e^{-\ii \omega_L t} \} = \int\limits_{-\infty}^{\infty} \dd t \e^{-\Gamma t} \theta (t) \e^{-\ii \omega_L t} \e^{\ii \omega t}\\
 =& \int\limits_{0}^{\infty} \dd t \e^{[-\Gamma + \ii (\omega - \omega_L)] t} = -\frac{1}{-\Gamma + \ii ( \omega - \omega_L)} = \frac{\Gamma + \ii (\omega - \omega_K)}{\Gamma^2 + (\omega - \omega_L)^2}\\
  \rightarrow \tilde E_3(\omega) =& \tilde E_1 (\omega) + \tilde E_2(\omega) = \frac{2\Gamma}{\Gamma^2 + (\omega - \omega_L)^2}
\end{align}
This shows, that these three fields with different temporal shapes, all have the same Lorentzian spectrum $I(\omega) = \vert E(\omega)\vert^2$.
These fields differ in the imaginary parts in frequency representation. It is obvious that time reversal in time domain is related to complex
conjugation in frequency domain ($E_1(t)$ is the time reversed version of $E_2(t)$). Fig. \ref{fig:statcorr} compares the second order
correlation functions for two different stationary fields, a thermal field (Fig. \ref{fig:thermfield}) and a phase diffusing field (Fig. \ref{fig:difflaserfield}), having
identical first order correlation functions. This underlines the importance of higher order correlation functions when characterizing light
fields~\cite{Leuchs86,GlauberHouches}.

\begin{figure}
 \center
 \includegraphics[height=300pt]{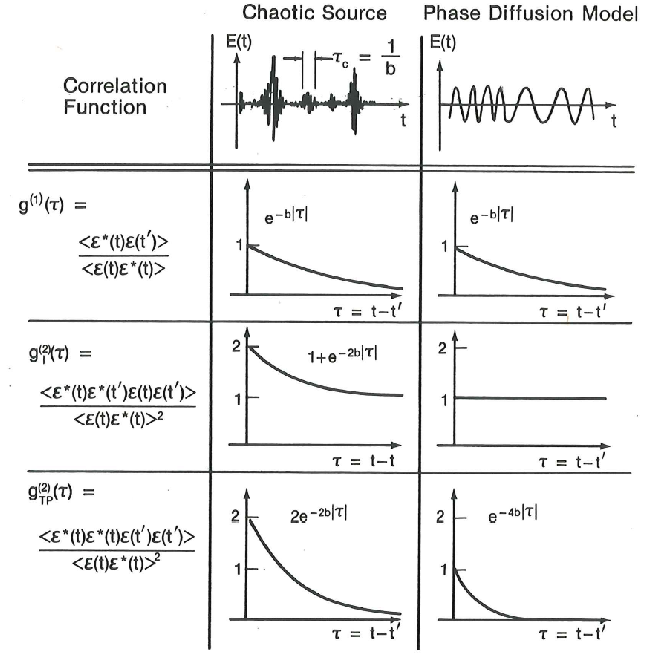}
 \caption{Influence of photon statistics on correlation functions (LEU86).}\label{fig:statcorr}
\end{figure}

\begin{figure}
\center
\captionbox{Chaotic source - thermal field.\label{fig:thermfield}}{\includegraphics[height=100pt]{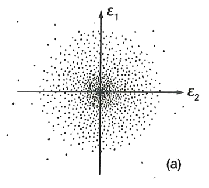}}
\hspace{0.1\textwidth}
\captionbox{Phase diffusing laser field.\label{fig:difflaserfield}}{\includegraphics[height=100pt]{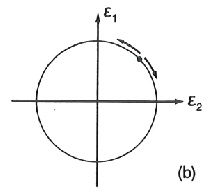}}
\end{figure}
\end{paragraph}

\chapter{Quantum optics of a single mode}

%%%%%%%%%%%%%%%%%%%%%%%%%%%%%%%%%%%%%%%%%%%%%%%%%%%%%%%%%%%%%%%%%%%%%%%%%%%%%%%%%%%%%%%%%%%%%%%%%%%%%%%%%%%%%%%%%%%%%
%%%%  3rd part 
%%%%%%%%%%%%%%%%%%%%%%%%%%%%%%%%%%%%%%%%%%%%%%%%%%%%%%%%%%%%%%%%%%%%%%%%%%%%%%%%%%%%%%%%%%%%%%%%%%%%%%%%%%%%%%%%%%%%%

In this chapter, the quantum properties of a \emph{single} light mode are described. We will focus on Gaussian states
of a light mode, i.\,e. states for which the Wigner function has a two dimensional Gaussian shape. Experimental setups
to measure these quantum properties will be explained. Moreover, we will give an overview about quasi-probability
distributions and review correlation functions. We will begin with the general description of the phase space for
a quantum state in analogy to the harmonic oscillator since the excitation within each mode is quantized.

\section{Phase space of a single mode light field}

\begin{figure}[h!]
 \center
 \includegraphics[height=60pt]{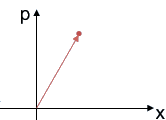}
 \caption{Phase space diagram, $X \equiv X(0)$, $P \equiv \frac{P(0)}{m  \omega}$.}
 \label{fig:osciclassic}
\end{figure}

The energy per mode of the light field is quantized, such as it is in the case of the harmonic oscillator. The \textbf{classical harmonic oscillator} is described as a pure sine wave:

\begin{align}
 x(t) = x(0)\cos (\omega t) + \frac{p(0)}{m \omega} \sin (\omega t).
\end{align}
Its energy is continuous and position and momentum are precise values i.\,e. values without any uncertainty (see Fig.~\ref{fig:osciclassic}).

\begin{figure}[h!]
 \center
 \includegraphics[height=70pt]{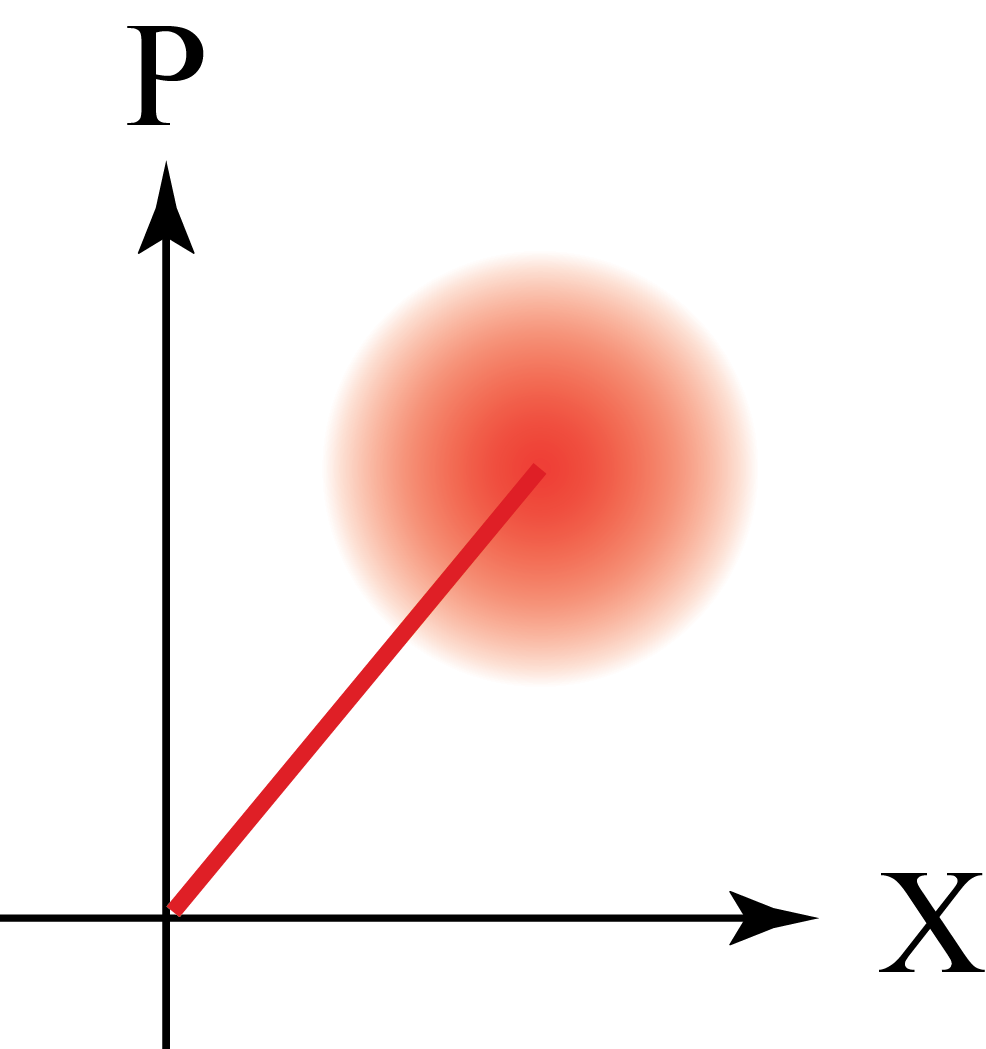}
 \caption{Phase space diagram. As we will see below, the shaded area represents a phase space distribution function.}
 \label{fig:osciquantum}
\end{figure}

The \textbf{quantum harmonic oscillator} instead is described by a diffuse sine wave with discrete energy values but uncertain position and momentum values (see Fig.~\ref{fig:osciquantum}).
The uncertainty (variance of the state) is  described by $\gen{\Delta X^2} \neq 0$ and $\gen{\Delta P^2} \neq 0$. The energy levels of the quantum harmonic oscillator are given by:
\begin{align}
 E_n = \left( n + \frac{1}{2} \right) \hbar \omega,
\end{align}
where the lowest energy level has positive value. The quantized description of a light mode is mathematically equivalent. In the case of a light field, the variables $X$ and $P$ are two conjugate field operators. In a specific case, they may represent the amplitude and the phase quadrature of the light field in a single mode.

A quantum optical field can be expressed in different ways and with different variables, \textbf{discrete} variables and \textbf{continuous} ones. The wave function of a single-mode quantum field can e.g. be written as a superposition of photon number states (Fock-basis) which corresponds to a discrete variable description. For an infinite dimensional complex Hilbert space, we write a quantum state as
\begin{align}
 \ket{\Psi} = \sum\limits_{i=0}^{\infty} \alpha_i \ket{i},
\end{align}
where $\ket{i}$ represent photon number states. One specific example is a state description by discrete dichotomic variables:
\begin{align}
 \ket{\Psi} = \alpha_0 \ket{0} + \alpha_1 \ket{1} = \sum\limits_{i=0}^1 \alpha_i \ket{i},
\end{align}
where $\ket{\Psi}$ is called a \emph{qubit}. This state consists of no more than one photon.\\
In the continuous case, there are the continuous variables $X$ and $P$ which span the phase space of the mode\footnote{An excellent introduction to the topic is given by W.~P.~Schleich~\cite{Schleich}}. %TODO: check reference
\begin{figure}[h!]
 \center
 \includegraphics[height=80pt]{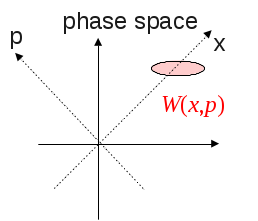}
\end{figure}\\
The values of $X$ and $P$ can e.g. be estimated by the Wigner function $W(X,P)$, a particular phase space distribution function
\begin{align}
 W(X ,P) = \frac{1}{2\pi\hbar} \int\limits_{-\infty}^{\infty} \dd \xi \exp \left( -\frac{\ii p \xi}{\hbar} \right) \Psi^* \left(X-\frac{1}{2} \xi \right) \Psi \left(X+\frac{1}{2} \xi \right),
 \label{Eq:Wigner}
\end{align}
where $\Psi(x) = \langle x \vert \Psi \rangle$ holds.
It is instructive to look at Wigner's original paper\footnote{The function later named after Wigner comes out of the blue. The only hint at how this function was found is given by Wigner in footnote 2. Asked later of what this other purpose was for which he and Szillard used this function, Wigner is reported to have answered that he made this up. His fellow countryman Szillard was looking for a job and Wigner thought such a footnote might help him.}\cite{WignerX}.
\\
\\
\indent Different types of continuous quantum variables are the field quadratures $\hat{X}$ and $\hat{P}$ (canonical operators) and the Stokes variables which describe the polarization of the quantum field. It is practical to define the field operators $\hat a$ and $\hat a^{\dagger}$ which are related to the canonical ones as follows:

\begin{align}
 \hat a =& \hat X + \ii \hat P\\
 \hat a^{\dagger} =& \hat X - \ii \hat P,
\end{align}
or vice versa
\begin{align}
 \hat X =& \frac{1}{2} (\hat a + \hat a^{\dagger})\label{eq:X}\\
 \hat P =& \frac{1}{2\ii} (\hat a - \hat a^{\dagger}).
\end{align}

For the field operators $\hat a$ and $\hat a^{\dagger}$ the commutation relation writes:
\begin{align}
 [\hat a, \hat a^{\dagger}] = \hat a \hat a^{\dagger} -  \hat a^{\dagger} \hat a = 1,
 \label{Eq:acomm}
\end{align}
as well as
\begin{align}
 \hat a \ket{n} = & \sqrt{n} \ket{n-1}\\
 \hat a^{\dagger} \ket{n} = & \sqrt{n+1} \ket{n+1}.
\end{align}
\\
They are also related to the photon number operator $\hat{n}=\hat a^{\dagger} \hat a$:
\begin{align}
 \hat a^{\dagger} \hat a \ket{n} = & n \ket{n},
 \label{Eq:n}
\end{align}
where $\ket{n}$ are the Fock states.\par

\begin{figure}[h!]
 \center
 \includegraphics[height=100pt]{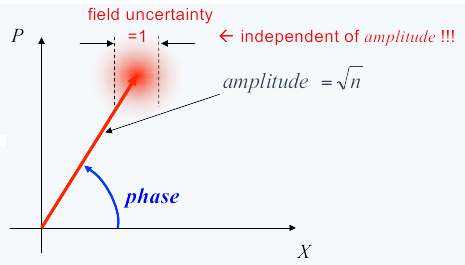}
 \caption{Phase space diagram for light field.}
 \label{fig:cohstate}
\end{figure}

A single quantum mode in general is described by one pair of variables $\hat X, \hat P$, where their mean varies continuously and their variances obey the uncertainty relation.
Furthermore, $\hat X$ and $\hat P$ are non commuting operators\footnote{The corresponding commutator for $\hat{X}$ and $\hat{P}$ is $[\hat X, \hat P]=\frac{\ii}{2}$.}. For equal
and minimum uncertainty of $\hat X$ and $\hat P$ we have a \emph{coherent state}, which defines the most ``classical'' quantum state of light (Fig.~\ref{fig:cohstate}), with amplitude and
phase fairly well defined simultaneously, as much as quantum effects allow for.

A coherent state $\ket{\alpha}$, ideally emitted by a laser, is an eigenstate of the field operator $\hat a$, namely
\begin{align}
 \hat a \ket{\alpha} = \alpha \ket{\alpha}.
\end{align}
It is possible to expand the coherent state $\ket{\alpha}$ in its Fock-basis \ket{n}:

\begin{align}
 \ket{\alpha} = \sum^{\infty}_{n=0}c_n \ket{n}
\end{align}
It holds
\begin{align}
 \hat a \ket{\alpha} = \sum^{\infty}_{n=0} c_n  \hat a \ket{n} = \sum^{\infty}_{n=1} c_n \sqrt{n} \ket{n-1} = \sum^{\infty}_{n=0} c_{n+1} \sqrt{n+1} \ket{n} \overset{!}{=}  \alpha \sum^{\infty}_{n=0}c_n \ket{n}
\end{align}
It follows:

\begin{align}
 c_n = \frac{\alpha}{\sqrt{n}} c_{n-1}
\end{align}
 and
\begin{align}
 c_n = \frac{\alpha^n}{\sqrt{n!}} c_{0}.
\end{align}
Normalizing the expression
\begin{align}
\sum^{\infty}_{n=0} \left| \frac{\alpha^n}{\sqrt{n!}} c_{0} \right|^2 \overset{!}{=}  1
\end{align}
gives the coherent state $\ket{\alpha}$, written in its Fock-basis \ket{n}:
\begin{align}
 \ket{\alpha} = \e^{-\frac{1}{2} \vert \alpha\vert^2} \sum\limits_n \frac{\alpha^n}{\sqrt{n!}} \ket{n}.
\end{align}

\begin{figure}[h!]
 \center
 \includegraphics[height=100pt]{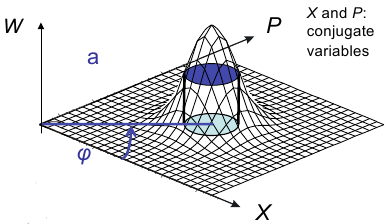}
 \caption{$\hat X$ and $\hat P$ are quadratures of the field with $\hat X = \frac{1}{2} (\hat a + \hat a^{\dagger})$ and $\hat P = \frac{1}{2\ii} (\hat a - \hat a^{\dagger})$.}
 \label{fig:WignerCoh}
\end{figure}

The Wigner function of a coherent state is given by
\begin{align}
 W_{\alpha} (\beta) = \frac{2}{\beta} \e^{-2\vert \alpha - \beta \vert^2}
\end{align}
with $\alpha=X_0+iP_0$ defining the position of the maximum of the Wigner function and $\beta = X + iP$ being the phase space vector. The Wigner function of a coherent state represented in phase space is visualized in Fig.~\ref{fig:WignerCoh}.

\section{Wigner function and quasi-probability distributions in general}

A quantum state can be fully described by different concepts, for example the wave function, the density operator or the following quasi-distributions: the Glauber-Sudarshan
distribution $P(\alpha)$, the Wigner distribution $W(\alpha)$ and the Husimi-Kano distribution $Q(\alpha)$. They all contain the entire information about the state but are different
mathematical descriptions of the quantum phase space and, more importantly, represent different ways of measuring a quantum state.\\
The wave function can be easily written for pure states, for example coherent states, Fock states, squeezed states, or the vacuum state, but a mixed or thermal state cannot be
expressed by its wave function. Therefore, the representation of a state by help of a quasi-probability distribution is often more convenient.
\\
\\
In order to describe a measurement process, we can choose the description with the field operators $\hat a$ and $\hat a^{\dagger}$. Depending on the measurement, there are different types of ordering the field operators:

\begin{itemize}
\item \textbf{normal} ordering $\hat a^{\dagger} \hat a$: \textbf{direct detection} (photo diode / camera)\\

\item \textbf{symmetric} ordering $\frac{1}{2}(\hat a^{\dagger} \hat a + \hat a \hat a^{\dagger})$: \textbf{homodyne detection}\\
\item \textbf{anti-normal} ordering $\hat a \hat a^{\dagger}$: \textbf{heterodyne detection} (or double-homodyne detection)\\
\end{itemize}

\textbf{Direct detection} corresponds to the \textbf{Glauber-Sudarshan distribution $P(\alpha)$}. The operators are \textbf{normally ordered} which can be explained with the annihilation operator acting first
diminishing the photon number by 1, when one photon is detected~\cite{GlauberHouches}. For a coherent state, the $P$-distribution is a $\delta$-distribution with $P(\alpha)=\infty$ at the mean value of the state. For a thermal
state, the $P$-distribution is a Gaussian function, and for Fock states and squeezed states, it evolves to a distribution which is singular and contains finite (Fock) and infinite (squeezed) derivatives of
the $\delta$-distribution~\cite{Schleich}. Thus, it might be complicated to use this description in general.\par

\begin{figure}[h!]
 \center
 \includegraphics[height=150pt]{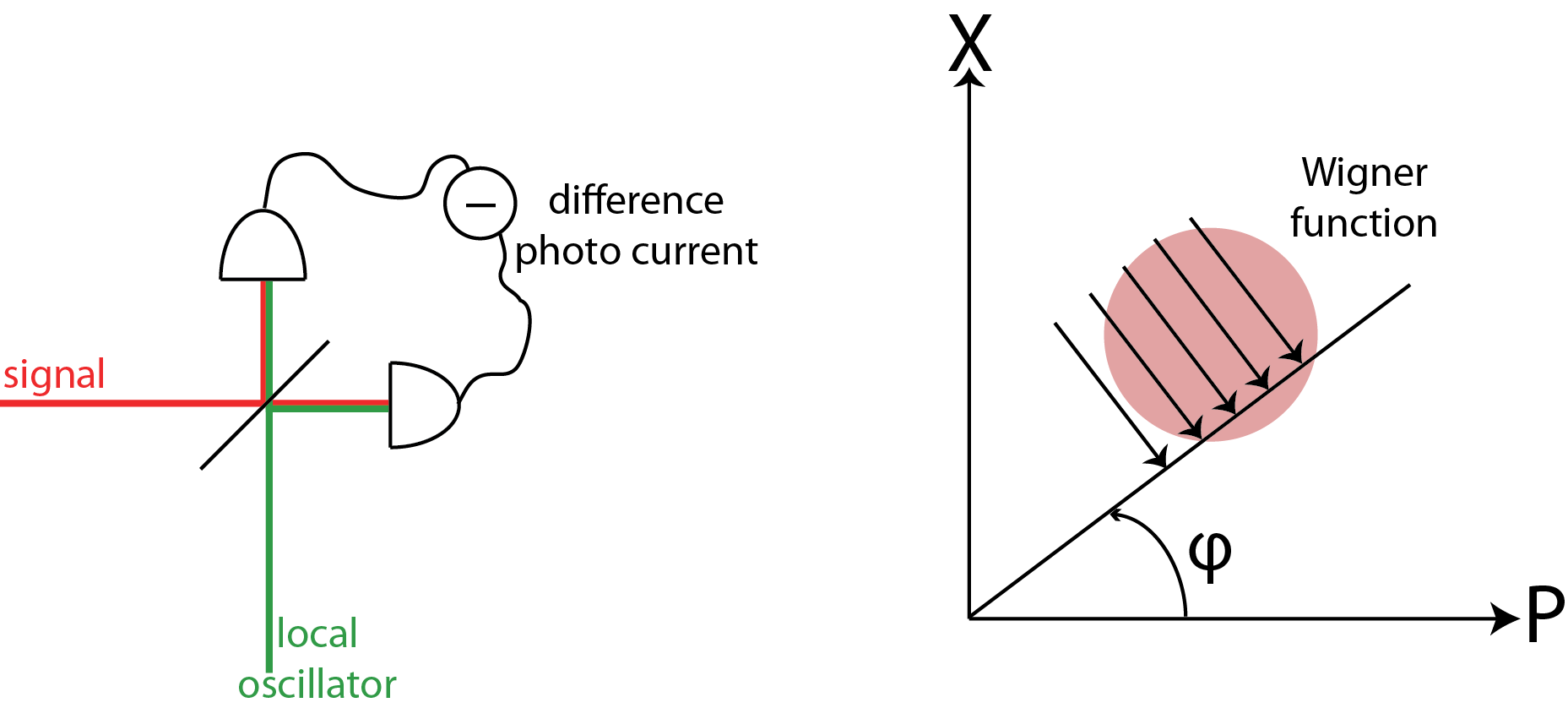}
 \caption{Homodyne detection and principle of Wigner function reconstruction.}
 \label{fig:homoB}
\end{figure}

The \textbf{Wigner distribution $W(\alpha)$}, which was already mentioned for a coherent state in Eq.~\ref{Eq:Wigner}, is formally the $P$-distribution convoluted with the vacuum state.  Experimentally, it can
be measured by \textbf{homodyne detection}, with the setup shown in Fig.~\ref{fig:homoB}.
To measure the Wigner function, the angle $\varphi$, which is the phase shift between the local oscillator and the signal, has to be varied from $0$ to $\pi$. At each angle $\phi$, the homodyne measurement yields
a histogram which corresponds to a projection of the Wigner function onto the axis which is defined by $\varphi$. The phase space distribution is then obtained by tomographic reconstruction~\cite{Raymer,Wallentowitz},
namely an inverse Radon transform.

For the operators describing the properties of the Wigner function, such as $\hat{X}^2+\hat{P}^2$, one has to arrange the field operator accordingly, for this case \textbf{symmetric ordering}. The Wigner function has
the advantage that it has a Gaussian shape for several commonly used states such as coherent states, squeezed states, thermal states, and many more.

\begin{figure}[h!]
 \center
 \includegraphics[height=100pt]{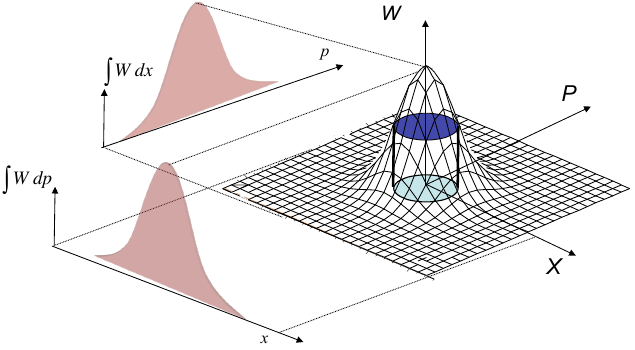}
 \caption{Example: vacuum state $W_0(\alpha) = \frac{2}{\pi} \e^{-2\vert\alpha\vert^2}$.}
 \label{fig:vacstate}
\end{figure}

\begin{figure}[h!]
 \center
 \includegraphics[height=100pt]{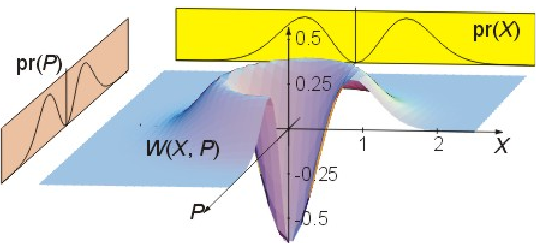}
 \caption{Photon number state in phase space with $W_1(\alpha) = \frac{2}{\pi} (4\vert\alpha\vert^2 -1)\e^{-2\vert\alpha\vert^2}$.}
 \label{fig:photnumstate}
\end{figure}

The Wigner function for the vacuum state is depicted in Fig.~\ref{fig:vacstate} and for a photon number state in Fig.~\ref{fig:photnumstate}. As can be seen in the latter, a phase space distribution function may have
negative values. Therefore, it is called a \emph{quasi-probability distribution}. The \emph{marginal distributions} instead are positive, i.e. any projection onto a line involving a one-dimensional integration such
as in the case of homodyne detection.

\begin{figure}[h!]
 \center
 \includegraphics[height=150pt]{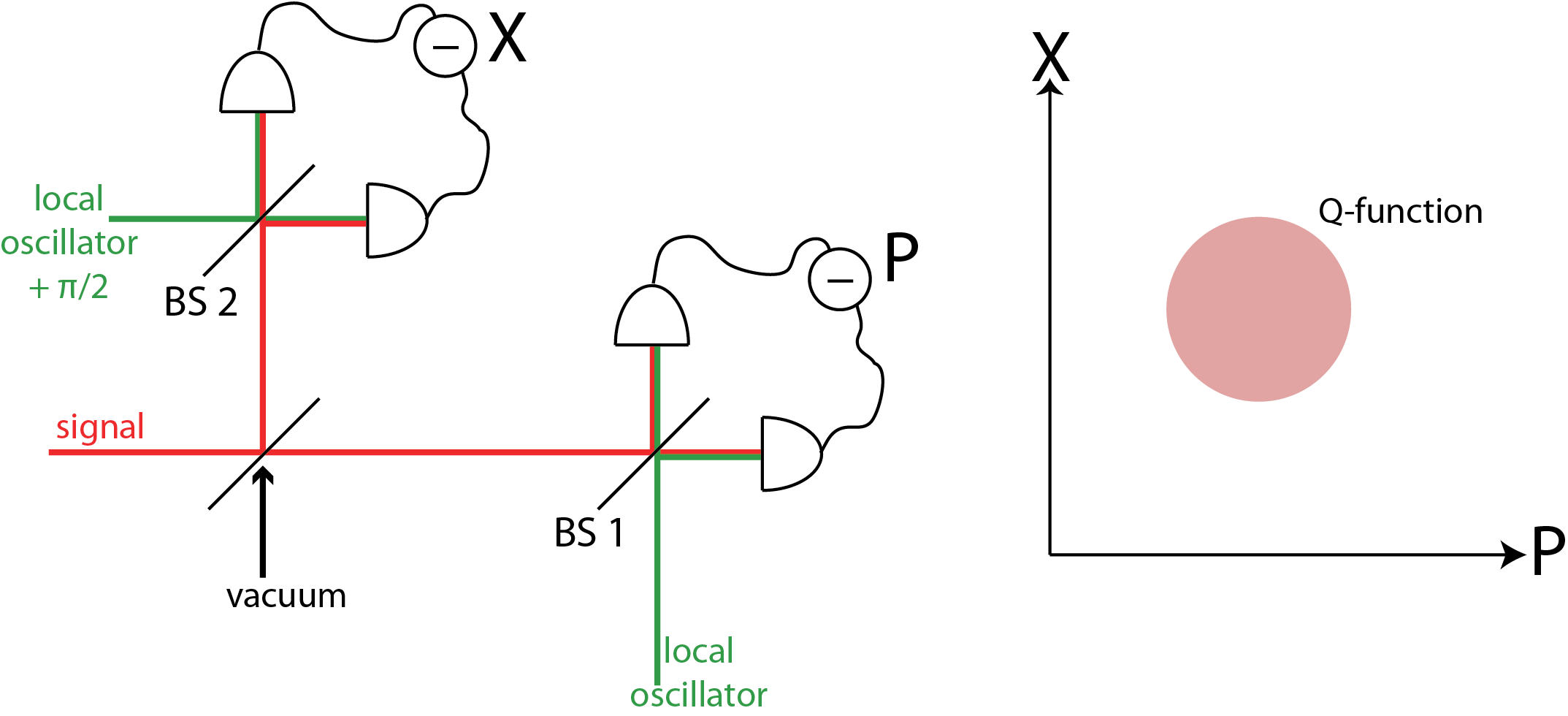}
 \caption{Heterodyne measurement setup.}
 \label{fig:heteroB}
\end{figure}

Yet another convolution, the convolution of the Wigner function with the vacuum state, gives us access to the \textbf{Husimi-Kano distribution $Q(\alpha)$}. In the experiment, we
also have to add a second time the vacuum state which results in the \textbf{heterodyne or double-homodyne measurement} setup, shown in Fig.~\ref{fig:heteroB}. We have directly access to the continuous
variables $\hat{X}$ and $\hat{P}$, but the setup is also more complicated. It involves two more beam splitters and a second local oscillator which is shifted by $\pi/2$ according to the first one. The problem of simultaneous measurement of two conjugate variables was discussed early by Arthurs and Kelly~\cite{Arthurs}. The Q-function is based on \textbf{anti-normal ordering} of the field operators.
\\
\\
All quasi-probability distributions can be calculated from any other one via convolution or deconvolution with the vacuum state. However, deconvolution is often a mathematically not very stable operation because the experimentally determined Wigner or Husimi-Kano functions are effected by fundamental quantum noise and technical noise. So it would be the best to directly measure the P-function which is, unfortunately, in general not straight forward. There is however a way to determine the Wigner function based on the measurement of the photon number distribution $P(\alpha)$ using a photon number resolving detector~\cite{Banaszek,Laiko,Leibfried,Lutterbach,Lutterbach2,NoguesHaroche}. The sum $\sum_n (-1)^n\cdot P(n) = W(0,0)$ gives the Wigner function at the origin (0,0). By first displacing the distribution in phase space, $W(X,P)$ can be mapped. But even this normal ordering based measurement yields $P(\alpha)$ only by the way of $W(X,P)$. In Table~\ref{t:quasi}, an overview of the different quasi-distributions and the corresponding measurement setups can be found.

\begin{table}[h!]
\caption{Overview of quasi-probability distributions }
	\centering
	
	\begin{tabular}{|l||c|c|c|}
	\hline
	& & & \\
			Ordering & normal & symmetric & anti-normal\\
			& & & \\ \hline
			& & & \\
			Energy & $\bra{n} \hat{a} ^{\dagger} \hat{a} \ket{n}$ & $\bra{n} \frac{1}{2} \left(\hat{a}^{\dagger} \hat{a} + \hat{a} \hat{a}^{\dagger} \right) \ket{n}$ & $\bra{n}\hat{a} \hat{a}^{\dagger} \ket{n}$ \\
			 & = n & = n + $\frac{1}{2}$ & = n + 1 \\ 
			& & & \\			 
			 \hline
			 & & & \\
			Detection  & direct detection: & homodyne: & double-homodyne: \\
	scheme & click detector, & 4-port detection & 8-port detection \\ 
		&	 photon number  & & \\
		
		&	 resolving & & \\
		& & & \\
			 \hline
			 
			 & & & \\
			
Determining & reconstruction by & tomographic & phase space \\ 
phase space			& deconvoluting the & reconstruction & distribution directly  \\ 
			distribution & Wigner function & from homodyne & measured with\\
	& & data & heterodyne \\		 
			& & & detection\\
			& & & \\
			 \hline
			 
			 & & & \\			 
			 
		Corresponding	 & P - distribution & Wigner - function & Q - function \\ 
		representation	& & &\\
			& & & \\\hline
			
		\end{tabular}
			\label{t:quasi}
\end{table}

\begin{paragraph}{Superposition of coherent states}\quad\\
A more complex example for a Wigner function with negative eigenvalues is a \emph{cat state}:
\begin{align}
  \ket{\varPsi_{\mathrm{even cat}}} = \frac{1}{\sqrt{2\left(1+\e^{-2\vert\alpha\vert^2}\right)}}(\ket{\alpha} + \ket{-\alpha})
 \end{align}
 which is depicted in Fig.~\ref{fig:catstate}. It is the superposition of two coherent states. 
\begin{figure}[h!]
 \center
 \includegraphics[height=150pt]{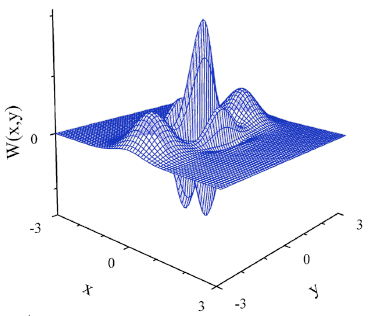}
 \caption{Wigner function of a cat state.}
 \label{fig:catstate}
\end{figure}
By means of this example, the analogy of the quantum-phase space to diffraction optics is easily understandable. In Fig.~\ref{fig:diffraction}, top view of a 3D-Wigner function of the cat state is shown. Integration along the X-axis gives two maxima which can be compared to the optical field through a double-slit. When integrating along the P-axis instead, we find an interference pattern, as it is observed on a screen far away from a double-slit. In the double-slit experiment, the interference pattern and the field distribution of the double-slit are mathematically connected by the Fourier transform. Thus, we can understand by means of analogy, that the conjugate variables $X$ and $P$ are likewise connected by Fourier transform. If we integrate along another axis with a certain skewed angle in phase space, we obtain all different interference patterns which can be found when observing the light field at different distances from the double-slit.
\begin{figure}[h!]
 \center
 \includegraphics[height=120pt]{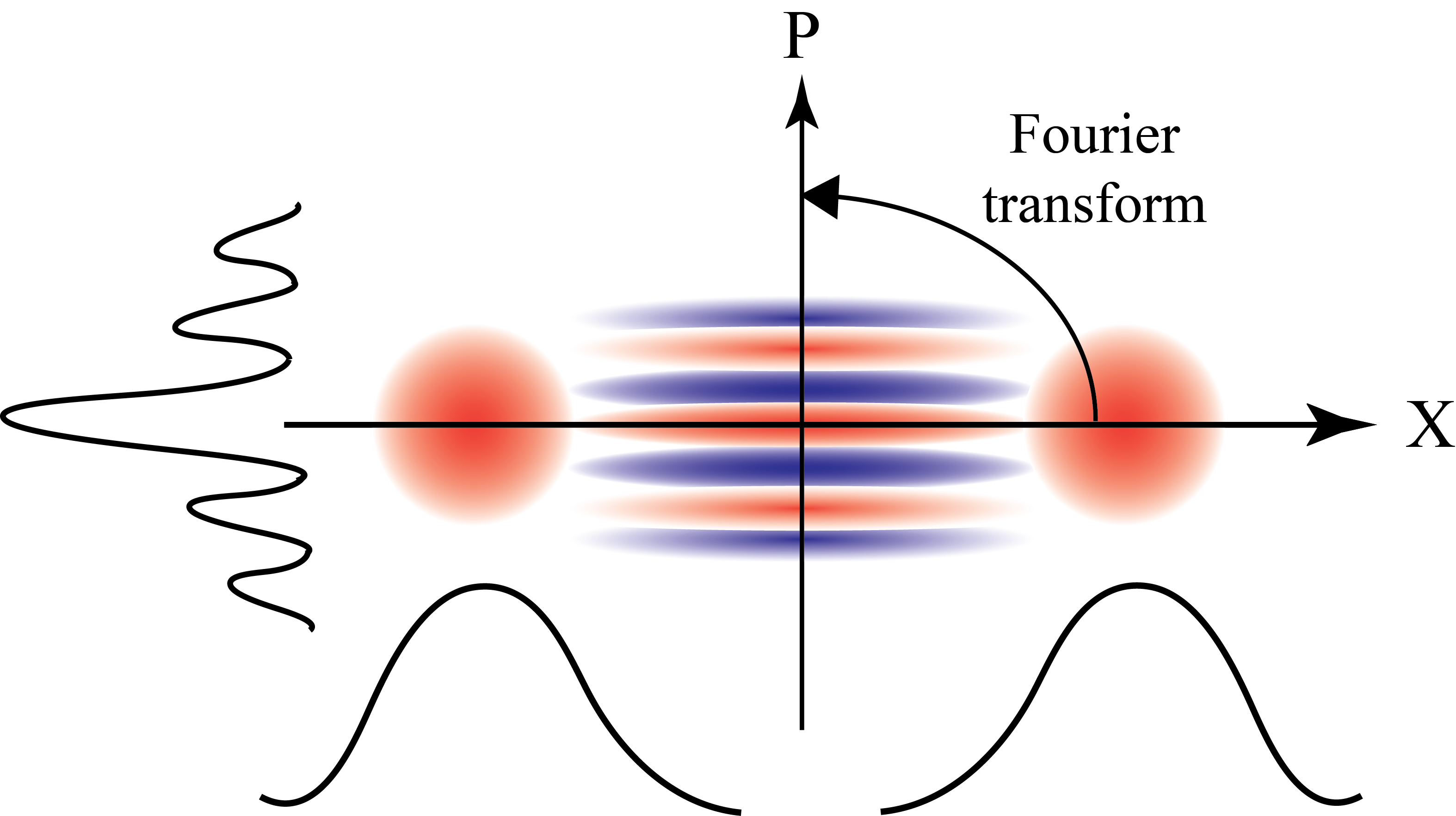}
 \caption{Top view of the Wigner function of a cat state, showing also the marginal distributions. Note the close analogy to diffraction in classical optics.}
 \label{fig:diffraction}
\end{figure}
\end{paragraph}

\begin{paragraph}{Thermal state}
The Wigner function is a helpful tool to represent also more complicated states, such as mixed states which cannot be described by a wave function. As an example we discuss a thermal state. The Wigner function of the vacuum with added thermal noise is:
\begin{align}
 W_{\mathrm{th}} (\alpha) = \frac{1}{\pi (\gen{n} + 1/2)} \e^{-\frac{1}{\gen{n}+1/2} \vert\alpha\vert^2}
\end{align}
The phase space representation of such a thermal state is shown in Fig.~\ref{fig:thermalstate}. It is similar to a coherent state but the width is larger (higher variance) because additionally to the quantum noise, also classical or thermal noise is added. 
\begin{figure}[h!]
 \center
 \includegraphics[height=150pt]{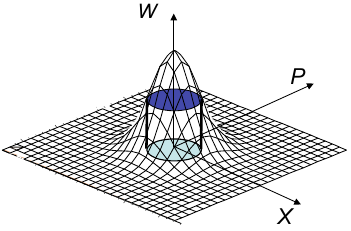}
 \caption{Phase space representation of a thermal state.}
 \label{fig:thermalstate}
\end{figure}\\ 
\end{paragraph}

\section{Detection}

\begin{figure}[h!]
 \center
 \includegraphics[height=150pt]{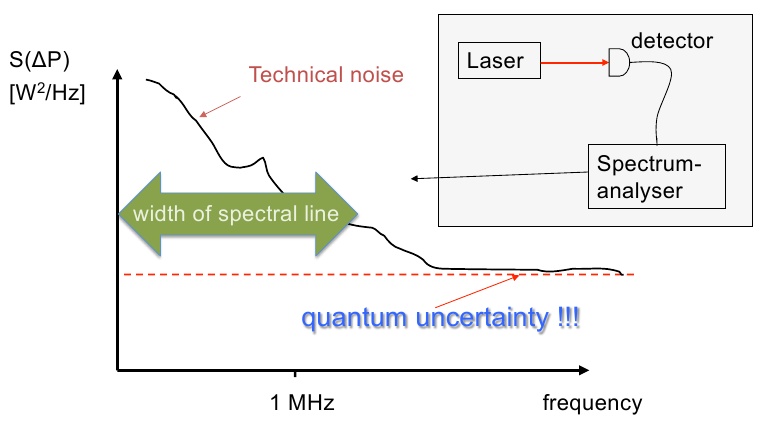}
 \caption{Spectrum of the power fluctuations of laser light.}
 \label{fig:fluc}
\end{figure}

When measuring a light field, we need to take into account the specifications of the detection scheme and the laser source. The spectral density of a light field can be measured by a spectrum analyzer:

\begin{align}
 S_I(f) = \int\limits_{-\infty}^{\infty} \underbrace{\langle I(t) I(t+\tau) \rangle}_{\equiv G^{(2)} (\tau)} \cos ( 2\pi f \tau) \dd \tau,
\end{align}

\noindent First of all, the electronic noise of the spectrum analyzer and the detector have to be low enough if one wants to detect signals in the quantum regime.\footnote{For a given electronic amplifier, the gain is inversely proportional to the bandwidth: the higher the bandwidth, the smaller is the gain, and hence the sensitivity of the detector.}
\\In the lower frequency range, thermal noise from the laser cavity is added to the quantum noise which is detrimental to measurements in the quantum regime (Fig.~\ref{fig:fluc}). For quantum noise, the variance of the photon number fluctuations is proportional to the mean intensity of the signal
\begin{align}
\gen{\Delta n^{2}}=\gen{n},
\end{align}
corresponding to Poisson statistics of the photon number as detailed below. Instead, for thermal noise, the variance on the photon number is proportional to the squared mean intensity of the signal
\begin{align}
\gen{n^{2}}=2\gen{n}^{2}+\gen{n},
\end{align}
yielding the root mean squared photon number fluctuation:
\begin{align}
\gen{\Delta n^{2}}=\gen{n^2}-\gen{n}^2=2\gen{n}^{2}+\gen{n}-\gen{n}^2=\gen{n}^{2}+\gen{n}.
\end{align}
\\
\begin{figure}[h!]
 \center
 \includegraphics[height=150pt]{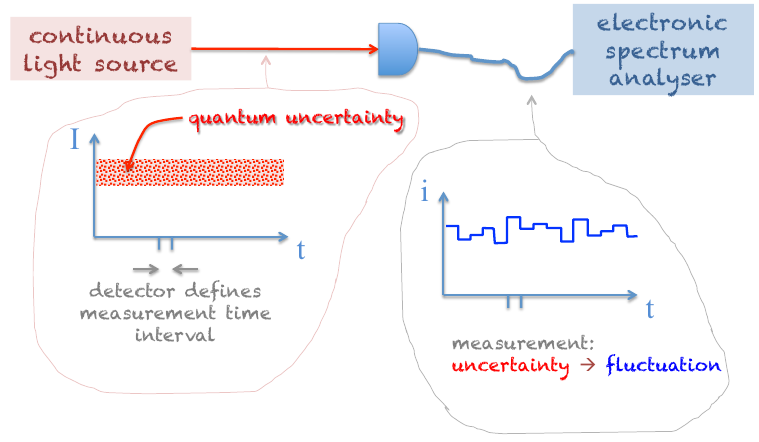}
 \caption{Quantum uncertainty in measurement}
 \label{fig:uncertainty}
\end{figure}
For high photon numbers the quadratic term  $\gen{n}^{2}$ is dominating and the variance is proportional to the square of the mean value:
\begin{align}
\gen{\Delta n^{2}} \approx \gen{n}^{2}
\end{align}
Dependent on the laser source, thermal noise can go up to 1-10\,MHz. Thus, quantum optics measurements have to be carried out in a frequency range above this limit.

A quantum von-Neumann measurement projects the state on one of its eigenstates. Note that a coherent state with its intrinsic quantum uncertainty is
a stationary state. But measuring it in a defined time interval projects the state to a fixed value which fluctuates over time. The quantum uncertainty is thus translated
to fluctuations in time via the measurement process (Fig.~\ref{fig:uncertainty}).
\vspace{10pt}

\begin{paragraph}{Direct detection}
If one converts the signal photons directly into electrons in a photo detector without any admixture of auxiliary light beams, this detection process is called ``direct detection'' (Fig.~\ref{fig:photodiode}). 

\begin{figure}[h!]
 \center
 \includegraphics[height=70pt]{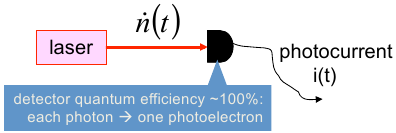}
 \caption{Direct detection via a photo diode}
 \label{fig:photodiode}
\end{figure}
Coherent states refer to ideal laser radiation when ignoring additional classical noise and the inherent phase diffusion. We assume that the photon number is $\gen{n} \gg 1$. For the mean value of the photon number, one gets:

\begin{align}
 n=& \gen{n} + \Delta n,\quad \gen{\Delta n} = 0
\end{align}
But what is the variance of the photon number $\gen{\Delta n^2}$? We can write the photon number as (see Fig.~\ref{fig:dirdetcoh}):

\begin{align} 
\gen{n} + \Delta n =& (\gen{X(\theta)} + \Delta X(\theta))^2\\
  =& \gen{X(\theta)}^2 + 2\gen{X(\theta)} \Delta X(\theta) + \Delta X^2 (\theta)
\end{align}
From $\gen{X}^2 = \gen{n}$, $2 \gen{X} \cdot (\Delta X) = \Delta n$ and neglecting $\Delta X^2$ follows that
\begin{align}
 \gen{\Delta n^2} = 4 \gen{n} \frac{1}{4} = \gen{n}
\end{align}

\noindent Thus, the photon noise in a laser beam underlies \emph{Poissonian statistics}, as stated above. Here it becomes clear that this Poisson statistics is the results of the uncertainty in phase space being independent of the quadrature amplitude.

\begin{figure}[h!]
 \center
 \includegraphics[height=120pt]{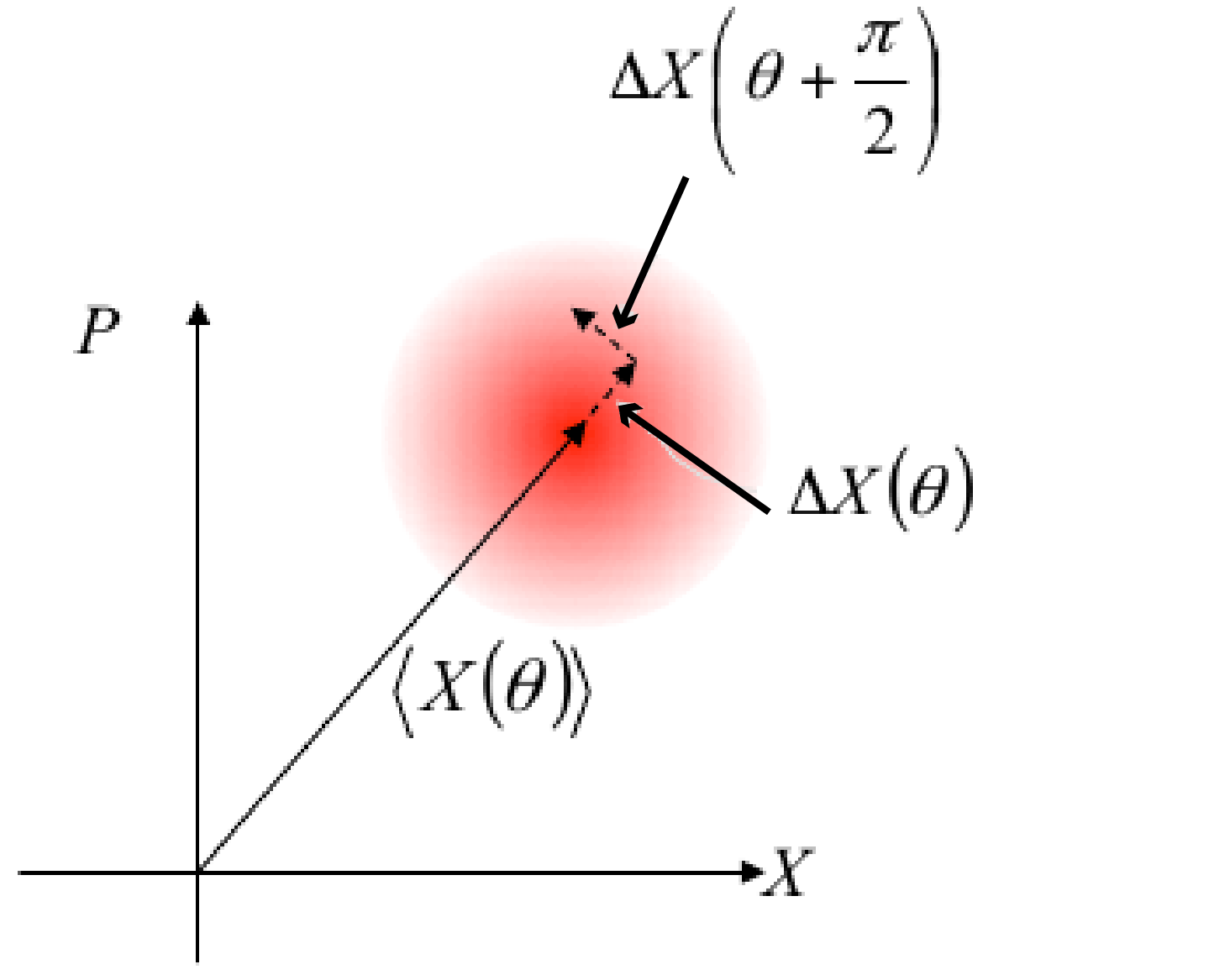}
 \caption{Coherent state in phase space. $\theta$ indicates the direction of the displacement in phase space.}
 \label{fig:dirdetcoh}
\end{figure}

The P-distribution, which is the underlying representation of the quantum state for direct detection, cannot be derived directly with help of a photon number resolving setup~\cite{Stenholm}. It has to be reconstructed from the Wigner function which itself can be measured by homodyne detection, as explained in the next paragraph. But the reconstruction causes many problem with real data due to
singularities. Nevertheless, the P-distribution is used to verify
nonclassicality by negativity. With data from a homodyne
measurement, an estimation of this distribution (e.\,g. by using
so-called pattern functions) may be sufficient~\cite{Kiesel}.

\end{paragraph}

\begin{paragraph}{Homodyne detection}
The convolution of the signal state with the vacuum state will give us access to the Wigner function. The corresponding experimental setup to this convolution is homodyne detection. A sketch can be found in Fig.~\ref{fig:homodyneLeuchs}. As mentioned previously, the integration of the Wigner function along a certain angle in 2D phase space will yield a 1D marginal distribution. This marginal distribution is measured in homodyne detection, the angle being determined by the phase of the coherent auxiliary beam called the local oscillator. The experimentally determined marginal distributions taken for various angles are the input data needed for tomographic reconstruction~\cite{Raymer}.
\begin{figure}[h!]
 \center
 \includegraphics[height=120pt]{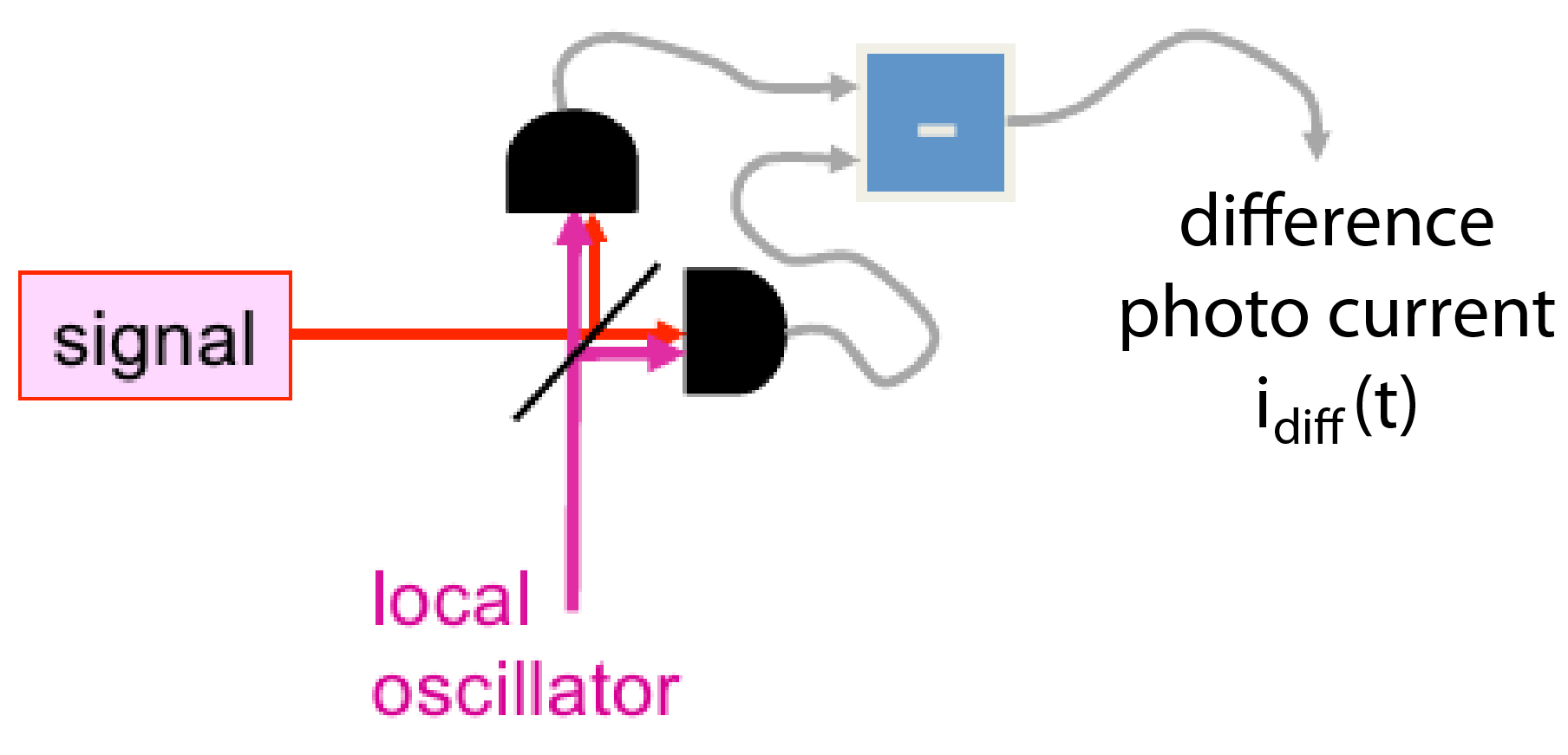}
 \caption{Homodyne detection. The local oscillator describes a mode in an intense coherent state.}
  \label{fig:homodyneLeuchs}
\end{figure}

Let us assume a strong local oscillator (Fig.~\ref{fig:lo})
\begin{align}
 \sqrt{\gen{n_{\mathrm{lo}}}} = \gen{X_{\mathrm{lo}} (\phi)} \gg X_{\mathrm{signal}},\Delta X
\end{align}
Then we can neglect all terms, quadratic in the small signal for the difference photo current: 

\begin{align}
 i_{\mathrm{diff}} \propto\;& [ \gen{X_\mathrm{lo}(\varphi)} + \Delta X_\mathrm{lo} (\varphi) + \gen{X_{\mathrm{signal}}(\theta)} \cos (\varphi - \theta) + \Delta X_{\mathrm{signal}} (\varphi)]^2\\
  \;&- [ \gen{X_\mathrm{lo}(\varphi)} + \Delta X_\mathrm{lo} (\varphi) - \gen{X_{\mathrm{signal}}(\theta)} \cos (\varphi - \theta) - \Delta X_{\mathrm{signal}} (\varphi)]^2\\
  \approx\;& 4 \gen{X_\mathrm{lo} (\varphi)} \cdot [\gen{X_\mathrm{signal} (\varphi)} + \Delta X_\mathrm{signal} (\varphi)],
\end{align}
where the minus sign inside the brackets in the second line is the result of the phase change imposed by the beam splitter (for details, see next chapter). By variation of $\varphi$ we can
measure projections of the signal on all directions, providing the input data for \emph{quantum state tomography}~\cite{Raymer}.

\begin{figure}[h!]
 \center
 \includegraphics[height=200pt]{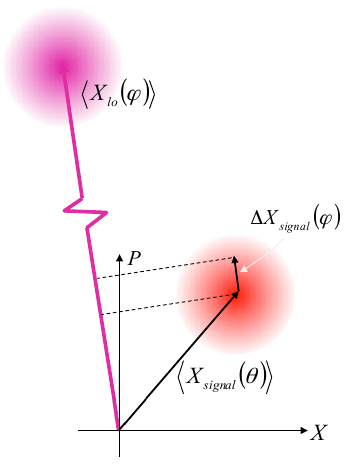}
 \caption{Local oscillator and signal in the phase space.}
  \label{fig:lo}
\end{figure}

Another approach to reconstruct the Wigner function of a quantum state was presented using data obtained in direct detection with photon number resolving detectors \cite{Banaszek,Laiko,Leibfried,Lutterbach,Lutterbach2,NoguesHaroche}. Using direct detection, the probability of each photon number $P(n)$ is estimated by measurement and it has been found that the sum over these $P(n)$ with alternating signs gives the value of the Wigner function at position (0,0) in phase space:

\begin{align}
 \sum\limits_n \left(-1\right)^n P(n) = W (0,0)
\end{align}

Shifting the quantum state via an asymmetric beam splitter with a local oscillator prior to detection - much the same as in homodyne detection - one can also reconstruct the Wigner function in the whole phase space. Note that the measurement required the convolution with a vacuum state (symmetric ordering) and not only direct detection (normal ordering), hence the procedure yields the Wigner function.
\end{paragraph}

\begin{paragraph}{Heterodyne detection}
For many applications, such as quantum key distribution, it is convenient to directly measure the Q-function of the quantum states. This can be performed with a heterodyne measurement or also so-called double-homodyne measurement or eight port homodyne detection, see~\cite{Freyberger}. For this purpose, a second local oscillator (LO) is inserted into the setup (see Fig.~\ref{fig:heteroB}) which is phase shifted by $\pi/2$ in order to measure both quadratures $X$ and $P$ simultaneously. In real experiments it is difficult to stabilize the phase between the LO at beamsplitter 1 and the LO + $\pi/2$ phase shift at beamsplitter 2. Therefore, another degree of freedom of the optical mode is often taken into account: the polarization (see Fig.~\ref{fig:osciquantum2}). Only one local oscillator is mixed to the signal, as orthogonally polarized beams. With help of a quarter wave plate, the relative phase between the orthogonally polarized signal and local oscillator field is changed by $\pi/2$, entering one of the two polarization beam splitters in order to measure the two conjugate quadratures at the two detector pairs. This setup is used e.g. as receiver in quantum key distribution systems~\cite{Heim,Imran}.

\begin{figure}[h!]
 \center
 \includegraphics[height=150pt]{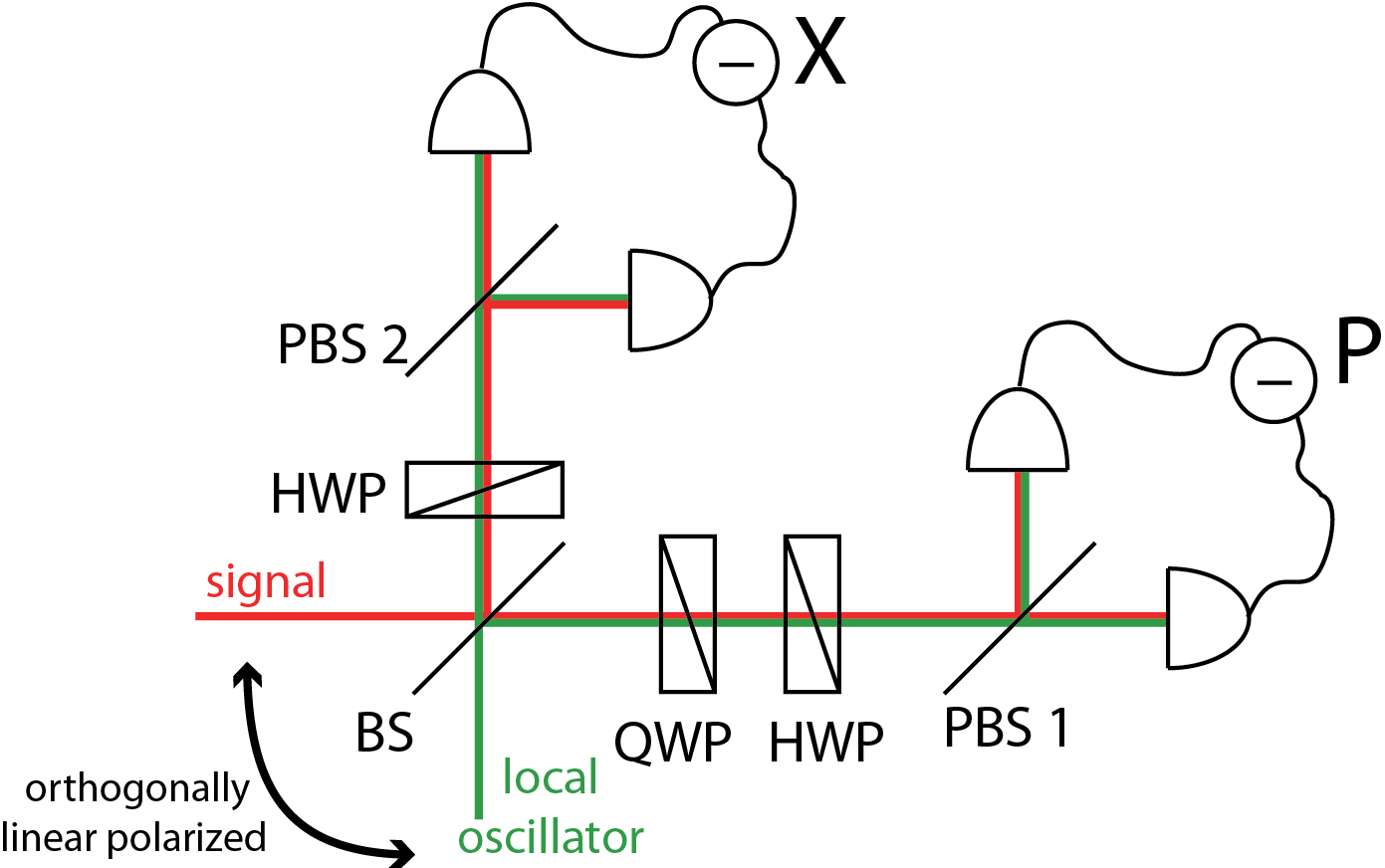}
 \caption{Heterodyne or double-homodyne measurement with help of the polarization of the light field. BS: beamsplitter; PBS: polarization beam splitter; HWP: half wave plate, used for aligning the polarization of signal and LO; QWP: quarter wave plate, aligned with one of the polarization directions of signal or LO.}
 \label{fig:osciquantum2}
\end{figure}
\end{paragraph}

\section{Squeezing the quantum uncertainty}
A single mode state of a light field can generally be described as an expansion of photon-number states, known as Fock-basis:
\begin{align}
 \ket{\varPsi} = \sum\limits_{n=0}^{\infty} c_n \ket{n},
\end{align}
with n-photon \emph{Fock-state} $\ket{n}$. The vacuum state is written as $\ket{0}$ and its uncertainty for both quadratures $\hat{X}$ and $\hat{P}$ is:
\begin{align}
  \gen{\Delta \hat{X}^2} = \frac{1}{4} = \gen{\Delta \hat{P}^2},
 \end{align}
whereas the mean value is zero: $\gen{\hat{X}} = \gen{\hat{P}} = 0$.
\\
\noindent Let us consider a state which consists of the vacuum state and a small admixture of a one-photon state $\ket{1}$, written as $\ket{\varPsi}  = (\ket{0}+\varepsilon \ket{1}) \frac{1}{\sqrt{1+\left|\varepsilon\right|^2}}$ with $\varepsilon \ll 1$. The mean value of this state can be calculated as:
 \begin{equation}
 \gen{\hat{X}} =  \bra{\varPsi} \hat{X} \ket{\varPsi} = \frac{1}{2(1+\left\vert\varepsilon\right\vert^{2})} (\varepsilon + \varepsilon^*) \approx \frac{1}{2} (\varepsilon + \varepsilon^*)
 \end{equation}
 with help of Eq.~\ref{eq:X}. Analogically it follows:
 \begin{equation}
 \gen{\hat{P}} = \frac{1}{2\ii(1+\left|\varepsilon\right|^{2})} (\varepsilon - \varepsilon^*) \approx \frac{1}{2\ii} (\varepsilon - \varepsilon^*)
 \end{equation}
Deriving the variance of both quadratures and neglecting terms of $\left|\varepsilon\right|^{2}$, we obtain:

\begin{align}
\gen{\Delta \hat{X}^2}  = &  \bra{\varPsi} {\hat{X}}^2 \ket{\varPsi}-\bra{\varPsi} \hat{X} \ket{\varPsi}^2 \approx \frac{1}{4}
\label{Eq:Varx}
\end{align}
\begin{align}
\gen{\Delta \hat{P}^2}  = &  \bra{\varPsi} {\hat{P}}^2 \ket{\varPsi}-\bra{\varPsi} \hat{P} \ket{\varPsi}^2 \approx \frac{1}{4}.
\label{Eq:Varp}
\end{align}

We see that, in this case, the variance is independent of the parameter $\varepsilon$. This means, that the small admixture of a one-photon-state causes a shift of the vacuum state without changing its shape in phase space, i.e. its uncertainty.
\begin{figure}[h!]
 \center
 \includegraphics[height=100pt]{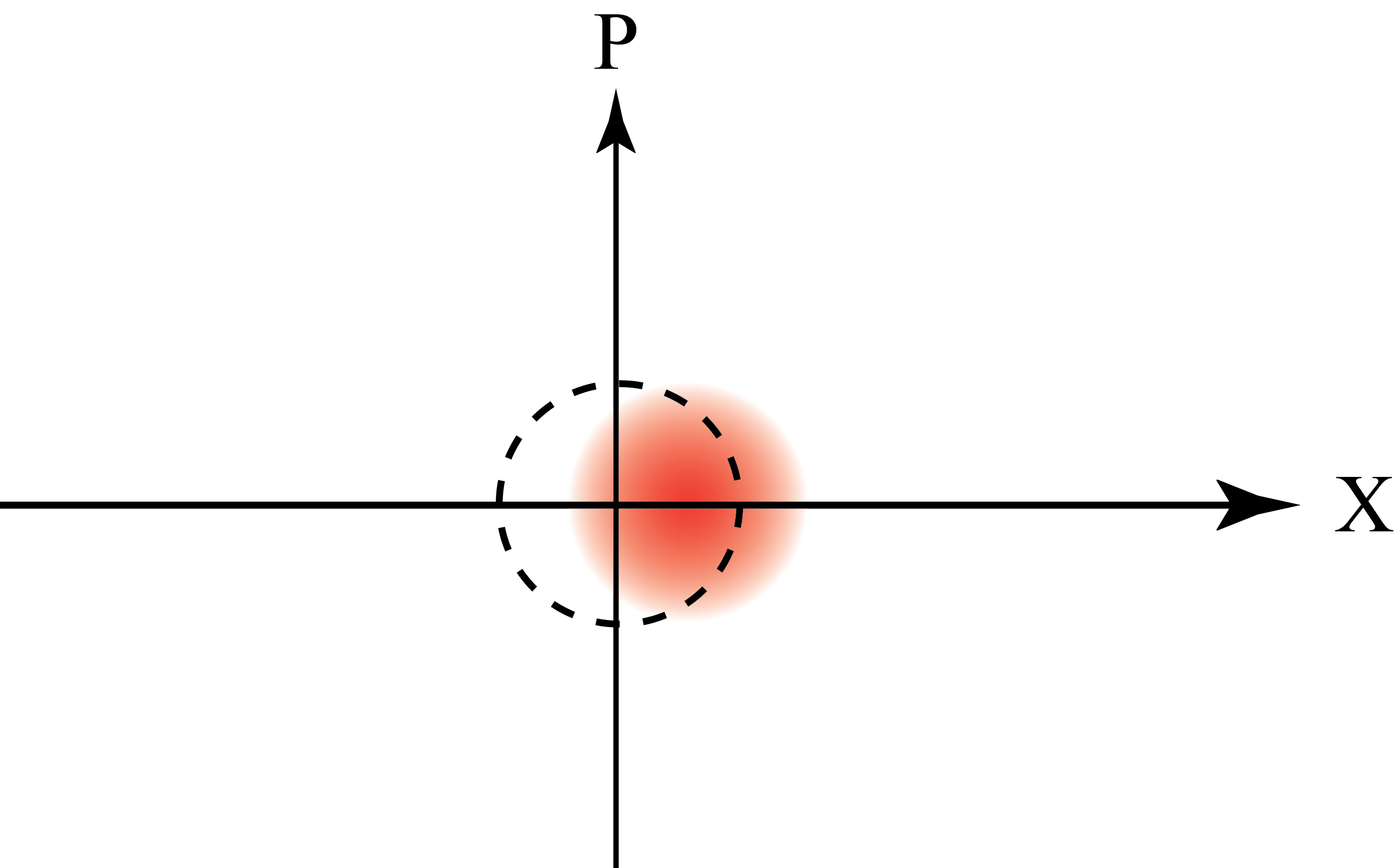}
\end{figure}
\\
\\
Proceeding in the same way for a small admixture of a two-photon-state $\ket{2}$, written as $\ket{\varPsi}  = (\ket{0}+\varepsilon \ket{2}) \frac{1}{\sqrt{1+\left|\varepsilon\right|^2}}$ with $\varepsilon \ll 1$, gives a different result. The mean values are calculated to be:
\begin{align}
 \gen{\hat{X}} &=  \bra{\varPsi} \hat{X} \ket{\varPsi} \approx 0\\
  \gen{\hat{P}} &=  \bra{\varPsi} \hat{P} \ket{\varPsi} \approx 0,\\
 \end{align}
 but the variance becomes (again neglecting terms of $\left|\varepsilon\right|^{2}$):
 \begin{align}
\gen{\Delta \hat{X}^2}  = &  \bra{\varPsi} {\hat{X}}^2 \ket{\varPsi}-\bra{\varPsi} \hat{X} \ket{\varPsi}^2 \approx
\frac{1}{4}(1+\sqrt{2}(\varepsilon+\varepsilon^*))\\
\gen{\Delta \hat{P}^2}  = &  \bra{\varPsi} {\hat{P}}^2 \ket{\varPsi}-\bra{\varPsi} \hat{P} \ket{\varPsi}^2 \approx
\frac{1}{4}(1-\sqrt{2}(\varepsilon+\varepsilon^*)).
\end{align} 

Thus, adding a two-photon-state does not change the mean amplitude of a state but only the variance of both quadratures in an inverse way: when $\hat{X}$ decreases, $\hat{P}$ will increase and vice versa. One quadrature is squeezed at the expense of a larger variance of the other quadrature.
\begin{figure}[h!]
 \center
 \includegraphics[height=100pt]{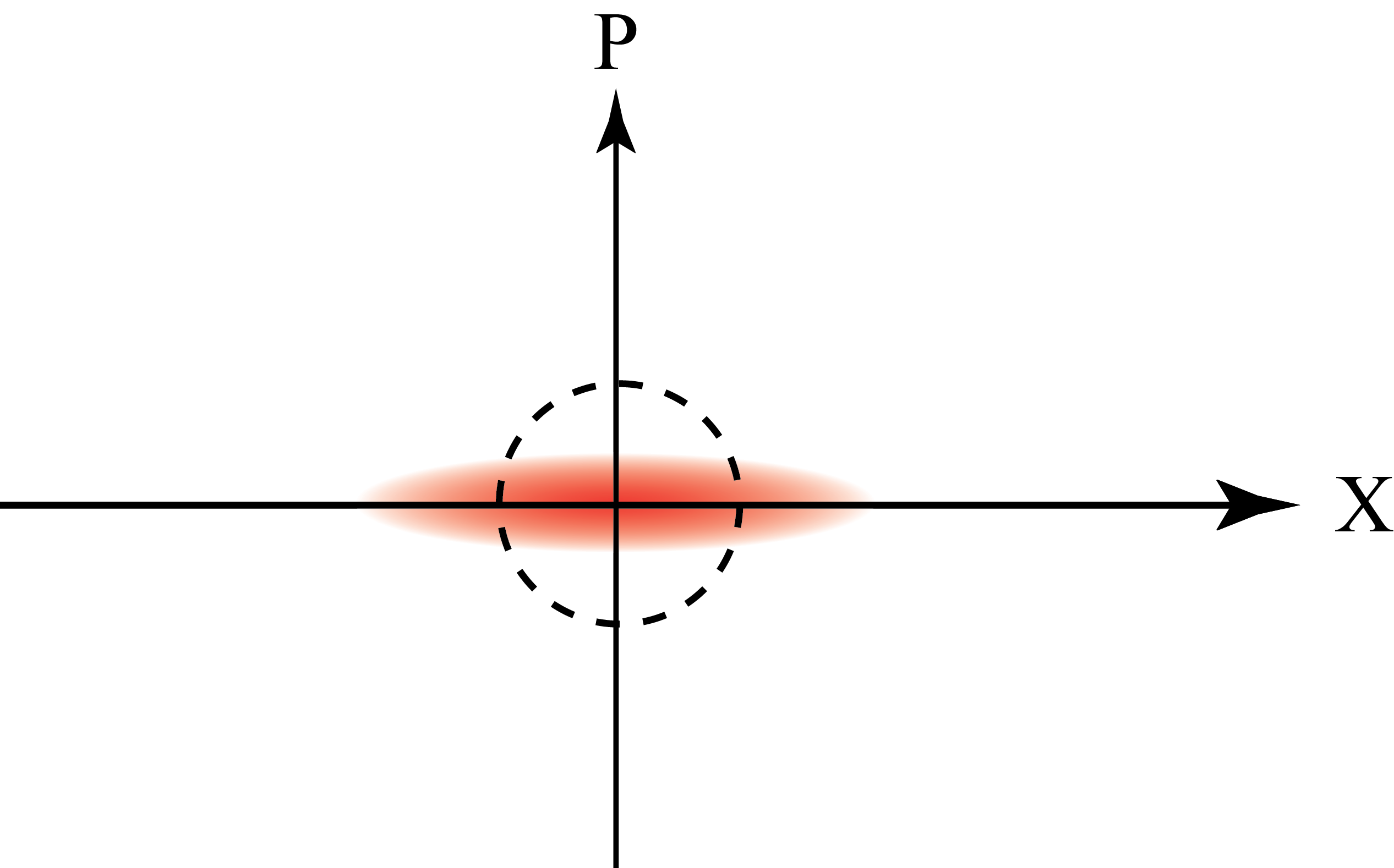}
\end{figure}
\\
We see that the generation of such a state, a \textit{squeezed} state, is based on a process where the simultaneous creation of two \textbf{two photons} is involved, such as parametric down conversion or other nonlinear effects. Note also, that the area in phase space stays constant, with 
\begin{align}
\gen{\Delta \hat{X}^2} \gen{\Delta \hat{P}^2} = 1-2(\varepsilon+\varepsilon^*)^2
\end{align}
in the approximation we used throughout this derivation, namely neglecting terms at order $\varepsilon ^2$. Without this approximation the area will still be constant.

\begin{figure}[h!]
 \center
 \includegraphics[height=120pt]{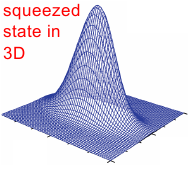}
 \caption{Wigner function of a squeezed state. The height represents the amplitude of the Wigner function and the base plane is the phase space spanned by the two quadratures.}
 \label{fig:wignersqueezed}
\end{figure}

The Wigner function of a squeezed state is depicted in Fig.~\ref{fig:wignersqueezed}.
In the Hamilton operator, squeezing follows from the ``$({\hat {a}}^\dagger)^2$''-term:
\begin{equation}%TODO: check 
 \hat H = \hbar \omega \left(a^{\dagger} a + \frac{1}{2} \right) + \hbar \gamma ({\hat {a}}^\dagger)^2\\
 \end{equation}
 The Hamilton operator determines the time evolution operator $U=\e^{\frac{\ii \hat H t}{\hbar}}$. Under this time evolution operator the vacuum state evolves as follows:
 \begin{align}
  \e^{\frac{\ii \hat H t}{\hbar}} \ket{0} \approx &
 \left( 1+ \ii \omega \left(\hat a^{\dagger} \hat a + \frac{1}{2}\right)t + \ii \gamma ({\hat {a}}^\dagger)^2 \cdot t \right) \ket{0}\\
 = & \left(1+ \frac{\ii \omega t}{2}\right)\ket{0} + \ii \gamma t \sqrt{2}\ket{2}
\end{align}
The latter equation suggests that squeezing can always be observed in a nonlinear interaction containing a quadratic term in the field operator but one has to take the losses into account which are detrimental for the observation of squeezing.

\begin{figure}[h!]
 \center
 \includegraphics[height=150pt]{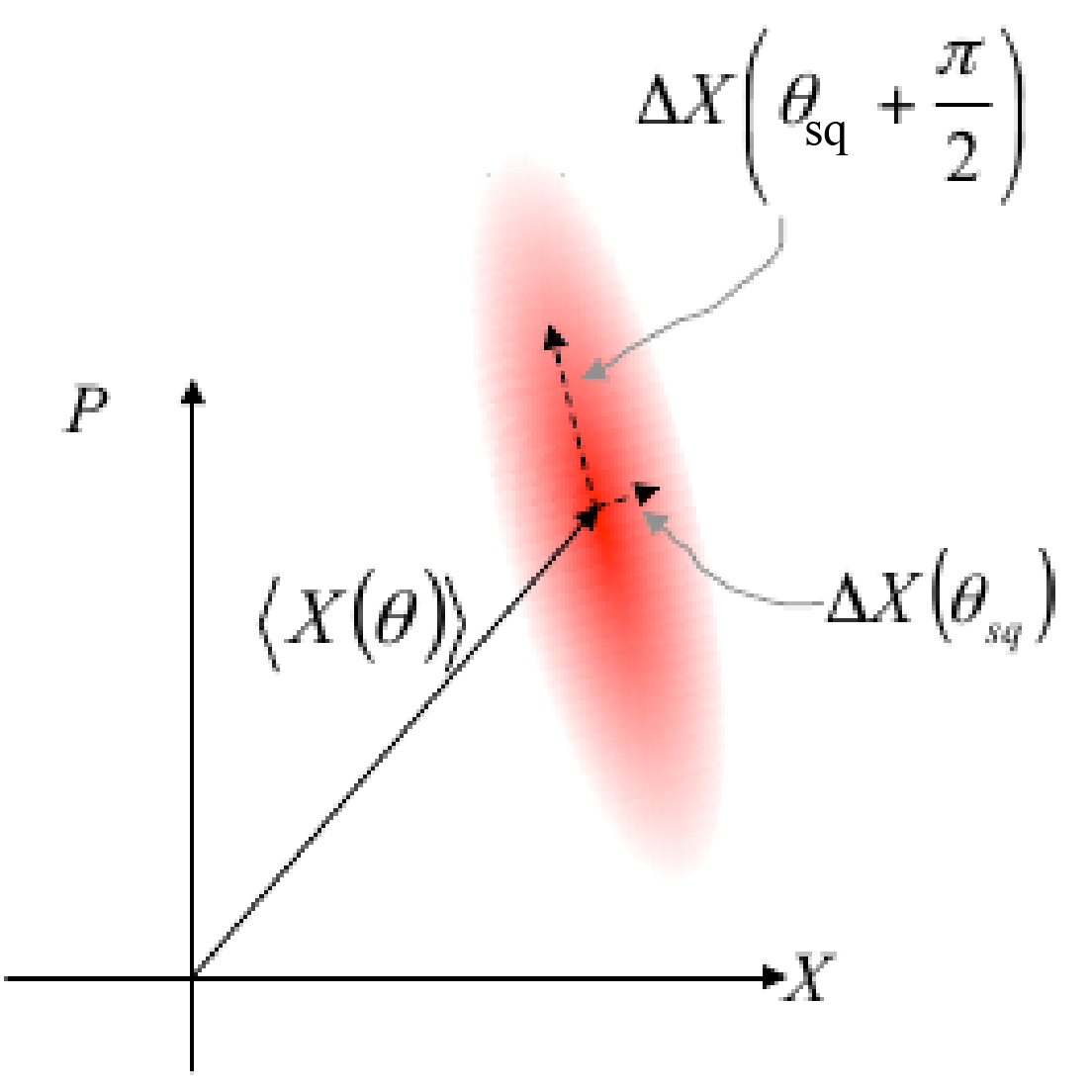}
 \caption{Squeezing along the squeezing angle}
 \label{fig:squeezed2}
\end{figure}

%% slide 5
\vspace{10pt}
\begin{paragraph}{Types of squeezing}
In general, the squeezing axis can be oriented under a skewed angle with respect to the quadratures axes. If the center of gravity is displaced away from the origin, this displacement defines the mean excitation. It is often convenient to use rotated quadratures $X(\theta)$, $P(\theta)$ such that one represents the amplitude and the other one the phase quadrature (see Fig.~\ref{fig:squeezed2}):
 \begin{align}
  \gen{\Delta X^2(\theta)} \cdot \gen{\Delta X^2 (\theta + \pi/2)} \geq \frac{1}{16}
 \end{align}
Squeezing along the direction $\theta_{\mathrm{sq}}$ means:
 \begin{align}
  \gen{\Delta X^2(\theta_{\mathrm{sq}})} < \frac{1}{4}\quad\mathrm{and}\quad \gen{\Delta X^2 (\theta_\mathrm{sq} + \pi/2)} > \frac{1}{4}.
 \end{align}
which allows a classification of three different squeezing operations:
 \begin{itemize}
  \item \textbf{skewed squeezing} if $\theta_{\mathrm{sq}} \neq \theta, \theta+\pi/2$, such as the Kerr effect~\cite{Sizmann} (compare Fig. \ref{fig:squeezed2}) .
  \item \textbf{amplitude squeezing} if $\theta_\mathrm{sq} = \theta$, such as the Kerr effect in combination with an asymmetric fibre-Sagnac interferometer~\cite{Sizmann,Kitagawa,Schmitt,Corney}.
  \item \textbf{phase squeezing} if $\theta_\mathrm{sq} +\frac{\pi}{2} = \theta$, such as with an optical parametric amplifier, four wave mixing~\cite{Bachor}.
  \end{itemize}  
\end{paragraph}

%% slide 13
\section{Intensity correlations}

The correlation functions are generally derived from the wave function, which requires the knowledge of the wave function of a state which is not always straight forward or even possible, for example for a thermal state. In the last section, we convinced ourselves that the Wigner function is a convenient representation of a quantum state in phase space, especially for the common states, pure or mixed, for which the Wigner function is Gaussian (thermal, coherent, squeezed). The Wigner-function is measured by a homodyne setup, projecting the state onto one axis, where the projection is always positive. Thus, it would be practical to also use the Wigner function for calculating the correlation functions. 
\begin{figure}[h!]
 \center
 \includegraphics[height=25pt]{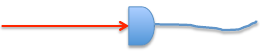}
\end{figure}
\\
Let us first start with a direct detection setup. The detection process requires \emph{normal ordering}. Hence, all dagger operators have to be on the left! The normal ordering is justified in direct detection where the annihilation operator diminishes the photon number by 1, when one photon is detected.
\\
\\
Given an initial state $\ket{i}$ and a final state $\ket{f}$, the probability amplitude for the detection of a photon at time $t$ is
\begin{align}
 \bra{f} \hat a(t) \ket{i},
\end{align}
whereas the probability amplitude for detecting a second photon at time $t'$ conditioned on the first detection at time $t$ is
\begin{align}
 \bra{f} \hat a(t') \hat a(t) \ket{i}.
 \label{Eq:f}
\end{align}
Using this probability amplitude, the probability of a correlation, i.e. for jointly detecting one photon at time $t$ and the other photon at time $t'$, can be derived by taking the absolute squares of Eq.~\ref{Eq:f} and summing over all final states $\ket{f}$:
\begin{align}
 \sum\limits_f \vert \bra{f} \hat a(t') \hat a(t) \ket{i} \vert^2 
 =& \sum\limits_f \bra{i} \hat a^{\dagger} (t) \hat a^{\dagger} (t') \ket{f}\bra{f} \hat a (t') \hat a(t) \ket{i}\\
 =& \bra{i} \hat a^{\dagger} (t) \hat a^{\dagger} (t') \left(\sum\limits_f \ket{f} \bra{f} \right) \hat a (t') \hat a(t) \ket{i} \\
 =& \bra{i} \hat a^{\dagger} (t) \hat a^{\dagger} (t') \hat a(t') \hat a(t) \ket{i}.
\end{align}
% 
%% slide 8
The ordering is thus determined by the type of measurement.
Using the correspondence between the electric fields and the field operators $\hat a$ and $\hat a^{\dagger}$ gives
\begin{align}
 E E^* \rightarrow& \hat a^{\dagger} \hat a\\
 E^* E E^* E \rightarrow& \hat a^{\dagger} \hat a^{\dagger} \hat a \hat a.
\end{align}
One recognizes the classical correlation function of chapter 2, with the difference that now the ordering matters. 

\begin{figure}[h!]
 \center
 \includegraphics[height=250pt]{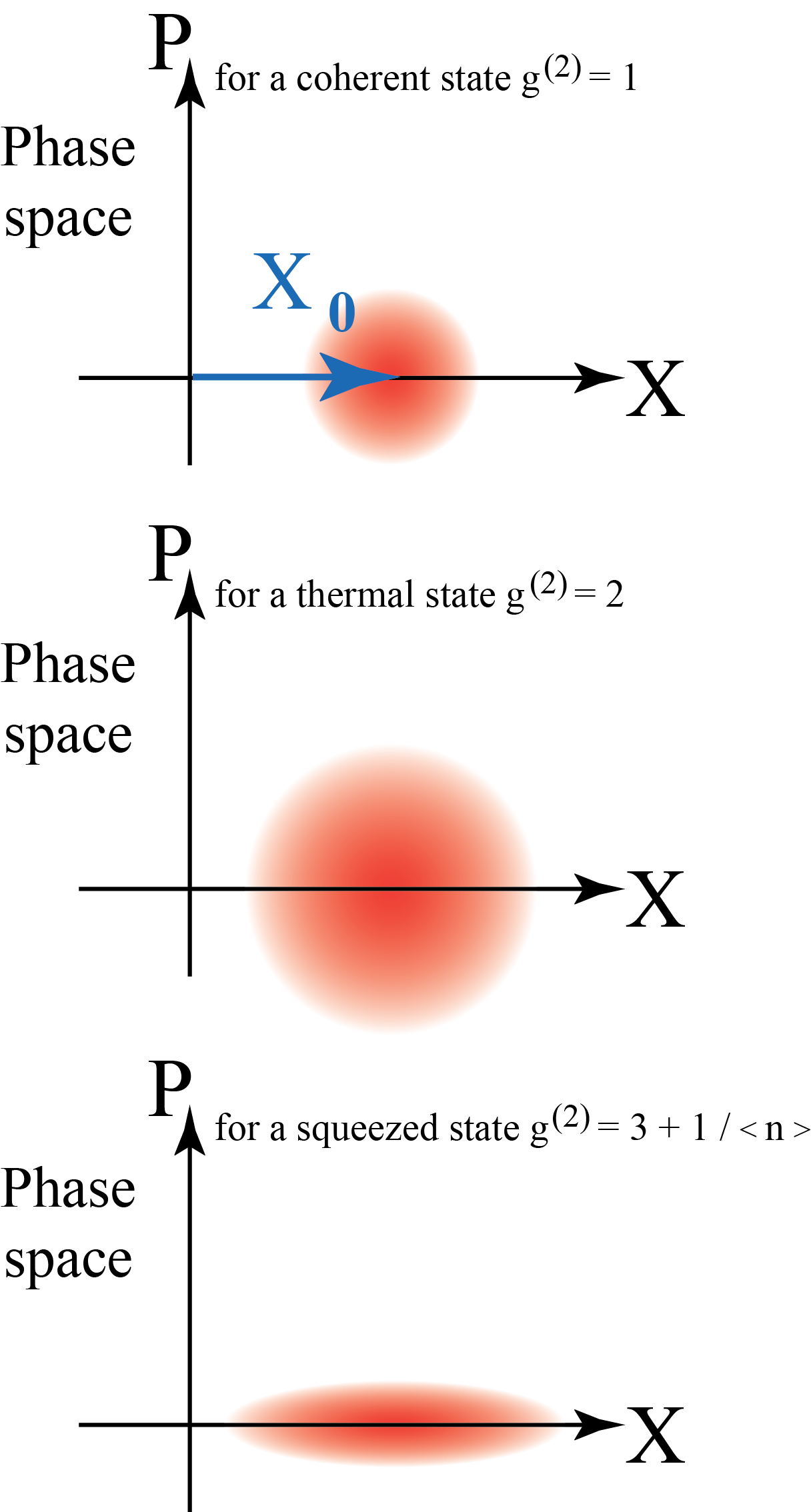}
 \caption{g$^{(2)}$ - correlation function for a coherent, a thermal and a squeezed state.}
 \label{fig:g2_2}
\end{figure}

Let us take the example of the intensity correlation function at $E=0$:
\begin{align}
 g_I^{(2)} = \frac{\gen{\hat a^{\dagger} \hat a^{\dagger} \hat a \hat a}}{\gen{\hat a^{\dagger} \hat a}^2}.
\end{align}
The notion for normal ordering is indicated by colons: 
\begin{align}
 \gen{: \hat a^{\dagger} \hat a \hat a^{\dagger} \hat a:} = \gen{\hat a^{\dagger} \hat a^{\dagger} \hat a \hat a}
\end{align}
If we erroneously assumed that the intensity correlation is given by $\gen{\hat{n}^2}/\gen{\hat{n}}^2$, we can quickly see the problem when applying this to a coherent state, for which $\gen{\Delta \hat{n}^2}=\gen{\hat{n}}$. Using this photon number variance we would get:
\begin{align}
\frac{\gen{\hat{n}^2}}{\gen{\hat{n}}^2} = 1 + \frac{1}{\gen{\hat{n}}}.
\end{align}
But the $\tau=0$ value of the intensity correlation function of a coherent state is 1. The error is that we did not normally order $\hat{n}^2$ as is appropriate for direct detection.
For the proper correlation function we have:
\begin{align}
 g_I^{(2)} = \frac{\gen{:\hat n^2:}}{\gen{\hat n}^2} = \frac{\gen{\hat a^{\dagger} \hat a^{\dagger} \hat a \hat a}}{\gen{\hat a^{\dagger} \hat a}^2}
 \end{align}
Since $\hat a\ket{\alpha}=\alpha \ket{\alpha}$ and $\bra{\alpha} \hat{a}^{\dagger} = \bra{\alpha} \alpha ^*$ for a coherent state, we can conclude the correct value for the correlation function:
 \begin{align}
 \frac{\gen{\hat a^{\dagger} \hat a^{\dagger} \hat a \hat a}}{\gen{\hat a^{\dagger} \hat a}^2} = \frac{ \alpha ^* \alpha ^* \alpha \alpha }{(\alpha ^* \alpha)^2} \rightarrow g_I^{(2)} = 1.
\end{align}
For the intensity correlation function defined for direct detection, we always have to use normal ordering, which corresponds to the P-distribution. But if we want to use the Wigner-function, as stated at the beginning of this paragraph, we need field operators that are ordered \textit{symmetrically}. We will derive in the following the g$^{(2)}$ - correlation function based on the Wigner function, for a coherent, a thermal and a squeezed state (see Fig.~\ref{fig:g2_2}). From text books, we already know the results~\cite{Loudon,Barnett,Knight,Walls}. These are the references to which the alternative derivation discussed will have to be compared\footnote{One of the authors (GL) is preparing a publication on the topic together with Wolfgang P. Schleich.}.

% 

% slide 14
We first calculate the photon number operator while using symmetric ordering, that is we sum the terms with
all possible different orderings and divide by the number of terms. As a result, we obtain a relation between
the photon number operators in normal and symmetric ordering which is indicated by ``Sym'':

\begin{align}
 \text{Sym}(\hat n) = \frac{1}{2} (\hat a \hat a^{\dagger} + \hat a^\dagger \hat a) = \hat n + \frac{1}{2}\\
 \Rightarrow \hat n = \text{Sym}(\hat n) - \frac{1}{2}
\end{align}
The operators $\hat{a}$ and $\hat{a}^{\dagger}$ have been replaced with help of the commutator:
$ \left[ \hat{a}, \hat{a}^{\dagger} \right] = 1 = \hat{a} \hat{a}^{\dagger} -  \hat{a}^{\dagger} \hat{a} $. In order to obtain $\text{Sym}(\hat{n}^2)$, we need to take into account the symmetrical product of four field modes, twice the $\hat a$ and $\hat a^\dagger$ operators in all possible constellations which results in:
\begin{align}
 \text{Sym}(\hat{n}^2) =& \frac{1}{6} (\hat a^\dagger \hat a^\dagger \hat a \hat a + \hat a \hat a \hat a^\dagger \hat a^\dagger +
 \hat a \hat a^\dagger \hat a \hat a^\dagger + \hat a^\dagger \hat a \hat a^\dagger \hat a + \hat a^\dagger \hat a \hat a \hat a^\dagger +
 \hat a \hat a^\dagger \hat a^\dagger \hat a)
 \label{Eq:symn2}
 \end{align}
The factor is 1/6, since we have 6 different terms. This approach can be generalized to:
\begin{align}
\gen{\text{Sym}(\hat{n})^k}=\frac{1}{\left(
\begin{array}{c}
2k\\
k\\
\end{array}
\right)}\gen{\hat{a}^k \hat{a}^{\dagger k} + \ldots}
\end{align}
but this is not straight forward at all. One can imagine Eq.~\ref{Eq:symn2} as the average of all possible constellations of $\hat a$, $\hat a$, $\hat a^\dagger$ and $\hat a^\dagger$. The reader can refer to a theoretical paper for details about the product of operators in symmetric ordering~\cite{Case}. 
Using again the known operator constellations such as Eq.~\ref{Eq:acomm} and Eq.~\ref{Eq:n}, the non-symmetric components can be rewritten as: 

 \begin{align}
 \Rightarrow  \text{Sym}(\hat{n}^2) =& \hat a^\dagger \hat a^\dagger \hat a \hat a +2 \hat a^\dagger \hat a + \frac{1}{2}\\
 \Rightarrow \text{Sym}(\hat{n}^2) =& \hat n^2 + \hat n + \frac{1}{2}
\end{align}
Note that $\text{Sym}(\hat{n}^2)$ is not the simple square of $\text{Sym}(\hat{n})$! This would give: $\text{Sym}(\hat{n})^2=\hat n^2 + \hat n + \frac{1}{4}$.
Hence, the second-order correlation function can be written with the photon-number operator in symmetric ordering:

\begin{align}\label{eqn:g20}
 g^{(2)}(0) = \frac{\gen{\hat a^\dagger \hat a^\dagger \hat a \hat a}}{\gen{\hat a^\dagger \hat a}^2} = 
 \frac{\gen{\hat n^2} - \gen{\hat n}}{\gen{\hat n}^2} = \frac{\gen{\text{Sym}(\hat n^2)} - 2 \gen{\text{Sym}(\hat n)} + \frac{1}{2}}{\left( \gen{\text{Sym}(\hat n)} - \frac{1}{2}\right)^2}
\end{align}

%% slide 15

With help of this expression, we will derive g$^{(2)}(0)$ for different states. We already stressed that the Wigner function corresponds to symmetric ordering of the field operators. This fact greatly
facilitates the calculation of the expectation values of symmetrized powers of the photon number operator, such that one
just has to perform an integration, averaging the Wigner function. In phase space, the photon number $n$ is related
to the sum of the squares of two orthogonal quadratures:

\begin{align}
\text{Sym}(\hat{n}) \quad \rightarrow \quad X^2+P^2
\end{align}
and the expectation value of a power of $\hat{n}$ is

\begin{align}
\gen{\text{Sym}(\hat{n}^k)}=\int (X^2+P^2)^k W(X,P)\dd X\dd P.
\label{Eq:Sym}
\end{align}
The evaluation of this integral is straight forward and for large $k$, Eq.~\ref{Eq:Sym} is very convenient. The calculation is particularly simple if the Wigner function is a Gaussian $W(X,P)=N\cdot e^{-aX^2-bP^2}$. To give an example:
\begin{align}
\gen{\text{Sym}(\hat{n})}=N\int(X^2+P^2) e^{-aX^2-bP^2}\dd X\dd P=\gen{X^2+P^2}.
\end{align}
In analogy to Eqs.~\eqref{eqn:phi4} and \eqref{eqn:xn} we find

\begin{align}
\gen{x^4}=3\gen{x^2}^2
\end{align}
Therefore, we can rewrite $\text{Sym}(\hat{n}^2)$:
\begin{align}
\gen{\text{Sym}(\hat{n}^2)} =& \gen{( X^2 + P^2)^2} = \gen{ X^4} + 2 \gen{ X^2  P^2} + \gen{ Y^4}\\
  =& \gen{ X^4} + 2\gen{X^2} \gen{P^2} + \gen{P^4}\\
  =& 3\gen{X^2}^2 + 2\gen{X^2} \gen{P^2} + 3\gen{P^2}^2
  \label{Eq:rec}
\end{align}
Let us first consider a \textbf{squeezed} vacuum state where the variances are the following (refer to Eq.~\ref{Eq:Varx} and Eq.~\ref{Eq:Varp}):
\begin{align}
 \gen{ X^2} = \frac{\varepsilon}{4}\quad \mathrm{and}\quad \gen{ P^2} = \frac{1}{4 \varepsilon},
\end{align}
with $\varepsilon$ being the squeezing parameter.
Thus:
\begin{align}
 \gen{\text{Sym}(\hat n^2)} = 3 \frac{\varepsilon^2}{16} + 2\frac{1}{16} +3\frac{1}{16\varepsilon^2} = \frac{3}{16}\left(\varepsilon^2 +\frac{1}{\varepsilon^2}\right) + \frac{1}{8}
\end{align}
and with $\gen{\text{Sym}(\hat n)} = \frac{1}{4} \left(\varepsilon + \frac{1}{\varepsilon}\right) \Rightarrow \left(\varepsilon^2 + \frac{1}{\varepsilon^2}\right) = 16 \gen{\text{Sym}(\hat n)}^2 -2$
\begin{align}
 \Rightarrow \gen{\text{Sym}(\hat n^2)} = 3 \gen{\text{Sym}(\hat n)}^2 - \frac{1}{4}
\end{align}
\noindent The intensity correlation function given in Eq.~\ref{eqn:g20} becomes
\begin{align}
 g^{(2)} (0) =&\frac{3\gen{\text{Sym}(\hat n)}^2 - 2 \gen{\text{Sym}(\hat n)} + \frac{1}{4}}{\left(\gen{\text{Sym}(\hat n)} - \frac{1}{2}\right)^2}\\
  =& \frac{3 \gen{\text{Sym}(\hat n)}^2 - 3 \gen{\text{Sym}(\hat n)} + \frac{3}{4} + \gen{\text{Sym}(\hat n)} - \frac{1}{2}}{\left(\gen{\text{Sym}(\hat n)} - \frac{1}{2}\right)^2}\\
  =& 3 + \frac{1}{\gen{\text{Sym}(\hat n)} - \frac{1}{2}}
\end{align}
For experimental evidence for the factor 3, see Refs.~\cite{Boitier,Iskhakov}.

% slide 17

For a \textbf{thermal} state, we know that $\gen{\Delta \hat{X}^2}  = \gen{\Delta \hat{P}^2}$ and with Eq.~\ref{Eq:rec} and $\text{Sym}(\gen{ \hat{n}})  = 2 \gen{\hat{X}^2}$ we can write:

\begin{align}
 \gen{\text{Sym}(\hat{n}^2)} = 2\gen{\text{Sym}(\hat{n})}^2
\end{align}
and the intensity correlation function given in Eq.~\eqref{eqn:g20} becomes
\begin{align}
 g^{(2)}(0) = 2.
\end{align}

%% slide 18
For a \textbf{coherent} state $X= X_0 + \Delta X$, $P=\Delta P$ with $\gen{\Delta \hat X^2}=\frac{1}{4}$ and $\gen{\Delta \hat P^2} =\frac{1}{4}$ we find
\begin{align}
 \gen{\text{Sym}(\hat n)} =& \gen{ X_0^2 + \Delta X^2 + 2  X_0 \Delta X + \Delta P^2} =  X_0^2 + \frac{1}{2}\\
 \gen{\text{Sym}(\hat{n}^2)} =&  X_0^4 +2  X_0^2 + \frac{1}{2} = \gen{\text{Sym}(\hat n)}^2 + \gen{\text{Sym}(\hat n)} - \frac{1}{4}
\end{align}
and the intensity correlation function given in Eq.~\eqref{eqn:g20} becomes
\begin{align}
 g^{(2)}(0) = 1.
\end{align}

\begin{figure}[h!]
 \center
 \includegraphics[height=150pt]{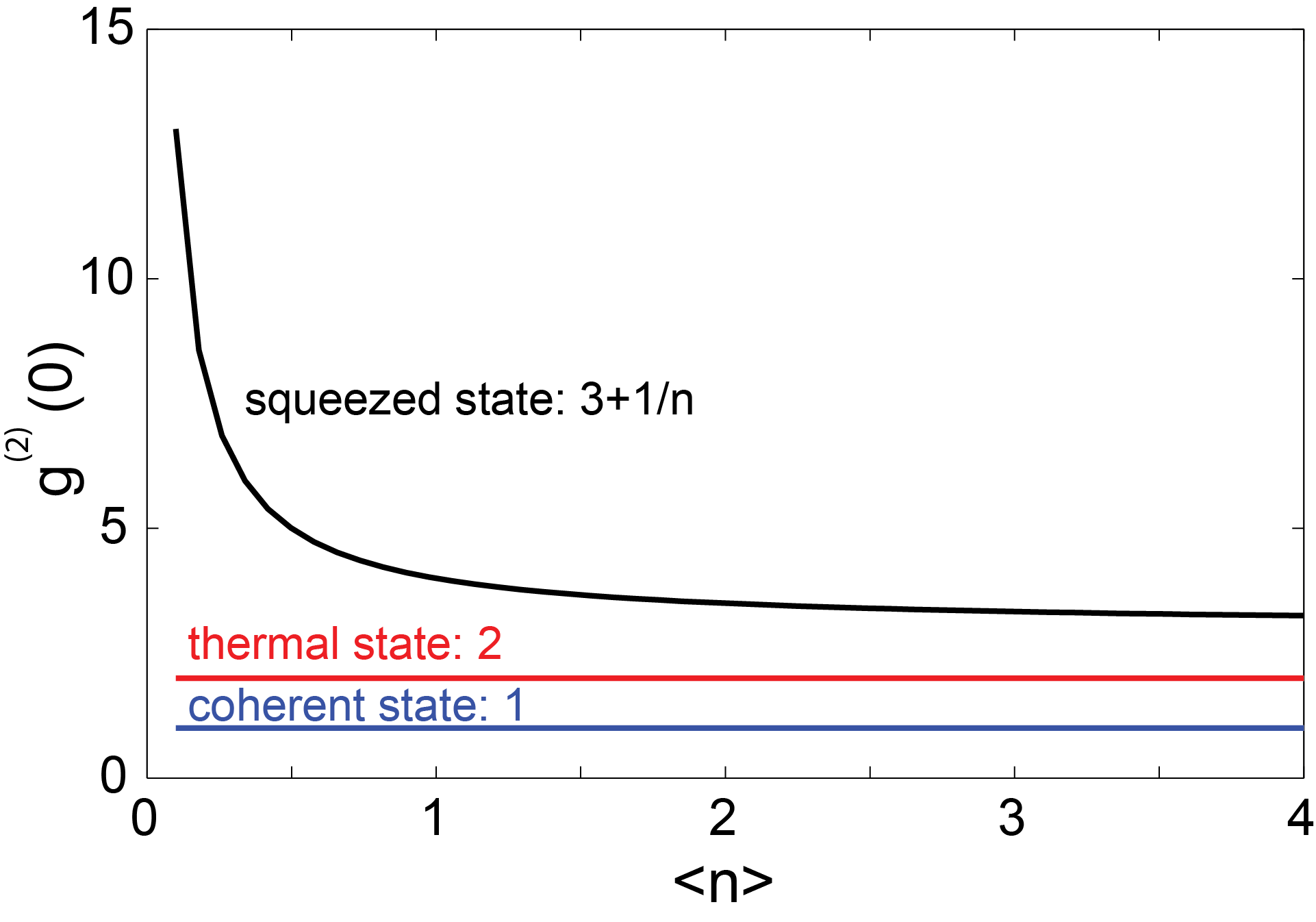}
 \caption{Correlation function $ g^{(2)}(0)$ for a coherent state, squeezed state and thermal state.}
 \label{fig:corr2}
\end{figure}

These cases are shown in Fig.~\ref{fig:corr2}. Obviously it is not possible to define a consistent value of $g^{(2)}(0)$ for the vacuum state because different
ways of approaching the limit suggests different values. This may seem disturbing at first sight but it is not, because it is impossible to do direct detection
of the vacuum state. The general procedure of calculating quantities such as the intensity correlation function through averaging with help of the Wigner function
is generally applicable and not limited to Gaussian states.

\chapter{Quantum optics of several modes}

In this chapter, the interaction between modes will be considered. Modes can differ in their spatial and their temporal properties. An important component for the interaction is the beam splitter which will be discussed in detail at the beginning. Moreover, the interference of a carrier with vacuum frequency sidebands as an essential aspect of squeezing the quantum uncertainty will be explained. Finally, we will have a look on the Bogoliubov transformation and how it can describe different processes such as squeezing, amplification, phase conjugation and attenuation in a unified way.

\section{Continuous variables and beam splitters}

Quantum systems can often be described by a stochastic model. For certain quantum states, a stochastic description is completely sufficient,
for others it cannot be used due to their specific nature. A famous example is the Bell inequality \cite{Bell}, a violation of which in experiment is a clear sign that a stochastic model will not work. However, a coherent state with its mean value presents a ``simple'' quantum state whose Wigner
function is not negative and it can easily be described with a Gaussian state. Hence, a stochastic interpretation is appropriate and practical for the
purpose of describing several quantum modes and their correlation.
\begin{figure}[h!]
 \center
 \includegraphics[height=140pt]{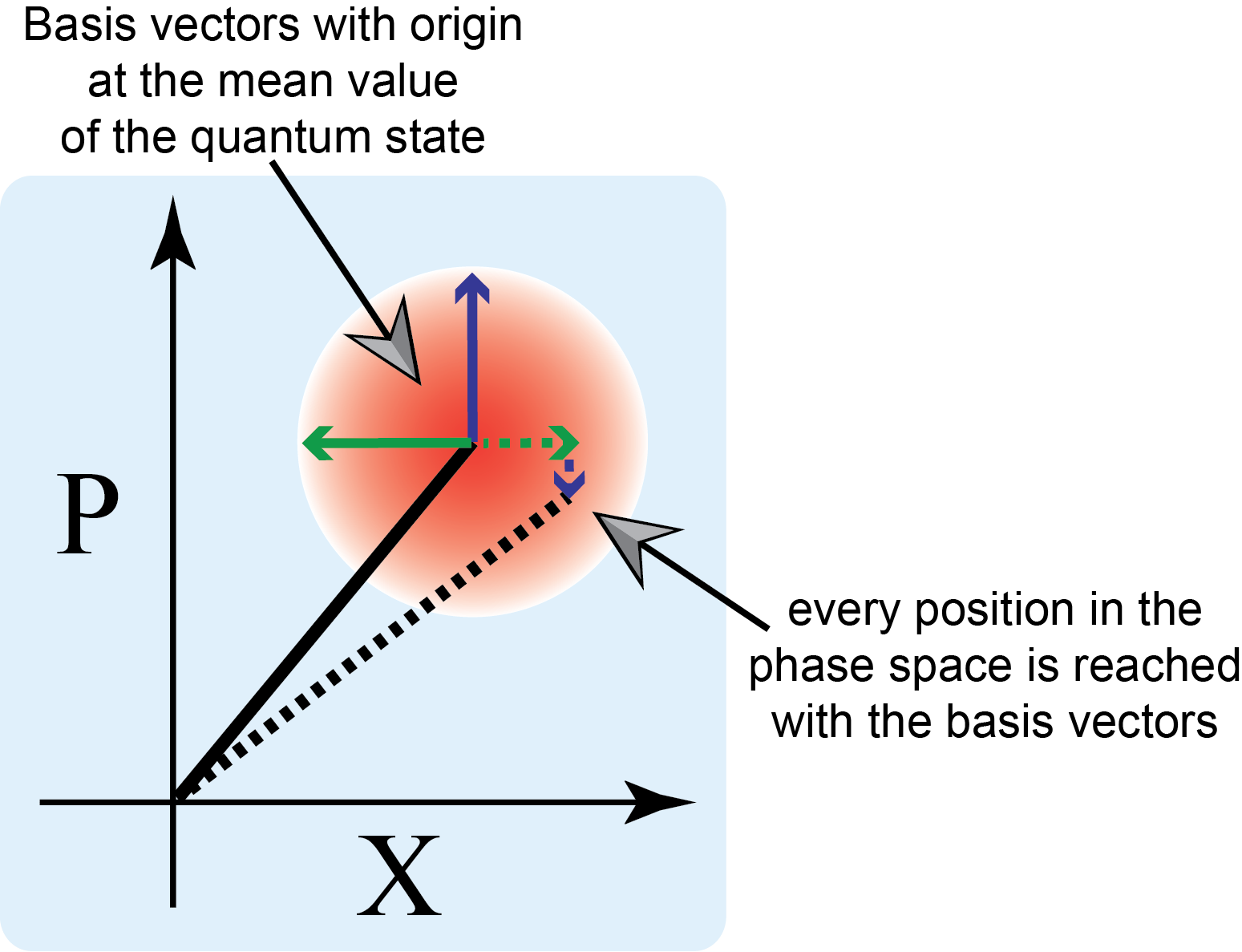}
 \caption{$\leftarrow, \uparrow$ are the basis vector. Multiplied by \emph{stochastic Gaussian variables} they span the \emph{phase space distribution}.}
 \label{fig:basis}
\end{figure}
For a field in a coherent state the uncertainties in amplitude and phase direction are the same and the contour is circular. Figure~\ref{fig:basis} shows a phase diagram where the contour of the Wigner function at half maximum is indicated. Let us now introduce two orthogonal arrows which span the circular region of uncertainty of
the field. In order to mimic the full uncertainty area one has to form a linear superposition of these two arrows with stochastic coefficients of Gaussian statistics. For the coherent state e.g. there will be no correlation between the two stochastic variables. The 2-dimensional phase space is hence spanned by two independent variables $X$ and $P$ with all possible values of the Gaussian distribution. When we multiply the basis vectors with a Gaussian distribution, we get all positions in the phase space with a certain probability. One such possible position is indicated in Fig.~\ref{fig:basis} with the dashed line. This approach holds in particular for all states with Gaussian Wigner function.

\subsection{Beam splitter - phase relations}

 \begin{figure}[h!]
    \center
    \includegraphics[height=100pt]{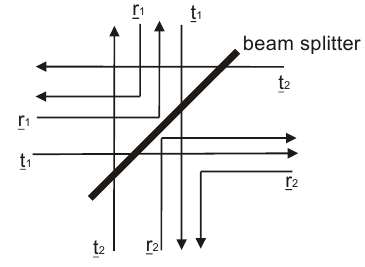}
    \caption{Transmission and reflection at a beamsplitter.}
    \label{fig:BS1}
  \end{figure}
\noindent A beam splitter has two input and two output ports~\cite{Haensch}. In a single mode picture each of these
ports is associated with a spatial mode of the quantized electromagnetic field. The corresponding
field operators are $\hat a_{\mathrm{in}1}$, $\hat a_{\mathrm{in}2}$, $\hat a_{\mathrm{out}1}$, and $\hat a_{\mathrm{out}2}$. They are related by the transmission and reflection coefficients $r_1$, $r_2$, $t_1$, and $t_2$ (see Fig.~\ref{fig:BS1}):

\begin{align}
\hat a_{\mathrm{out}1} \quad=\quad & t_1 \hat a_{\mathrm{in}1}+r_1 \hat a_{\mathrm{in}2} \\
\hat a_{\mathrm{out}2} \quad=\quad & r_2 \hat a_{\mathrm{in}1}+t_2 \hat a_{\mathrm{in}2}
\end{align}

Note that the coefficients are complex valued to account for both amplitude and phase.
\begin{figure}[h!]
    \center
    \includegraphics[height=100pt]{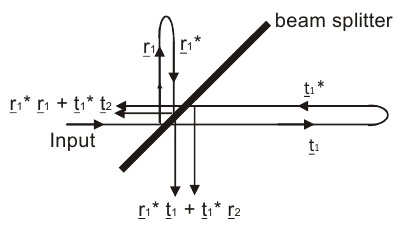}
    \caption{Scenario for time reversal, transmission and reflection at a beamsplitter.}
    \label{fig:BS2}
  \end{figure}
The general relation between the corresponding coefficients for the electric fields has been derived by G. G. Stokes (1849) in a remarkable paper which he based upon the principle of time reversal symmetry~\cite{Stokes}. He considered one incoming ray being split by the beam splitter. Then he argued that if the two rays exiting the beam splitter are time reversed they will have to interfere upon re-arrival at the beam splitter such as to generate the time reversed incoming beam (Fig.~\ref{fig:BS2}). In full generality, Stokes allowed for the reflection and transmission coefficients $r_1$, $r_2$, $t_1$, and $t_2$ to be all different in both amplitude and phase. Stokes' argument leads to the following formulae when using the complex notation (which he did not use). In this notation time reversal corresponds to taking the complex conjugate of the wave amplitude. We offer a short excursion into the correspondence between time reversal and phase conjugation at the end of this paragraph. Time reversal means that all the light goes back to input port 1 and no light out of input port 2. This leads to the following expressions at both inputs of the beam splitter (on the left and bottom in Fig.~\ref{fig:BS2}):
 \begin{align}
  1\quad=\quad& r_1^* r_1 + t_1^* t_2\\
  0\quad=\quad& r_1^* t_1 + t_1^* r_2
 \end{align}

We rewrite $t_k = \left| t \right| \e^{\ii \tau_k}, r_k = \left|r \right| \e^{\ii \rho_k}, k=1,2$. Taking into account energy conservation $\left| r_k^2 \right| + \left| t_k^2 \right| = 1$, for $k =1,2$, we conclude that $t_1 =  t_2  =  t $ and $\left| r_1 \right| = \left|r_2 \right| = \left| r \right|$ and therefore:
 \begin{align}
  \tau_1 - \tau_2 \quad=\quad 0 + 2m\pi,  \quad m \in \mathbb{N}
  \end{align}
  and
   \begin{align}
   0\quad=\quad& \left|r\right| \e^{-\ii \rho_1} \cdot  \left|t\right| \e^{\ii \tau_1} + \left|t\right| \e^{-\ii \tau_1} \cdot  \left|r\right| \e^{\ii \rho_2}\\
   \Rightarrow \e^{-\ii \rho_1 + \ii \tau_1} \quad=\quad & - \e^{-\ii \tau_1 + \ii \rho_2}\\
   \Rightarrow
  2 \tau_1 - \rho_1 - \rho_2 \quad=\quad & \pi + 2m\pi, \quad m \in \mathbb{N}.
 \end{align}
\\
Without loss of generality, we can claim $\tau_1 = 0$ and $m = -1$, so:
\begin{align}
\rho_1+\rho_2 = \pi
\end{align}
\\This leads to two cases:
 \begin{enumerate}
  \item Symmetric beam splitter, $\rho_1 = \rho_2 = \pi/2$, e.\,g. a lamella, a beamsplitting cube
  \item Asymmetric beam splitter, $\rho_1=0, \rho_2=\pi$, e.\,g. a glass plate and a coated plate
 \end{enumerate}
 \vspace{0.5cm}
 In the following we will use the first version. For further references see~\cite{Hamilton}.

\paragraph{Short excursion into time reversal} 

Let us assume an electro-magnetic wave with real and imaginary parts:
\begin{align}
E(r,t)=Ae^{\ii\phi}e^{\ii\omega t - \ii \vec{k}\vec{r}} + c.c.
\end{align}
\noindent If we apply a time reversion operator $\hat{T}$, this changes the sign of each $t$:
\begin{align}
\hat{T} \left(E(r,t)\right)=& \hat{T}\left(Ae^{\ii\phi}e^{\ii\omega t - \ii\vec{k}\vec{r}} + A^{*}e^{-\ii\phi}e^{-\ii\omega t + \ii\vec{k}\vec{r}}\right) =\\
=& Ae^{\ii\phi}e^{-\ii\omega t - \ii\vec{k}\vec{r}} + A^{*}e^{-\ii\phi}e^{\ii\omega t + \ii\vec{k}\vec{r}}=\\
=& A^{*}e^{-\ii\phi}e^{\ii\omega t + \ii\vec{k}\vec{r}} + c.c.
\end{align}
The conjugate amplitude $A^{*}e^{\ii\vec{k}\vec{r}}$ is now with the term oscillating as $e^{\ii \omega t}$ and $Ae^{\ii\vec{k}\vec{r}}$ with the complex conjugate. Thus, we can easily understand that time reversal corresponds to phase conjugation in the mathematical description.

\subsection{Beam splitter - correlation of the quantum uncertainty}
\begin{figure}[b]
 \center
 \includegraphics[height=150pt]{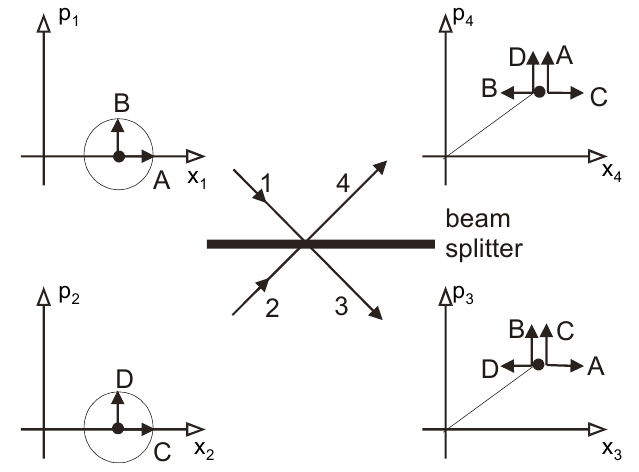}
 \caption{Beam splitter with two coherent input states in the pictorial presentation.}
 \label{fig:BS3}
\end{figure}
The action of a beam splitter on states of light described by a positive valued Wigner function will be to transfer each arrow from an input port to both output ports with reduced amplitudes, keeping in mind that in the end each arrow will have to be multiplied with its own stochastic variable. In the model we have to properly take into account the Stokes relations. If the same stochastically varying input arrow contributes to two output ports one may expect correlations between these two output fields. We will show that this is not the case for every quantum state.

To get used to the arrow description we consider the interference of two coherent states at a beam splitter. Coherent states have a circular region of uncertainty in phase space and as a result all four arrows describing the two coherent states have the same size. The situation is shown graphically in Fig.~\ref{fig:BS3}. Note that the angles in phase space correspond to the classical optical phase of the light beams. The Stokes relations are obeyed by associating a 90$^\circ$ phase shift to each reflection, i.e. the factor ``i'', and 0$^\circ$ phase shift to each transmission. In practice this corresponds to a symmetric beam splitter such as a lamella or a beam splitting cube. The amplitude reduction is not shown, for simplicity. The four input arrows $A$, $B$, $C$, and $D$ determine the field uncertainties in the two output ports 3 and 4. The amplitude uncertainty in output 3 is determined by the projections of all arrows onto the amplitude direction: $A + B + C - D$. Note that $C$ and $D$ reach output 3 by reflection and get a phase shift by 90$^\circ$. Each arrow would then still have to be multiplied with its individual stochastic coefficient. This and the additional amplitude reduction due to the projection is again not shown as it will be the same factor for all arrows in the cases considered here. Likewise the amplitude uncertainty at output 4 is determined by $A - B + C + D$.

Although the uncertainties in both output ports are governed by the same four arrows, they are \textbf{not correlated}. The reason for this lack of correlation can be traced back to the sum of two statistically independent stochastic variables and their difference being again statistically independent. Output 3 is governed by the sum of two statistically independent variables $(A+C)+(B-D)$ and output 4 is determined by their difference $(A + C) - (B - D)$. To be specific, one and the same particular value at output 3 can be the result of infinite many sums of $(A+C)$ and $(B-D)$ values. At output 4 the difference of the same values will span a wide range and not result in just one value. Consequently, there is no correlation between the two outputs although they are determined by the same quantities $A$, $B$, $C$, and $D$.

If, however, the two input states were amplitude squeezed states with close to zero amplitude uncertainty $A$, $C\approx 0$, and correspondingly larger uncertainty in the phase direction, the amplitude uncertainties of the two output ports would be anti-correlated, port 3: ``$(B-D)$'' and port 4 ``$-(B-D)$''. The uncertainties in the phase directions are likewise correlated as can be seen by going through similar arguments for the orthogonal projection\footnote{Actually, one squeezed input field is enough to obtain quantum correlations between the two output fields but the correlations will be less strong.}~\cite{viii}.

The variables $\hat X$ and $\hat P$ do not commutate because:

\begin{align}
\left[\hat X,\hat P\right]=&\hat X\hat P -  \hat P\hat X =\\ =& \frac{1}{4\ii}(\hat a^2 - \hat{a}\hat{a}^\dagger + \hat{a}^\dagger\hat{a} - (\hat{a}^\dagger)^2 - \hat{a}^2 - \hat{a}\hat{a}^\dagger + \hat{a}^\dagger\hat{a} + (\hat{a}^\dagger)^2)=\\
=&\frac{1}{4\ii}(-2\hat{a}\hat{a}^\dagger + 2\hat{a}^\dagger\hat{a})=\frac{\ii}{2}
\end{align}

Nevertheless, we can combine the two output amplitude operator $\hat X_3$ and $\hat X_4$ to $\hat X_3+\hat X_4$ and the phase amplitude operators $\hat P_3$ and $\hat P_4$ to $\hat P_3-\hat P_4$. We calculate the commutator of these combined operators and obtain:

\begin{align}
\left[\hat X_3+\hat X_4,\hat P_3-\hat P_4\right]=\left[\hat X_3,\hat P_3\right] - \left[\hat X_4,\hat P_4\right] - \left[\hat X_3,\hat P_4\right] + \left[\hat X_4, \hat P_3\right] = 0
\end{align}
with
\begin{align}
\left[\hat X_3,\hat P_3\right] = \frac{\ii}{2}, \quad
\left[\hat X_4,\hat P_4\right] = \frac{\ii}{2}, \quad
\left[\hat X_3,\hat P_4\right] = 0, \quad
\left[\hat X_4,\hat P_3\right] = 0 
\end{align}

Thus, the operators $(\hat X_3 + \hat X_4)$ and $(\hat P_3 - \hat{P}_4)$ commutate! We can hence follow that the uncertainty for these operators can be simultaneously well defined, the variances being zero:

\begin{align}
\gen{(\Delta(\hat{X}_3+\hat{X}_4))^2}=0  \quad \text{and}\quad  \gen{(\Delta(\hat{P}_3-\hat{P}_4))^2}=0
\end{align}
\noindent Note that for the single quadratures, the following Heisenberg uncertainty relation is valid:
\begin{align}
\gen{(\Delta \hat{X_i})^2}\gen{(\Delta \hat{P_i})^2}  \geq \frac{1}{16} \quad \text{for i=1,... 4}
\end{align}

The fact that the uncertainty for the combination of two operators vanishes seems to be paradox, for which it is also called the \textbf{Einstein-Podolsky-Rosen paradox (EPR)}, since it
violates a basic quantum mechanical assumption, the Heisenberg uncertainty relation. This ``Gedankenexperiment'' was formulated in 1935 questioning that quantum theory is not
complete~\cite{EPR}. At that time, they supposed that special relativity theory forbids that while measuring two states at a distance decide instantaneously their projection to the
same (or anti-correlated) state (``spooky'' action at a distance) and proposed additional ``local hidden variables''. Only in 1964, John Bell solved this conflict, formulating his
famous Bell theorem, where he proved that it is possible to distinguish by experiment between local theories which claim unobservable local properties, and non-local theories~\cite{Bell}.
These theoretical assumptions have been confirmed in the 1970s~\cite{Fre72,Aspect}. 
\\
\\
We conclude this paragraph with a short overview about different interactions at a beam splitter. As discussed, the interference of two coherent states $\ket{\alpha}$ (or a coherent
state and the vacuum state $\ket{0}$) gives uncorrelated outputs at port 3 and 4. This finds an application for example in cryptography when using post selection.\\
\begin{figure}[h!]
 \center
 \includegraphics[height=34pt]{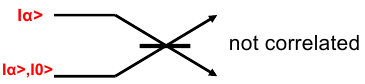}
 \end{figure}\\
When squeezing the two input states (or preparing them with other nonlinear processes) and letting them interact linearly at the beam splitter, the outputs of the beam splitter are
entangled, which is used e.g. for teleportation, secret sharing or quantum erasing~\cite{Barnett}.\\
 \begin{figure}[h!]
  \center
  \includegraphics[height=57pt]{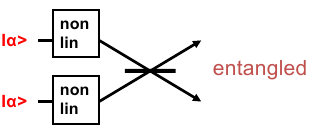}
  \end{figure}\\
Another way to obtain two entangled states, is by using a nonlinear interaction itself, for example a nonlinear crystal, as it is used in Ou et al.~\cite{Ou} for non-degenerate parametric amplification.\\
\begin{figure}[h!]
  \center
  \includegraphics[height=45pt]{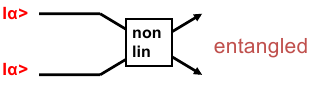}
  \end{figure}

%% slide 31
\section{Squeezing and sideband entanglement}
\begin{figure}[h!]
 \center
 \includegraphics[height=80pt]{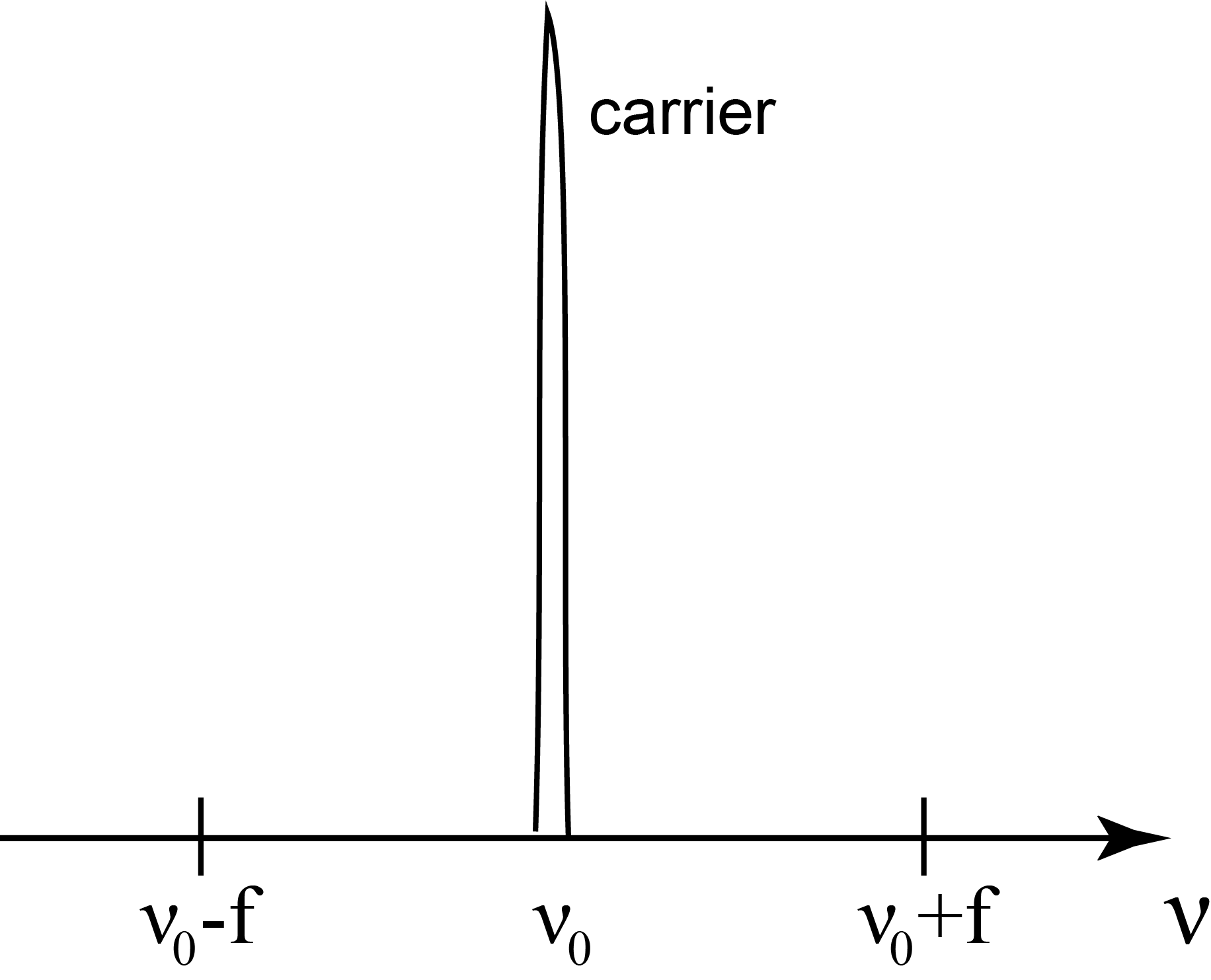}
 \caption{Carrier (laser signal, local oscillator) and sidebands at distance f.}
 \label{fig:Sidebands1}
\end{figure}

\begin{figure}[h!]
 \center
 \includegraphics[height=160pt]{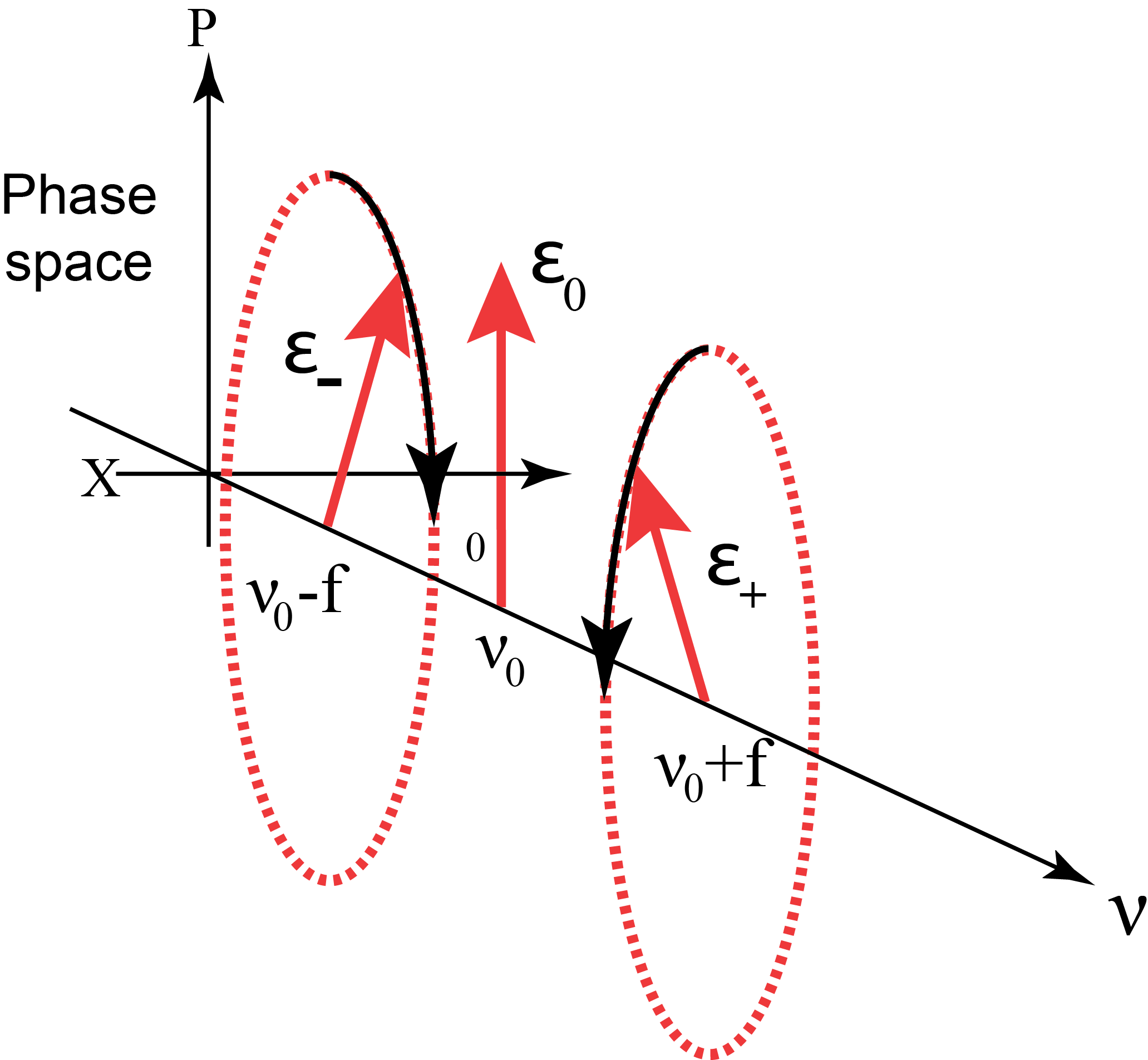}
 \caption{Phase space diagrams at upper and lower sidebands: measuring at the carrier frequency, the phase vectors of the sidebands rotate either clockwise or counterclockwise (higher or lower frequency).}
 \label{fig:Sidebands2}
\end{figure}

Let us go one step back to single mode squeezing. For amplitude squeezing, the situation is similar to the one in Fig.~\ref{fig:uncertainty}. The only difference is that the uncertainty band is reduced, resulting in sub-shotnoise photocurrent fluctuations. But there is an alternative way. Measuring the uncertainty or squeezing of a quantum state in the experiment can be realized with a spectral measurement on a spectrum analyzer. As mentioned before, the technical noise of the laser does not allow to measure close to the carrier frequency but above about 10\,MHz. Thus, we measure at a certain distance $f$ of the carrier frequency $\nu_0$ where an uncertainty equivalent to half a photon (zero point uncertainty) has to be assumed, in order to explain the shot noise as detailed below. In fact, two sidebands equally spaced above and below the carrier will contribute to the shot noise and equivalently we can easily imagine that quantum noise squeezing at the sidebands $\nu_0+f$ and $\nu_0-f$ arises from the interference of the carrier with the vacuum sidebands (see Fig.~\ref{fig:Sidebands1}). 
In Fig.~\ref{fig:Sidebands2}, the phase space of the carrier and two sidebands, which are shifted in frequency, are depicted. The fast rotation of the carrier is taken out and thus the phase space vectors of the two sidebands rotate in opposite directions, because one is at higher frequency and the other one at lower frequency than the carrier.

%% slide 32

\begin{figure}[h!]
 \center
 \includegraphics[height=160pt]{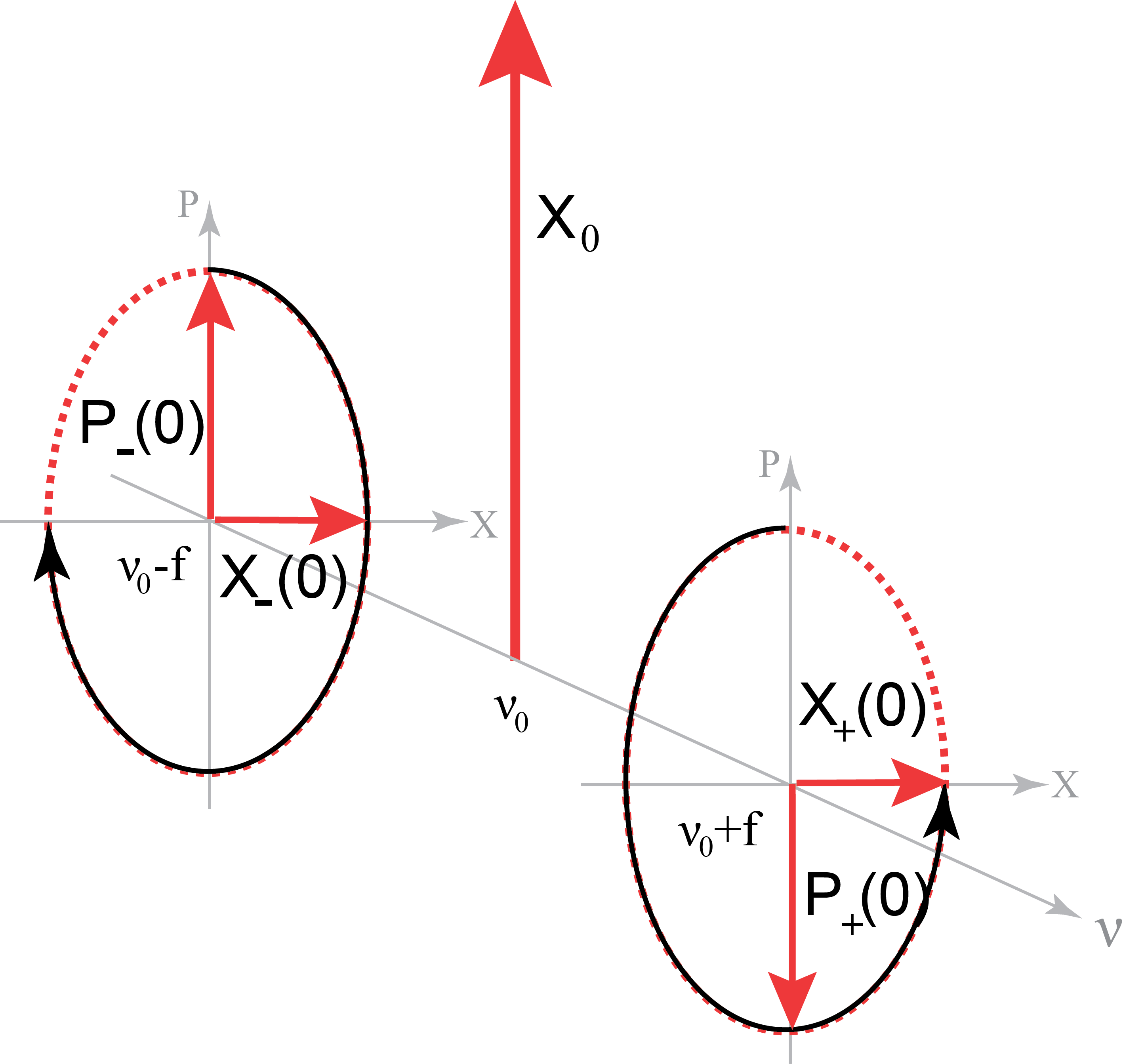}
 \caption{Correlation of phase space vector in two corresponding sidebands.}
 \label{fig:corrphasespa}
\end{figure}

The total electric field contains the carrier $\varepsilon_0$ and the upper and lower sideband with indices ``+'' and ``-''.
\begin{align}
 \varepsilon(t) \e^{\ii \omega_0 t} = \varepsilon_0 (t) \e^{\ii \omega_0 t} + \varepsilon_+ (t) \e^{\ii (\omega_0 + \Omega) t} + \varepsilon_- (t) \e^{\ii (\omega_0 - \Omega)t},
\end{align}
where $2\pi f \equiv \Omega$. The field operators read like:
\begin{align}
 a(t) =& a_0 + a_+ (t) \e^{\ii \Omega t} + a_- \e^{-\ii \Omega t}
 \end{align}
 and the canonical variables are (Fig.~\ref{fig:corrphasespa}):
 \begin{align}
 X_+(t) =& \frac{1}{2}\left( a \e^{\ii \Omega t} + a^\dagger \e^{-\ii \Omega t} \right)\\
 P_+(t) =& \frac{1}{2\ii}\left( a \e^{\ii \Omega t} - a^\dagger \e^{-\ii \Omega t} \right).
\end{align}
Analogous equations hold for $X_-$ and $P_-$.\par

%% slide 33
For $\Omega t = \pi/2$ we find
\begin{align}
 X_+\left(\frac{\pi}{2 \Omega}\right) =& \frac{1}{2} ( a \ii + a^\dagger (-\ii)) = -\frac{1}{2\ii} ( a - a^\dagger) = - P_+(0)\\
 P_+\left(\frac{\pi}{2 \Omega}\right) =& \frac{1}{2\ii} ( a \ii - a^\dagger (-\ii)) = X_+(0)\\
 X_-\left(\frac{\pi}{2 \Omega}\right) =& \frac{1}{2} ( a (-\ii) - a^\dagger \ii) = P_-(0)\\
 P_-\left(\frac{\pi}{2 \Omega}\right) =& \frac{1}{2\ii} ( a (-\ii) - a^\dagger \ii) = -X_-(0)
\end{align}
We find the total intensity with signal $X_0$ at $\omega_0$ as:

\begin{align}
 (X_0 + X_+(0)+X_-(0))^2 =& X_0^2 + 2X_0( X_+(0) + X_-(0)) + \ldots\\
 (X_0 + X_+\left(\frac{\pi}{2 \Omega}\right)+X_-\left(\frac{\pi}{2 \Omega}\right))^2 =& X_0^2 + 2X_0(- P_+(0) + P_-(0)) + \ldots
\end{align}
% 
%% slide 34
The noise is reduced if $X_+(0)+X_-(0) \rightarrow 0$ and $P_+(0)-P_-(0) \rightarrow 0$. In this case the two sidebands are entangled, in analogy to our discussion on the Einstein-Podolsky-Rosen paradox before. The result, possibly surprising at first sight, is that any single mode squeezed state is actually entangled. The entangled quantities are the two frequency sideband modes. This entanglement becomes apparent if one zooms into a spectral decomposition of the field. The correlation of sidebands in quantum optics was first proposed theoretically~\cite{Caves85} and was called ``two-mode squeezing''. Nowadays, the term ``two-mode squeezing'' is used for any modes that are correlated with each other, not only vacuum sidebands.

%% slide 35
\begin{figure}[h!]
 \center
 \includegraphics[height=90pt]{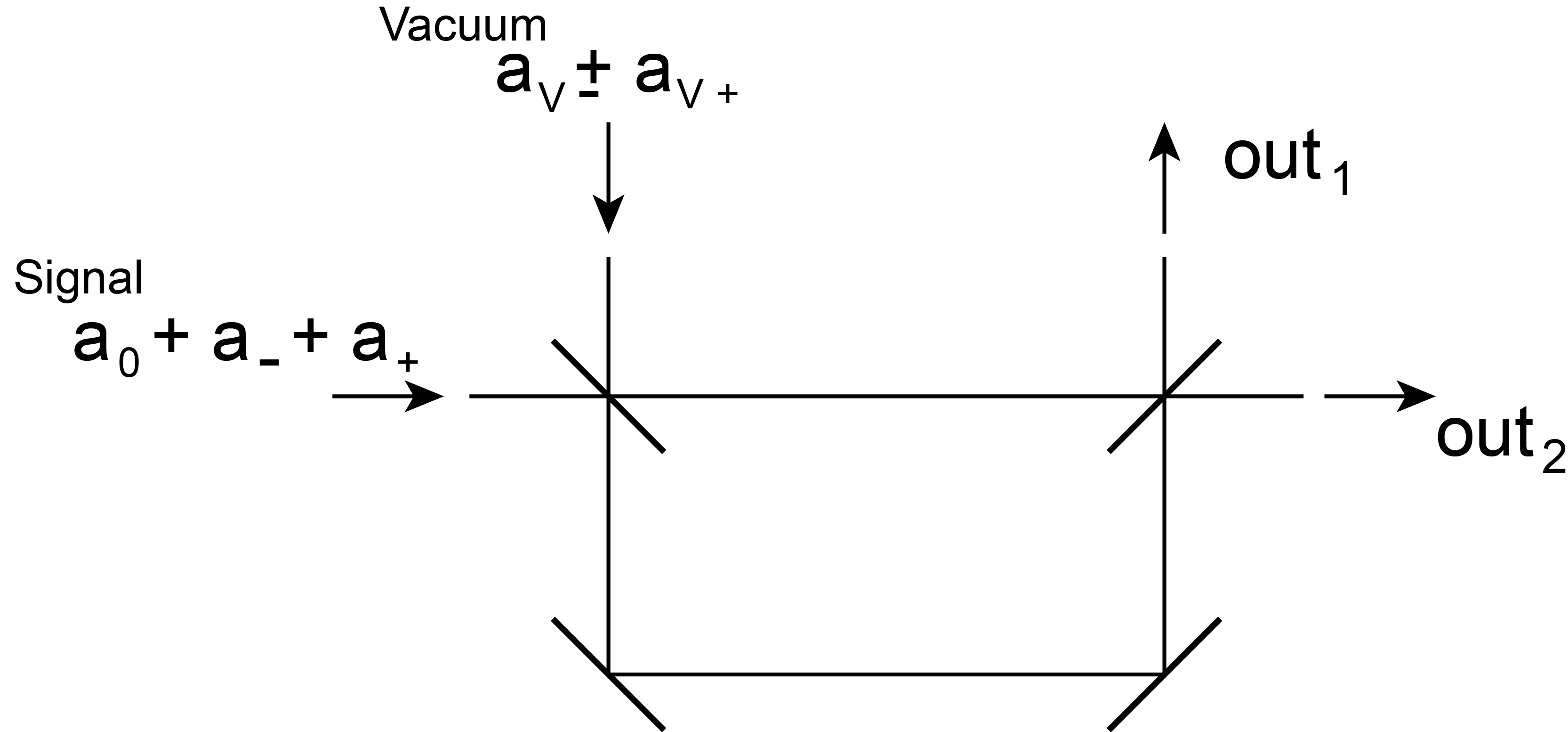}
 \caption{Setup of an unbalanced interferometer.}
 \label{fig:int2}
\end{figure}

There have been different approaches to measure the entanglement between these vacuum sidebands~\cite{Huntington,Hage}. The setup shown in Fig.~\ref{fig:int2}, an unbalanced interferometer, is one solution to measure the correlation of the sidebands. The signal, carrier plus the two entangled sidebands, is injected at one input of the first beam splitter and interferes with the vacuum state, entering the other input. We can write for the field at output 1 (and analogous for output 2):

\begin{align}
 \text{``out1''} =&\frac{1}{2} \left( a_0 + a_0 \e^{\ii \omega_{0} \frac{L}{c} } +\right.\\
 &+ (a_{+} + a_{V+}) + (a_{+} + a_{V+}) \e ^{\ii (\omega_{0} + \Omega)\frac{L} {c}}+\\
 &\left.+ (a_{-} + a_{V-}) + (a_{-} + a_{V-})\e ^{\ii(\omega_{0}-\Omega)\frac{L}{c}}\right)
\end{align}
where $L$ is the armlength difference. The goal is to separate both correlated sidebands at the outputs of the interferometer in order to measure the noise of the sidebands. Let us name the phase difference for the different frequencies as follows:
\begin{align}
 \Delta \phi = \Omega \frac{L}{c},\quad \Delta \varphi = \omega_0 \frac{L}{c}
\end{align}
Since $\Omega$ and $\omega_0$ are separated by several orders of magnitude, it is possible to choose the armlength $L$, up to a certain precision, such as to fulfill the conditions for constructive and destructive interference at the respective output:
\begin{align} 
 \Omega \frac{L}{c} =& \frac{\pi}{2} \rightarrow t=\frac{\pi}{2\Omega}\\
  \omega_0 \frac{L}{c} =& \frac{\pi}{2} + 2 \pi m, t=\frac{\pi}{2 \omega_0} + \frac{2 \pi m}{\omega_0},\quad m\in \N
\end{align}
The first condition determines L, the second equation $m$.

We can imagine that tuning the armlength $L$ will cause a fast rotation of the phase for $\omega_0$ and a slow rotation of the phase for $\Omega$. Therefore, we obtain at output 1:
%% slide 36
\begin{align}
 \text{``out 1''} =& \frac{1}{2} [(X_0 + X_+ + X_- + \ii P_+ + \ii P_- + \ldots ) + \\ 
 &\;\ii( X_0 - P_+ + P_- + \ii X_+ - \ii X_- + \ldots )]\\
 =& \frac{1}{2} [(X_0 + \ii X_0) +2X_- + 2\ii P_- + \ldots ]
\end{align}\par
\begin{figure}[h!]
 \center
 \includegraphics[height=90pt]{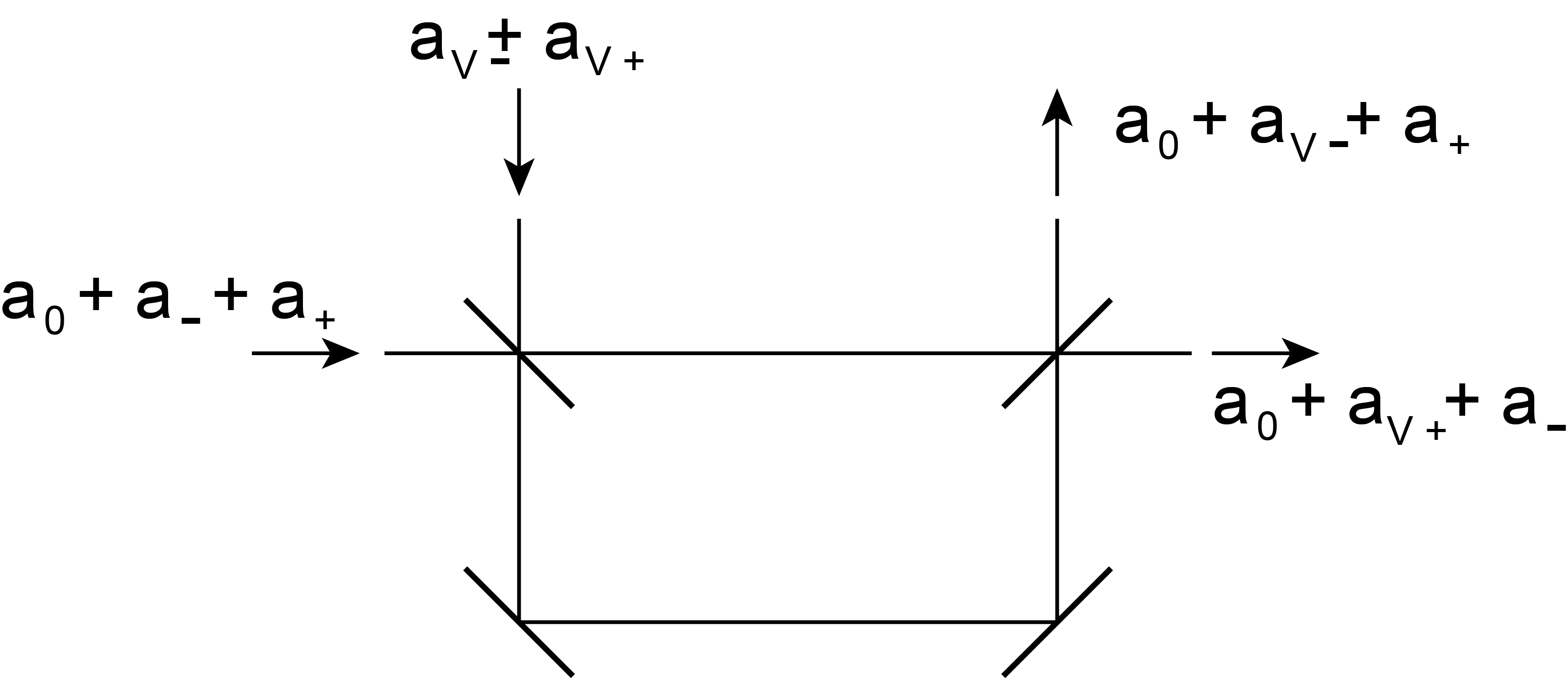}
 \caption{Setup of an unbalanced interferometer with separated sidebands at the outputs.}
 \label{fig:int1}
\end{figure}

The interferometer thus splits the sidebands coming from one input port, see Fig.~\ref{fig:int1}. It can be shown that the  noise for each of the separated entangled sidebands is higher than for the input beam with combined sidebands and even higher than it would be for a coherent state, which is an indicator for a correlation between the sidebands. Instead, if the difference signal of both outputs is measured, the noise is suppressed and even lower than the shot noise limit, hence we observe the squeezing as a result of the interplay between the two sidebands. We can keep in mind that subsystems of an entangled state possess each on its own increased noise and reveal the squeezed quantum noise only in the difference signal of both of them. When tracing over one subsystem, the other one is projected into a mixed state. Conversely, systems in pure states can never be entangled with another system.

\newpage
\section{Bogoliubov transformation}

In the last sections, we discussed the interaction at a beam splitter, where modes are split into several modes, or vice versa, and modes can interfere with each other. Generally speaking, the beam splitter introduces a Bogoliubov transformation. Several phenomena can be reduced to such a model, not only attenuation of a state, but also amplification of a state, the aforementioned squeezing and phase conjugation. In order to find a general formula for these two-mode interactions, we have a look at the different processes.

\subsection{Attenuation}
\begin{figure}
 \center
 \includegraphics[height=160pt]{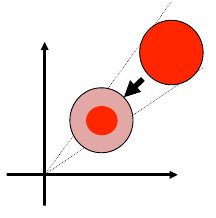}
 \caption{Attenuation: pure state $\rightarrow$ pure state.}
 \label{fig:att}
\end{figure}

Let us first consider the attenuation of a coherent state, which is a pure quantum state, depicted in phase space in Fig.~\ref{fig:att}. The coherent state has its uncertainty in both quadratures and while the amplitude of the state is reduced, the uncertainty does not decrease because of the Heisenberg uncertainty relation being state independent. The noise figure (NF) is defined as the relation of the signal-to-noise ratio (SNR) at the output to the SNR at the input:
\begin{align}
\text{NF}= \frac{\mathrm{SNR}_{\mathrm{out}}}{\mathrm{SNR}_{\mathrm{in}}} = \frac{\frac{\text{signal}_\text{out}}{\text{noise}_\text{out}}}{\frac{\text{signal}_\text{in}}{\text{noise}_\text{in}}}
\end{align}
and it increases for the attenuation of a pure, coherent state.

The field operator $\hat{a}$ corresponds to the electric field $E$ and for finding the field operator in case of attenuation, we can guess a new field operator $\hat{c} = \sqrt{\eta } \hat{a}$ in order to obtain a lower field amplitude. However, calculating the commutator gives:

\begin{align}
\left[ \hat{c}, \hat{c} ^{\dagger} \right] = \left[ \sqrt{\eta } \hat{a}, \sqrt{\eta } \hat{a}^{\dagger} \right] = \eta \left[ \hat{a}, \hat{a}^{\dagger} \right] = \eta \neq 1.
\end{align}
Thus $\sqrt{\eta } \hat{a}$ is not a valid field operator and one needs to add an ancillary operator $\hat{L}$:
\begin{align}
\hat{c}=\sqrt{\eta}\hat{a} + \hat{L}.
\end{align}
We derive again the commutator:
\begin{align}
\left[ \hat{c}, \hat{c} ^{\dagger} \right] =& \left[ \sqrt{\eta } \hat{a} + \hat{L}, \sqrt{\eta } \hat{a}^{\dagger} + \hat{L}^{\dagger} \right]\label{eqn:com_cc}\\
=&  (\sqrt{\eta} \hat{a} + \hat{L})( \sqrt{\eta} \hat{a}^{\dagger} + \hat{L}^{\dagger} ) - ( \sqrt{\eta} \hat{a}^{\dagger} + \hat{L}^{\dagger} ) ( \sqrt{\eta} \hat{a} + \hat{L}) \\
=& \eta \left[ \hat{a}, \hat{a} ^{\dagger} \right] +
\left[ \hat{L}, \hat{L} ^{\dagger} \right]=1
\end{align}
Under the condition:
\begin{align}
\left[ \hat{L}, \hat{L} ^{\dagger} \right]=1 - \eta,
\end{align}
$\hat{c}$ indeed fulfills the commutator relation required for a field operator.
It follows then, with $\eta < 1$, that the operator $\hat{L}$ is an annihilation operator and can be written as:

\begin{align}
\hat{L} =
\begin{cases}
 \sqrt{1-\eta} \cdot \hat{a} & \text{case 1},\\
 \sqrt{1-\eta} \cdot \hat{b} & \text{case 2}
\end{cases}
\end{align}
with $\hat{a}$ the same mode operator and $\hat{b}$ another mode operator. Case 1 is not possible because it cannot satisfy the commutator relation $\left[ \hat{c}, \hat{c} ^{\dagger} \right] = 1$ in general, except for $\eta = 0,1$. Hence, we can conclude that the field operator $\hat{c}$ in case of \textbf{attenuation} has to be written in the following form:
\begin{align}
\mathbf{Attenuation:} \quad \quad \hat{c}=\sqrt{\eta}\hat{a} + \sqrt{1-\eta}\hat{b}.
\end{align}

We note that attenuation comes always along with the interaction with a second mode, thus the noise addition by the second mode is the reason for the decrease of the signal-to-noise ratio~\cite{Fabre}. We also note that we recover the operator relation at the beam splitter.

% \vspace{40pt}
\subsection{Amplification and squeezing: phase insensitive and phase sensitive amplifier}

In case of amplification, we need to write the field operator $\hat c$ as before but with an amplification factor $G$:

\begin{align}
\hat{c}=\sqrt{G}\hat{a} + \hat{L}.
\end{align}
An ancilla mode operator is again needed for the same reasons as for the process of attenuation. The commutator can be derived analogously to Eq.~\eqref{eqn:com_cc}, with

\begin{align}
\left[ \hat{L}, \hat{L} ^{\dagger} \right]=1-G
\end{align}
and G$>$1, we rewrite:

\begin{align}
\left[ (\hat{L} ^{\dagger}),(\hat{L} ^{\dagger})^{\dagger} \right]=G-1
\end{align}

Thus, the operator $\hat{L} ^{\dagger}$ has to be an annihilation operator and $\hat{L}$ a creation operator and we can deduce again two cases~\cite{Caves}

\begin{align}
\hat{c} = \sqrt{G} \hat{a} + \hat{L}, \quad \hat{L} = \left\{ 
\begin{matrix}
\sqrt{G-1} \cdot b^{\dagger}, \quad\text{case 1} \\
\sqrt{G-1} \cdot a^{\dagger}, \quad\text{case 2}
\end{matrix}
\right\}
\end{align}
The operators $a^{\dagger}$ and $b^{\dagger}$ are creation operators, $\hat{a}$ the same optical mode as the signal and $\hat{b}$ another optical mode. Both cases are in principle possible as the commutator relation can be satisfied. Let us start with case 1, where an interaction with a second mode $\hat{b}$ takes place.

\subsubsection{Case 1}~\\
\begin{figure}[h!]
 \center
 \includegraphics[height=130pt]{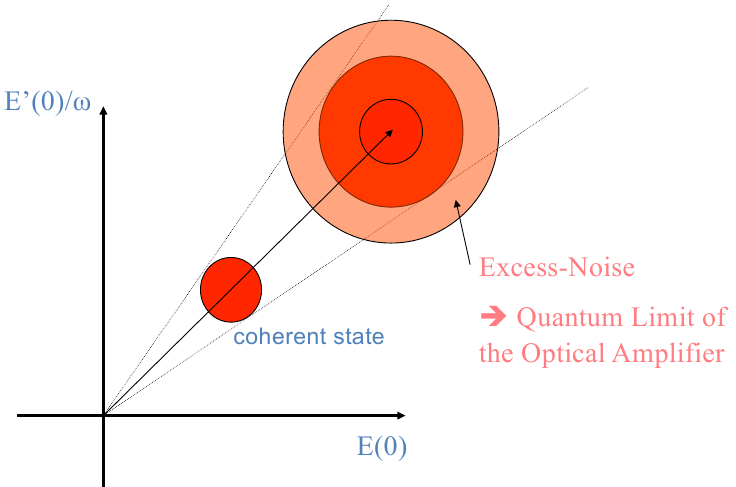}
 \caption{Amplification with a phase-insensitive amplifier.}
 \label{fig:ampins}
\end{figure}
For the field operator $\hat{a}$, we give the index ``a'' and remember the relation to the canonical phase space quadrature operators:
\begin{align}
 \hat{X}_a = \frac{1}{2} (\hat{a} + \hat{a}^{\dagger}), \quad \hat{P}_a = \frac{1}{2i} (\hat{a} - \hat{a}^{\dagger}).
\end{align}
For the field operator~\cite{Fabre}

\begin{align}
\hat{c}=\sqrt{G} \hat{a} + \sqrt{G-1}\hat{b}^{\dagger},
\end{align}
we can write for the continuous variable $\hat{X}_c$ and its square:
\begin{align}
\hat{X}_c =& \frac{1}{2} (\sqrt{G}(\hat{a} + \hat{a}^{\dagger}) + \sqrt{G-1}(\hat{b} + \hat{b}^{\dagger})) =  \sqrt{G} \hat{X}_a + \sqrt{G-1} \hat{X}_b 
\end{align}
\begin{align}
\hat{X}_c^2 = G \hat{X}_a^2 + 2\sqrt{G(G-1)}\hat{X}_a \hat{X}_b + (G-1) \hat{X}_b^2
\end{align}
For the mean values follows:
\begin{align}
\gen{\hat{X}_c^2 }=G \gen{\hat{X}_a^2 } + (G-1) \gen{\hat{X}_b^2} + 2 \sqrt{G(G-1)} \gen{\hat{X}_a} \gen{\hat{X}_b}
\end{align}
\begin{align}
\gen{ \Delta \hat{X}_c^2 } &= G (\gen{\hat{X}_a^2 }-\gen{\hat{X_a}}^2) + (G-1) (\gen{\hat{X}_b^2 }-\gen{\hat{X_b} }^2)
+\\
&+ 2 \sqrt{G(G-1)} \gen{\hat{X}_a} \gen{\hat{X}_b} - 2 \sqrt{G(G-1)} \gen{\hat{X}_a} \gen{\hat{X}_b}
\end{align}
\begin{align}
 \gen{\Delta \hat{X}_c^2 }=&G \gen{\Delta \hat{X}_a^2 } + (G-1) \gen{\Delta \hat{X}_b^2 }
\end{align}
with $\gen{\hat{X}_b}=0$. Analogous, we find for the other quadrature $\Delta \hat{P}_c$:

\begin{align}
\gen{\Delta \hat{P}_c^2 }=&G \gen{\Delta \hat{P}_a^2 } + (G-1) \gen{\Delta \hat{P}_b^2 }
\end{align}
Obviously, there is excess noise in both quadratures, see Fig.~\ref{fig:ampins}. The SNR before amplification is the signal over the noise:

\begin{align}
 \text{SNR}_{\mathrm{in}} = \frac{\text{signal}_\text{in}}{\text{noise}_\text{in}} = \frac{\gen{\hat{X_a}}^2}{\gen{\Delta \hat{X}_a^2 }}
\end{align}
The noise figure for this process, as defined before, can then be found as (and analogous for $\hat{P_a}$):

\begin{align}
\text{NF}=\frac{G \gen{\hat{X_a}}^2}{G\gen{\Delta \hat{X}_a^2 }+(G-1)\gen{\Delta \hat{X}_b^2 }}\cdot \frac{\gen{\Delta \hat{X}_a^2 }}{\gen{\hat{X_a}}^2}=\frac{G}{2G-1} \quad \text{if} \quad \Delta \hat{X_b^2}=\Delta \hat{X_a^2}
\end{align}
The condition applies if the signal is a coherent state and the ancillary mode is in the vacuum state. For high values of $G$, the noise figure tends to:

\begin{align}
\text{NF} \geq \frac{1}{2}
\end{align}
This is known as the 3\,dB quantum limit of an amplifier and means that noise is always added to both quadratures, as long as the amplification is phase insensitive. Thus, this case is called the \textbf{phase-insensitive amplifier}. It can be described with:

\begin{align}
\mathbf{Amplification:} \quad \quad \hat{c}=\sqrt{G} \hat{a} + \sqrt{G-1}\hat{b}^{\dagger},
\end{align}

For low gain $G>1$ the quantum limited noise figure is closer to one. It turns out that for continuous quantum variables, the optimal cloning operation, i.e. two clones of one original, can be implemented by a quantum optimized $G=2$ - amplifier, followed by a beam splitter. After being proposed theoretically ~\cite{Braunstein,Fiurasek} optimal cloning was implemented experimentally~\cite{Andersen}.
Quantum limited amplification was also demonstrated using a quantum electro-optic feed forward amplifier~\cite{Josse} based on the development of more general quantum electro-optic feed forward techniques~\cite{Lam}.

\subsubsection{Case 2}~\\
In the second case, we discuss the interaction with the complex conjugate of the same mode $\hat{a}^{\dagger}$:
\begin{align}
\hat{c}=\sqrt{G} \hat{a} + \sqrt{G-1}\hat{a}^{\dagger}
\end{align}
In order to obtain the noise figure NF, we derive again $\hat{X}_c$, $\gen{\hat{X}_c^2 }$ and $\gen{ \Delta \hat{X}_c^2 }$:

\begin{align}
\hat{X}_c =& \frac{1}{2} (\sqrt{G}(\hat{a} + \hat{a}^{\dagger}) + \sqrt{G-1}(\hat{a} + \hat{a}^{\dagger}))\\
          =& (\sqrt{G} + \sqrt{G-1}) \hat{X}_a\\
\gen{\hat{X}_c^2 }=&\left[ G + (G-1) + 2 \sqrt{G(G-1)} \right] \gen{\hat{X}_a^2 }\\
\gen{ \Delta \hat{X}_c^2 }=& \left[ G + (G-1) + 2 \sqrt{G(G-1)} \right] \gen{ \Delta \hat{X}_a^2 }\\
=& (\sqrt{G}+\sqrt{G-1})^2 \cdot \gen{ \Delta \hat{X}_a^2 }
\end{align}
The analogous expression holds for the other quadrature $\hat{P}_c$:

\begin{align}
\hat{P}_c =& (\sqrt{G} - \sqrt{G-1})\hat{P}_a\\
\gen{\hat{P}_c^2 }=&(\sqrt{G} - \sqrt{G-1})^2 \gen{\hat{P}_a^2 }\\
\gen{\Delta \hat{P}_c^2 } =& (\sqrt{G} - \sqrt{G-1})^2 \gen{\Delta \hat{P}_a^2 }
\end{align}

We are now able to calculate the noise figure for both quadratures and we obtain for
the quadrature $\hat{X}_c$:

\begin{align}
\text{NF}= \frac{(\sqrt{G} + \sqrt{G-1})^2 \gen{\hat{X_a}}^2}{(\sqrt{G} + \sqrt{G-1})^2 \gen{\Delta \hat{X}_a^2 }} \cdot \frac{\gen{\Delta \hat{X}_a^2 }}{\gen{\hat{X_a}}^2}=1
\end{align}
and the quadrature $\hat{P}_c$:
\begin{align}
\text{NF}= \frac{(\sqrt{G} - \sqrt{G-1})^2 \gen{\hat{P_a}}^2}{(\sqrt{G} - \sqrt{G-1})^2 \gen{\Delta \hat{P}_a^2 }} \cdot \frac{\gen{\Delta \hat{P}_a^2 }}{\gen{\hat{P_a}}^2}=1.
\end{align}

\begin{figure}[h!]
 \center
 \includegraphics[height=140pt]{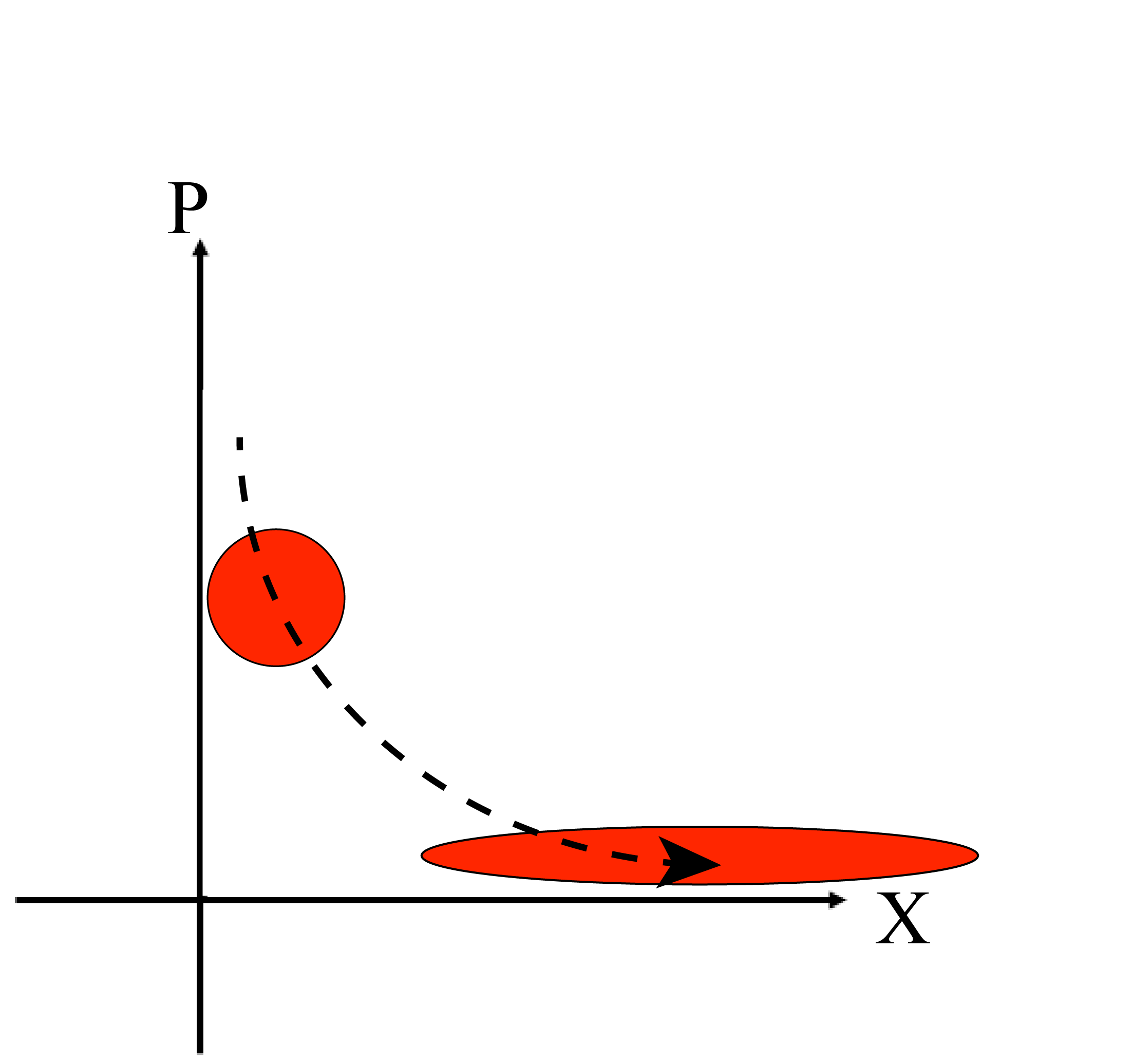}
 \caption{Amplification with a phase-sensitive amplifier.}
 \label{fig:ampsens}
\end{figure}

For this case, no additional noise is added to the state during amplification, the SNR remains the same throughout the process. However, this does not mean that the noise is the same in both quadratures (this would be nonphysical). In order to understand this fact, we look on the mean values of both quadratures:\\
Quadrature $\hat{X}_c$:
\begin{align}
\gen{\hat{X}_c}=(\sqrt{G}+\sqrt{G-1})\gen{\hat{X}_a}
\end{align}
Quadrature $\hat{P}_c$:
\begin{align}
\gen{\hat{P}_c}=(\sqrt{G}-\sqrt{G-1})\gen{\hat{P}_a}
\end{align}

For $\hat{X}_c$, the amplitude is increasing, the state is amplified. But since the SNR remains the same, the noise is also increasing. For $\hat{P}_c$, the opposite is the case. For high $G$, the amplitude tends to the minimum. Also here, the SNR does not change, which means that the noise in the direction of the quadrature $\hat{P}_c$ is reduced, even smaller than the shot noise limit (Fig.~\ref{fig:ampsens})! This is known as a \textbf{phase-sensitive amplifier} and corresponds to \textbf{squeezing}, as discussed before in Chapter 3. The general description is then:

\begin{align}
\mathbf{Squeezing:} \quad \quad \hat{c}=\sqrt{G} \hat{a} + \sqrt{G-1}\hat{a}^{\dagger}.
\end{align}

\subsection{Phase conjugation}

In order to obtain an operator for phase conjugation, we can assume the following form, where the use of $\hat{a}^{\dagger}$ is required:
\begin{align}
\hat{c}=\sqrt{G} \hat{a}^{\dagger} + \hat{L}
\end{align}
We use again the commutator relation
\begin{align}
\left[ \hat{c}, \hat{c}^{\dagger} \right] = G \cdot \left[ \hat{a} ^{\dagger}, \hat{a} \right] + \left[\hat{L}, \hat{L}^{\dagger} \right] = 1
\end{align}
With $[\hat{a}^{\dagger},\hat{a}]=-1$ and  $[\hat{c},\hat{c}^{\dagger}]=1$, it is:

\begin{align}
\left[\hat{L}, \hat{L}^{\dagger} \right] = G + 1
\end{align}
The ancilla operator $\hat{L}$ is thus an annihilation operator and so we can write:

\begin{align}
\hat{L}=\sqrt{G+1} \hat{b}
\end{align}
As before, we have also to take into account the possibility to use the same mode, such as $\hat{L}=\sqrt{G+1} \hat{a}$. But calculating the commutator $\left[ \hat{c}, \hat{c}^{\dagger} \right]$ for this case, will give -1, hence we need a second independent mode and we can write for the process of \textbf{phase conjugation} the following transformation~\cite{cerf}
\begin{align}
\mathbf{Phase}\text{ }\mathbf{conjugation:} \quad \quad \hat{c}=\sqrt{G} \hat{a}^{\dagger} + \sqrt{G+1}\hat{b}.
\end{align}
Again, we can calculate the variance of the quadrature operators at the output and obtain the following noise figure NF:
\begin{align}
\text{NF}=\frac{G}{2G+1}
\end{align}

\subsection{General form of the Bogoliubov transformation}

In the last paragraphs, we expressed the different processes, such as attenuation, amplification, squeezing and phase conjugation with the respective mode operators. We can combine them into one transformation, which is called the \textbf{Bogoliubov transformation} for two modes:

\begin{align}
\hat{c} = \alpha_1 \cdot \hat{a} + \alpha_2 \cdot \hat{b} + \alpha_3 \cdot \hat{a}^{\dagger} + \alpha_4 \cdot \hat{b}^{\dagger}
\end{align}
The respective processes are:

\begin{figure}[h!]
 \center
 \includegraphics[height=80pt]{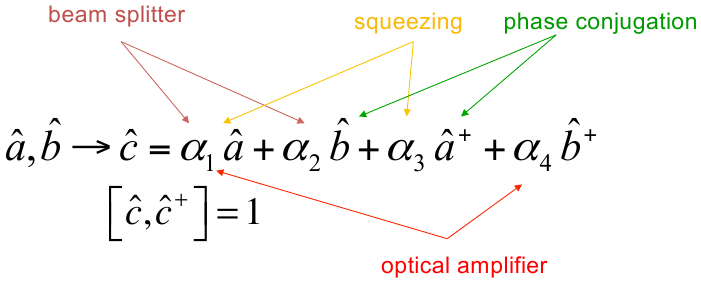}
\end{figure}

\begin{figure}[h!]
 \center
 \includegraphics[height=160pt]{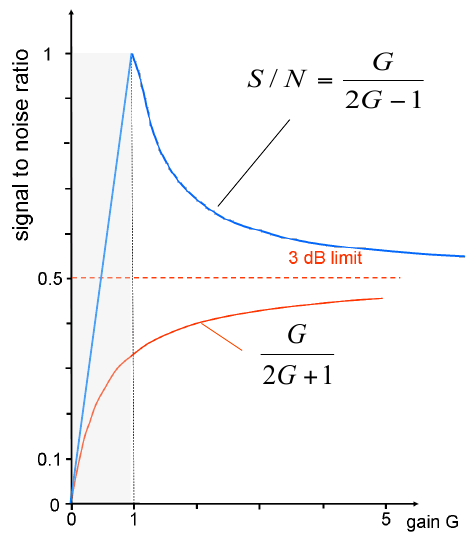}
 \caption{The blue line shows the behavior of $S/N$ over the gain for the ideal amplifier, the red line the performance of a phase conjugating mirror which
 equals the performance of the amplifier ``destructive measurement \& recreation''.}
 \label{fig:last}
\end{figure}

Plotting the noise figures as a function of gain for the different processes gives some insight (Fig.~\ref{fig:last}). When measuring the Q-function of a signal state (Fig.~\ref{fig:heteroB}),
one can amplify electronically the photocurrent measured simultaneously for the two quadratures $\hat X$ and $\hat P$ and modulate a laser with these amplified quadratures. This is the technique used
in electronic repeater stations in optical telecommunication. It is interesting to note that the noise figure for this electronic amplification of optical signals is identical to the noise
figure of the phase conjugation. It is furthermore worth noting that the limiting performance of the optimum optical quantum amplifier for large gain is essentially the same as the one of the electronic amplifier.

\end{document}